\documentclass[fleqn,usenatbib]{mnras}

\usepackage{newtxtext,newtxmath}
\usepackage{stackengine}
\usepackage{tikz}
\usepackage[export]{adjustbox}
\usepackage{ulem}

\usepackage{caption}

\usepackage{subfigure}

\usepackage[T1]{fontenc}
\usepackage{hyperref}

\DeclareRobustCommand{\VAN}[3]{#2}
\let\VANthebibliography\thebibliography
\def\thebibliography{\DeclareRobustCommand{\VAN}[3]{##3}\VANthebibliography}

\usepackage{graphicx}	
\usepackage{subfigure}
\usepackage{mwe}
\usepackage{amsmath}	

\title[GRB Classifier]{GRB Optical and X-ray Plateau Properties Classifier Using Unsupervised Machine Learning}

\author[Bhardwaj et al.]{Shubham Bhardwaj$^{1,2}$, Maria G. Dainotti$^{1,2,3}$\thanks{E-mail: maria.dainotti@nao.ac.jp}, 
Sachin Venkatesh$^{4}$, 
Aditya Narendra$^{5,6}$, 
Anish Kalsi$^{7}$, 
\newauthor Enrico Rinaldi,$^{8}$ 
and Agnieszka Pollo$^{5,9}$\thanks{The first four authors have equally contributed to the paper}
\\
\\
$^{1}$National Astronomical Observatory of Japan, 2 Chome-21-1 Osawa, Mitaka, Tokyo 181-8588, Japan\\
$^{2}$The Graduate University for Advanced Studies, SOKENDAI, Shonankokusaimura, Hayama, Miura District, Kanagawa 240-0193, Japan\\
$^{3}$Space Science Institute, 4765 Walnut St Ste B, Boulder, CO 80301, USA\\
$^{4}$School of Physics, Georgia Institute of Technology, North Avenue, Atlanta, GA 30332, USA\\
$^{5}$Astronomical Observatory of Jagiellonian University, Orla 171, 30-244 Krakow, Poland\\
$^{6}$Jagiellonian University, Doctoral School of Exact and Natural Sciences, 30-348 Krakow, Poland\\
$^{7}$Delhi Technological University, New Delhi, 110042, India\\
$^{8}$Interdisciplinary Theoretical \& Mathematical Science Program, RIKEN (iTHEMS),
2-1 Hirosawa, Wako, Saitama, Japan 351-0198\\
$^{9}$National Centre for Nuclear Research, 02-093 Warsaw, Poland\\
}

\date{Accepted XXX. Received YYY; in original form ZZZ}

\pubyear{2023}

\begin{document}
\label{firstpage}
\pagerange{\pageref{firstpage}--\pageref{lastpage}}
\maketitle

\begin{abstract}

The division of Gamma-ray bursts (GRBs) into different classes, other than the ``short” and ``long”, has been an active field of research. We investigate whether GRBs can be classified based on a broader set of parameters, including prompt and plateau emission ones. Observational evidence suggests the existence of more GRB sub-classes, but results so far are either conflicting or not statistically significant. The novelty here is producing a machine-learning-based classification of GRBs using their observed X-rays and optical properties. We used two data samples: the first, composed of 203 GRBs, is from the Neil Gehrels Swift Observatory (Swift/XRT), and the latter, composed of 134 GRBs, is from the ground-based Telescopes and Swift/UVOT. Both samples possess the plateau emission (a flat part of the light curve happening after the prompt emission, the main GRB event). We have applied the Gaussian Mixture Model (GMM) to explore multiple parameter spaces and sub-class combinations to reveal if there is a match between the current observational sub-classes and the statistical classification. With these samples and the algorithm, we spot a few micro-trends in certain cases, but we cannot conclude that any clear trend exists in classifying GRBs. These microtrends could point towards a deeper understanding of the physical meaning of these classes (e.g., a different environment of the same progenitor or different progenitors). However, a larger sample and different algorithms could achieve such goals. Thus, this methodology can lead to deeper insights in the future.

\end{abstract}




\section{Introduction}\label{sec:intro}

Gamma-ray bursts (GRBs) are highly energetic and luminous electromagnetic phenomena observed up to a very high redshift, $z$ $\sim$ 9.4 \citep{2011ApJ...736....7C}, and can be used as powerful probes of the high-$z$ Universe. Historically, GRBs are classified into short (SGRB) and long (LGRB), based on $T_{90}$ \citep{1993ApJ...413L.101K}, the prompt emission duration, namely the time in which 90$\%$ of the counts are detected from 5\% to 95\% of the total emission. SGRBs have $T_{90}$ $<$ 2 seconds, and LGRBs have $T_{90}$ $>$ 2 seconds. SGRBs are believed to originate from the merger of two compact objects: two neutron stars (NS-NS) or a neutron star and a black hole (NS-BH) \citep{1976PhDT.........3L, 1992ApJ...395L..83N, 1992ApJ...392L...9D, 1992Natur.357..472U, 1994MNRAS.270..480T, 2007PhR...442..166N, 2017ApJ...846L...5G, 2017ApJ...848L..14G, 2017ApJ...848L..12A}. On the other hand, LGRBs may originate from the collapse of massive stars \citep{woosley1993ApJ...405..273W, woosley1993, paczynski1998ApJ...494L..45P, Woosley2006ARA&A}. These two main classes have been explored in detail, but a plethora of observational sub-classes have also been identified in the literature. Here we list the classes and their connection with the progenitor or with the environment:

\begin{enumerate}
\item Intrinsically SGRBs (IS, having $T_{90}$/(1 + $z$) $<$ 2 seconds, \citep{Rossi2002, Levesque2010, Zhang2021NatAs}).

\item Very long GRBs (VL, having  $T_{90} > $ 500 seconds, \citep{Levan2014}).

\item SGRBs exhibiting extended emission (SEE, having $T_{90}$ $>$ 2 seconds like LGRBs, while simultaneously possessing spectral properties akin to those SGRBs, \citep{Norris2000, Norris2006, Levan2007, Norris2010}). It has been discussed by \cite{barkov2011model, Bucciantini2012MNRAS.419.1537B, 2020ApJ...896..166R, dichiara2021, Zhang2021NatAs} that the SEEs may originate from the same progenitor as the SGRBs. A plausible model for the extended emission is the bipolar outflow in rapidly spinning proto-magnetars formed after surviving the NS-NS merger.

\item SGRBs associated with a kilonovae (KNe), hereafter GRB-KNe, \citep{Valenti_2017, 2018ApJ...860...62G, 2020NatAs...4...77J, 2022Natur.612..223R}). The KNe suggests that the progenitor is a compact object merger, indicating that merger events can produce heavy elements because a shower of neutrons is produced in these collisions \citep{Tanvir_2017, 2017Natur.551...67P, 2017Natur.551...71T, Rossi2020, 2022Natur.612..223R}.

\item Ultra-long GRBs (UL, having $T_{90}$ $>$ 1000 seconds \citep{2004ApJ...611.1005G, stratta2013ultra, Gendre2013, Nakauchi2013ApJ...778...67N, zhang2014long,  Greiner2015Natur, Kann2018, Gendre2019}) are thought to originate from the low metallicity blue supergiant or the collapse of stars with much larger radii than those attributed to GRB progenitors \citep{Gendre_2013, Levan2014, Piro2014, Boër_2015, 2018ApJ...859...48P,aloy2021MNRAS.500.4365A}. Blue supergiants are the expected end lives of Population III stars with low metallicity (less than 10$^{-4}$). Thus, Population III stars could serve as the progenitor for UL, providing a unique opportunity to study these elusive objects. Indeed, according to \cite{Piro2014}, ULs usually originate in low-density wind environments of Population III-like progenitors and are surrounded by thermal cocoons of plasma. Another possibility is that the origin of UL should account for a substantial amount of available mass, distributed in a manner that can reproduce their very long duration \citep{Gendre_2013}. \cite{Levan2014} conclude that the collapse of stars of large radii, such as H-rich supergiants, may naturally explain ULs. This is the reason why in this paper, we also look at possible multiple clustering of ULs. 

\item X-ray flashes/X-ray rich (XRF/XRR, both are sub-classes of LGRBs with a greater value of Fluence in X-rays than in $\gamma$-rays, with XRFs having a greater X-ray Fluence compared to XRRs \citep{soderberg2004redshift, 2007PASJ...59..695A, chincarini2010MNRAS.406.2113C, 2016ApJS..224...20Y, bi2018statistical, 2019ApJ...884...59L}). One possibility for the origin of XRF/XRR could be that these GRBs are seen off-axis of the collimated GRB jet emission \citep{2009A&A...499..439G, 2019ApJ...871..123F}. However, whether a GRB should be off-axis to be an XRF is unclear. An example of a nearly on-axis XRF is the case of GRB 060218, which has a very wide jet opening angle of 80$^\circ$ \citep{dainotti2017a}. A counterexample is a case of GRB 080710 with the observational evidence of the jet off-axis \citep{2009A&A...508..593K}, but it is classified as a regular LGRB. So again, a clear distinction from a physical point of view has not been pointed out yet. On the other hand, we anticipate that some micro-trends that have been found show that XRR clusters together with regular LGRBs in our analysis.

\end{enumerate}

All these classes enumerated above are also important because of how to distinguish if a particular GRB obeys a correlation (e.g., the one between the isotropic X-ray luminosity at the end of the plateau emission and the rest frame time at the end of the plateau emission) becomes tighter when a given class of sub-samples is considered \citep{Dainotti2010}. In this regard, we can mention that, for example, the GRB-KNe are all placed below the GRB fundamental plane relation between the luminosity at the end of the plateau emission, the time at the end of the plateau emission in the rest frame, and the peak prompt luminosity \citep{dainotti2020a}. The reason why this happens is still debated, but again a completely independent, unsupervised method that accounts for the observed properties of the plateau in principle can help to solve this debate. Indeed, if, through unsupervised clustering, we find that these properties are emerging independently from the classes, this means that the plateau emission is a key property for revealing the classes in a completely independent way \citep{2014ApJ...785...74L, 2021ApJ...907...78F}. Another huge current debate in the literature is the secure association of GRBs with a supernova (SNe) of Type Ib/c \citep{1998Natur.395..663K, 2006Natur.442.1008C, 2006Natur.442.1014S, Kaneko_2007, dainotti2017a}. It is believed that all LGRBs must have an associated SNe \citep{Woosley2006ARA&A}. However, \cite{2014A&A...567A..29M} pointed out that a significant number of GRB-SNe associations at high-$z$ ($z$$\geq$1) remained unobserved due to the lack of sensitive search for brightening of the late GRB afterglow light curve (LC) due to the emergence of the SNe component in the overall emission. \cite{2006Natur.442.1014S, Guetta2007ApJ...657L..73G, 2022ApJ...938...41D, Rossi2022ApJ...932....1R} reported that only a 2-7\% of GRBs associated with SNe have been observed in connection to LGRBs. It is often discussed by \citet{2006Natur.444.1050D} and \citet{Fynbo2006} that there exists a third class of events for which the SNe Ib/c is not intrinsically associated with the GRBs because there is evidence of cases for which the association would have necessarily been seen (see the case of GRB 060614 and GRB 060505; \citep{Fynbo2006}). However, we have not seen the SNe associated with these two events. Because of the possibility of this third class, in this paper, we have also explored the combination of parameters in which the class of GRB-SNe has been considered separately from the regular, canonical LGRBs. In addition, to be even more cautious about the strength of the GRB-SNe association, we have considered additional subclasses. We rely on the \citet{2012grb..book..169H} classification, for which we can further divide the GRB-SNe classes in A, B, C, D, and E according to their spectroscopic association with the SNe Ib/c:

\begin{enumerate}
    \item SNe-A: strong spectroscopic evidence supports the connection between the observed SNe and the GRB.
    \item SNe-B: the LC displays a noticeable bump, and some spectroscopic evidence indicates an association between LGRB and SNe.
    \item SNe-C: a clear bump on the LC is consistent with the association between LGRB and SNe, but no spectroscopic confirmation exists.
    \item SNe-D: the LC shows a significant bump, but the inferred properties of the SNe are not entirely consistent with other LGRB-SNe associations, the bump is inadequately sampled, or the spectroscopic redshift of the GRB is unknown.
    \item SNe-E: the LC shows a bump of low significance or is inconsistent with other LGRB-SNe identifications, but there is a spectroscopic redshift of the GRB.
\end{enumerate}

It is important here to stress that a correlation between the luminosity of the plateau emission and its rest-frame duration has a slope of $-1.9 \pm 0.3$ when the A and B classes are considered. Thus, this slope is different from $-1.0 \pm 0.1$ when all LGRBs without the GRB-SNe are considered and different from $-1.5 \pm 0.3$ when all classes of GRB-SNe Ib/c are used. Motivated by these results, we decided to also investigate these sub-classes as separate from the rest \citep{dainotti2017a}.

In addition, from the theoretical point, the breakdown of the canonical X-ray LC yields four distinct power-law phases: an initial abrupt decay; a following, shallower than usual decay; a normal decay; and a late steeper decay. Although the very steep, normal, and late steep decays are typically linked to the end of the prompt phase, the standard synchrotron forward-shock model and the jet break can explain the existence of the plateau emission. Indeed, the plateau phase has been interpreted in a variety of ways, including energy injection \citep{2015ApJ...806..205D, 2006MNRAS.369.2059P}, stratified density environments \citep{Granot2006}, structured jets with different angular distributions \citep{2018PhRvL.120x1103L}, off-axis emission, variations in microphysical parameter values \citep{2020ApJ...896...25F, 2022ApJ...940..189F}, magnetic dissipation \citep{2007MNRAS.378.1043J}, dust scattering \citep{2007ApJ...660.1319S}, and gravitational microlensing \citep{2021ApJ...921L..30V}. A sub-relativistic explanation for the plateau phase has also been proposed (Cocoons, shock breakouts, etc.; \citep{2019ApJ...871..123F}).

The plateau phase, which does not satisfy the closure relations of the standard synchrotron model, is usually interpreted as continuous energy injection from a central engine - either the fallback accretion of matter onto a black hole \citep{2008MNRAS.388.1729K, 2009ApJ...700.1047C, 2011ApJ...734...35C} or the spin-down luminosity from a newborn magnetar \citep{1998ApJ...496L...1R, Sari_2000, 2001ApJ...552L..35Z, 2006MNRAS.372L..19F}.

Here, we address the problem of classes in terms of the observed and one rest-frame feature ($T^*_{90}$), and we attempt to clarify which features can highlight peculiar differences among classes and to what extent the imbalance in the number of sources in relation to these classes and the paucity of the sample can influence our results. With this very detailed classification with our unsupervised machine learning (ML) method, we strive to connect the classes and check if there is a trend in the clustering pointing to similar progenitors or the same progenitor with different environments. A possible explanation for the existence of different classes can be attributed to each GRB's underlying mechanism of origin. Thus, based on the GRB progenitors, \citet{Zhang2004IJMPA..19.2385Z, Zhang2009, zhang2014long} proposed that there are two types of GRBs: Type I and Type II. Type II includes LGRBs, XRFs, and ULs and is produced by the collapse of massive stars \citep{woosley1993ApJ...405..273W, woosley1993}. Type I contains SGRBs, SEEs, and ISs and originates from the mergers of compact binaries \citep{1976PhDT.........3L, 2007PhR...442..166N, 2017ApJ...848L..12A}. The novelty of our approach is that we here use the plateau emission, which has not been used previously in the literature.  The previous method adopted in the literature primarily relies on the prompt emission properties, which are summarized in the section below.

This paper is structured as follows. Section \ref{previous methods} discusses previous attempts in the literature, both with supervised and unsupervised ML methods. Section \ref{data sample} outlines the data sample in X-ray and optical. Section \ref{methodologies} details the numerical setup and the algorithm used for clustering. In Section \ref{results}, we summarize the results, and in Section \ref{discussion}, we discuss them. Finally, in Section \ref{summary and conclusion}, we draw our conclusions, and we highlight this analysis's utility for future studies.

\section{Previous methods adopted in the literature}\label{previous methods}

Here, we present a summary of the current methods found in the literature that focus specifically on the properties of prompt emissions in GRBs. The various conclusions about GRB classes and subclasses discussed below are drawn with both supervised and unsupervised ML methods. In the supervised ML approach, the utilization of labeled data is essential to train ML models, allowing them to discern underlying relationships within the data and make predictions accordingly. This approach is commonly employed for classification and regression tasks. On the other hand, unsupervised ML methods operate without the need for labeled data during training and are primarily utilized for identifying patterns in unlabelled data, such as clustering. Compared to other papers, the primary novelty in our study lies in the inclusion of plateau emission properties from X-rays and optical GRB LCs to classify GRBs using GMM. Furthermore, we explore a multi-dimensional space by considering 11 parameters for X-ray data and 10 parameters for the optical data that consist of both plateau and prompt emission properties. Indeed, the maximum number of parameters explored in the previous research has been three or four. These were mainly prompt emission properties such as $(T_{90}$), \textit{Fluence}, Peak flux, and spectral hardness \citep{2017MNRAS.469.3374C, chattopadhyay2018, 2019ApJ...887...97T}.

\subsection{Application of Supervised ML}

Here, we detail the evidence in the data of three classes. The most important property of prompt emission investigated extensively in the literature is $T_{90}$. Of notable mention is \citet{1998ApJ...508..757H}, who proposed the existence of a third ``intermediate'' class of GRBs which have a duration between 2 and 10 seconds. They reached this conclusion by fitting with three Gaussians the $\log(T_{90})$ distribution of 797 GRBs in the Burst and Transient Source Explorer (BATSE) 3B catalog (1991 April - 1994 September, \citep{1996ApJS..106...65M}). They compared the results with the $\chi^2$  probability and assessed the existence of a third class at the 98 $\%$ probability level. This conclusion was confirmed again by  \citet{horvath2002}, but using an extended sample of 1929 GRBs from the BATSE 4Br catalog (1991 - 2000, \citep{1999ApJS..122..465P}). When the Swift era approached, similar results pointing to the trimodal distribution were found by \citet{2008A&A...489L...1H} in a sample of 222 GRBs taken from the first \textit{Swift}-Burst Alert Telescope (BAT, \citep{2004SPIE.5165..175B}) catalog (2004 December - 2007 June \citep{2008ApJS..175..179S}). A year later, \cite{huja2009} used the same method of the $\chi^{2}$ fitting of \citet{1998ApJ...508..757H} with the three Gaussians and obtained the same result, namely the trimodal distribution using 388 GRBs from \textit{Swift}-BAT catalog (2004 November - 2009 February). Later, \citet{horvath2016} conducted an analysis using an even larger sample, consisting of more than double the number of GRBs (888 GRBs) compared to the previous analysis. The data for this analysis was sourced from the \textit{Swift}-BAT catalog spanning the period of November 2004 to September 2015. Importantly, \citet{horvath2016} analysis confirmed the findings of the earlier study. Thus, it appears that the trimodal distribution is statistically possible for GRBs observed both by \textit{Swift} and by BATSE.

Continuing on the discussion regarding which fit is the most appropriate for the $T_{90}$ distribution, \citet{zitouni2015} used 757 GRBs with known and unknown redshifts from the \textit{Swift}-BAT catalog (2004 December - 2014 February) and a subsample of these 757 GRBs composed of 248 GRBs with known redshifts. They showed that three log-normal functions better fit the $\log(T_{90})$ distribution in the \textit{Swift}-BAT data. They used $\chi^{2}$ statistics to arrive at this conclusion, similar to \citet{1998ApJ...508..757H}.

We here detail the literature where there is no evidence of the three classes. However, contrary to previous findings, \citet{zitouni2015} reported different results when they re-examined 2041 GRBs from the BATSE 4Br catalog (1991 - 2000). Their analysis showed that the $\log(T_{90})$ distribution prefers two log-normal fits using the same $\chi^{2}$ test as in \citet{1998ApJ...508..757H}. In the era of \textit{Fermi}, \cite{tarnopolski2015} also found that the $\log(T_{90})$ distribution in 1566 GRBs taken from \textit{Fermi}'s Gamma-ray Burst Monitor (GBM) catalog (2008 July - 2012 July \citep{2014ApJS..211...12G, 2014ApJS..211...13V}) is fundamentally bimodal using the standard $\chi^{2}$ test used by \citet{1998ApJ...508..757H}. In addition, \citet{2016NewA...46...54T} also found two Gaussian components in $\log(T_{90})$ distribution in a sample of 947 GRBs with known and unknown redshifts taken from a different catalog, the \textit{Swift}-BAT catalog (2004 December - 2015 September) and a sub-sample of 347 GRBs with known redshifts taken from these 947 GRBs. Instead of the $\chi^{2}$ test, the author used Maximum Likelihood, Akaike Information Criteria (AIC, \citet{1100705, sakamoto1986akaike}), and Bayesian Information Criteria (BIC, \citet{10.1214/aos/1176344136}) to compare the different models, where AIC and BIC are the statistical metrics for model comparison \footnote{Both AIC and BIC are used to penalize the extra number of parameters.}. The difference between AIC and BIC is that the penalty in the BIC test is harsher than that of the AIC test. Mathematically, AIC = -2\textit{ln(L)} + 2\textit{q}, and BIC = -2\textit{ln(L)} + \textit{qln(m)}, where \textit{q} is the number of free parameters in the model, \textit{m} is the number of observations, and \textit{L} is the likelihood. 
This study implied that the third intermediate class of GRBs is probably absent from this \textit{Swift}-BAT catalog based on the duration only. Continuing on this long-debated discussion, \citet{2016MNRAS.458.2024T} expanded the study by including the data from different catalogs, such as BATSE 4Br (1991 - 2000), \textit{Swift}-BAT (2004 December - 2015 April), and \textit{Fermi}-GBM (2008 July - 2012 July) comprising of 2041, 914, and 1593 GRBs, respectively. Using the Maximum Likelihood and AIC, the author found that for the \textit{Fermi}-GBM data, two skew-normal or two sinh-arcsinh distributions are more credible fits for the $\log(T_{90})$ distribution, as compared to the conventional three Gaussian fits. Meanwhile, in BATSE and \textit{Swift}-BAT $\log(T_{90})$ distributions, the author could not determine the better fit based on the small AIC difference between two sinh-arcsinh, two skew-normal, or the traditional three Gaussian fits. Thus, based on these discrepant results, the author discarded the presence of the third intermediate class of GRBs. These articles were followed up by \citet{2019MmSAI..90...45T} and \citet{2019ApJ...870..105T}, where the author used two parameters: hardness and $\log$($T_{90}$). In the first paper, the author used GRBs from the \textit{Swift}-BAT (2005 - 2015, \citep{2016ApJ...829....7L}), \textit{Konus}-Wind (1994 - 2010, \citep{2016ApJS..224...10S}), \textit{RHESSI} (2002 February - 2008 April \citep{2009A&A...498..399R}), and \textit{Suzaku}/Wide-band All-sky Monitor (WAM) (2005 August - 2010 December, \citep{2016PASJ...68S..30O}) satellites, and found no clear indications of the proposed third intermediate class based on AIC and BIC. While in the latter, the author investigated GRBs from the \textit{Fermi}-GBM (2008 July - 2014 July, \citep{2016ApJS..223...28N}) and BATSE 4Br (1991 - 2000) catalog and found that the two-component skewed Student-t distributions explain the data better based on AIC and BIC. Therefore, the existence of the third intermediate class was ruled out again. Following the same methodology \citet{2019ApJ...887...97T} used a more extensive data set from the BATSE 4Br (1991 - 2000) catalog and a larger parameter space: $\log(T_{90}$), hardness, and \textit{Fluence}, and different combinations of these parameters. The author found inconsistencies in the outcome based on AIC and BIC. Thus, the existence of a third class remained uncertain.

\citet{kulkarni2017} also showed similar results by making the model comparison between two (or three) log-normal components for $\log(T_{90})$ distribution. The difference between their analysis and previous ones is due to the different sample size chosen and the addition of the GRBs observed by BeppoSAX (1996 - 2002, \citep{2009ApJS..180..192F}) besides GRBs observed by BATSE 4Br, \textit{Swift}-BAT, and \textit{Fermi}-GBM. They also applied the Maximum Likelihood method but with a probability density function modeled with the superposition of \textit{k} log-normal distributions, with \textit{k} being the total number of GRB classes. They did not find any consistent result on whether a third GRB class is present, based on the $\chi^{2}$ test, AIC, and BIC. Consequently, the nature of the intermediate class of GRBs remains unclear and debated. Recently, \cite{Luo2022arXiv221116451L} trained the supervised ML method Extreme Gradient Boosting (XGBoost, \citep{10.1214/aos/1016218223}) to classify GRBs as Type I and Type II GRBs.  The model was trained on three different groups of features: prompt emission, afterglow, and host Galaxy, and the analysis denied the existence of the third intermediate GRB class based on the $T_{90}$ duration distribution. They used a sample of 1905 GRBs for the classification scheme collected from GRB Big Table \citep{2020ApJ...893...77W} and \href{https://www.mpe.mpg.de/~jcg/grbgen.html}{Greiner's GRB catalog}, ranging from April 1991 - July 2021.

\subsection{Application of Unsupervised ML}

Although we have discussed a plethora of supervised ML approaches, mainly based on testing several distributions and checking if this distribution is the best fit, unsupervised methods in the literature have tried to overcome the classification problem.

We here detail the evidence of three classes. One of the pioneering applications of unsupervised ML was demonstrated in \citet{mukherjee1998}. They took the prompt parameters, such as the spectral hardness, \textit{Fluence}, and peak flux, of 797 GRBs from the BATSE 3B catalog (April 1991 - September 1994) and applied two multivariate clustering procedures. First, a nonparametric hierarchical agglomerative clustering procedure (a technique based on the sequential merging of adjacent pairs of object clusters), and second, a parametric Maximum Likelihood model-based unsupervised clustering procedure with expectation maximization (EM) algorithm (an algorithm used to determine local Maximum Likelihood estimates, MLE, or Maximum A Posteriori estimates, MAP). With this, they presented evidence for the existence of three GRB classes.

Similarly, limiting the parameters in the analysis to only spectral hardness and $\log(T_{90})$, \citet{2006A&A...447...23H, 2010ApJ...713..552H} also confirmed the presence of a third class using 1956 GRBs from the BATSE 4Br catalog (1991 - 2000) and 325 GRBs from the \textit{Swift}-BAT catalog (2004 December - 2008 December, \citep{2008ApJS..175..179S}), respectively. Both the studies fitted Gaussian distributions on the duration-hardness plane using Maximum Likelihood with the EM algorithm. However, according to \citet{hakkila2000}, the duration and \textit{Fluence} of some faint LGRBs are detector biased and could lead to these intermediate class traits.

Continuing on the work of the classification, \citet{tsutsui2014} confirmed three subclasses of LGRBs by using two different properties: Absolute Deviation from the Constant Luminosity of their cumulative LC and the ratio of the mean counts to the maximum counts, which are independent of the distance and burst's jet opening angle. These two parameters were obtained from the LCs of 168 out of 769 bright \textit{Swift}-BAT LGRBs detected from December 2005 to May 2013. They carried out cluster analysis via the unsupervised ML method, the Gaussian Mixture Model (GMM, a probabilistic model generated from a mixture of Gaussian distributions), along with the EM algorithm. 

We here detail the lack of evidence of the three classes in the data. Summarizing the findings, the evidence supporting the existence of three distinct GRB classes remains inconclusive. \citet{zhang2016} conducted an analysis using GMM on multiple catalogs, including \textit{Swift}-BAT, BATSE, and \textit{Fermi}-GBM. They concluded that only \textit{Swift}-BAT data shows a three-component Gaussian distribution for $\log(T_{90})$, while BATSE and \textit{Fermi}-GBM show two-component Gaussian distribution for $\log(T_{90})$. Based on these, they could not find consistent results of a third ``intermediate” class of GRBs.

Recently, \citet{2022Ap&SS.367...39B} analyzed the BATSE (1991 - 2000) and \textit{Fermi}-GBM (2008 July - 2020 June, \citet{2020ApJ...893...46V}) datasets with 1934 and 2329 GRBs, respectively, by carrying out 2D clustering in the hardness vs $T_{90}$ plane using the Extreme Deconvolution Gaussian Mixture Model (XDGMM), an extension of GMM. It uses Gaussian mixtures for density estimation with an extreme deconvolution algorithm, a general algorithm designed to infer a d-dimensional distribution function from a heterogeneous, noisy, or incomplete dataset. They also used AIC and BIC to optimize the number of GRB classes. They found that both AIC and BIC prefer a two-component fit for the BATSE data, while for the Fermi data, AIC prefers a three-component fit, and BIC prefers a two-component fit. 

Recently, \citet{2023ApJ...951....4G} analyzed \textit{Swift}-BAT GRBs observed until July 2022 using t-distributed stochastic neighbor embedding (t-SNE), a technique that helps visualize the high-dimensional data by reducing the N-dimensional data space to 2D or 3D data space \citep{van2008visualizing}. They found two GRB subgroups and observed that SEE GRBs, as reported by various authors using different criteria, tend to cluster within specific regions of t-SNE maps.

Now, we summarize the results in the literature in which five classes have been pointed out. Using unsupervised ML methods, \citet{2017MNRAS.469.3374C, chattopadhyay2018} obtained five groups of GRBs from the BATSE 4Br catalog (1991 - 2000). \citet{2017MNRAS.469.3374C} use six parameters, namely $T_{90}$, $T_{50}$ (time by which 50$\%$ of flux arrives), $P_{256}$ (peak flux measured in bins of 256 ms), $F_{t}$ (total Fluence of GRB), $H_{32}$ (spectral hardness ratio using Fluences in the 50-100 and 100-300 keV), and $H_{321}$ (hardness ratio using Fluences in the 20-50, 50-100, and 100-300 keV), where the analysis was done using \textit{k}-means (an unsupervised learning algorithm that works by dividing the data points in \textit{k} number of different clusters) and GMM clustering on 1599 and 1929 BATSE GRBs, respectively. In the follow-up study, \citet{chattopadhyay2018} use nine parameters: $T_{90}$, $T_{50}$, $P_{256}$, $P_{64}$ (peak flux measured in bins of 64 ms) $P_{1024}$ (peak flux measured in bins of 1024 ms), $F_{1}$ (Fluence in 20-50 keV), $F_{2}$ (Fluence in 50-100 keV), $F_{3}$ (Fluence in 100-300 keV), $F_{4}$ (Fluence in $>$ 300 keV). They use multivariate \textit{t}-mixtures-model-based clustering (tMMBC, a model-based clustering using a multivariate probability distribution rather than Gaussian probability distribution) on 1599 BATSE 4B GRBs.

\section{Data sample}\label{data sample}

 In this work, we have used two sets of GRB data. The first set is taken from \cite{Srinivasaragavan2020} and contains X-ray properties. The second set is the optical data from \cite{2022ApJS..261...25D}. The unique aspect of these two data sets is that all the GRBs show plateau emission in their LCs, the flat portion in the GRB afterglow. This enables us to obtain the respective data sets' X-ray and optical plateau properties.

As mentioned previously, to classify GRBs, we consider a ten-dimensional parameter space for the optical data and an eleven-dimensional parameter space for the X-ray data. In addition to plateau properties, both data sets contain prompt and afterglow emission properties. A detailed explanation for these two data sets is given in subsections \ref{xray data sample} and \ref{optical data sample}, respectively. 

\begin{figure*}
\begin{minipage}\textwidth
    \textbf{(a)}
    \centering
        \includegraphics[width=0.38\textwidth]{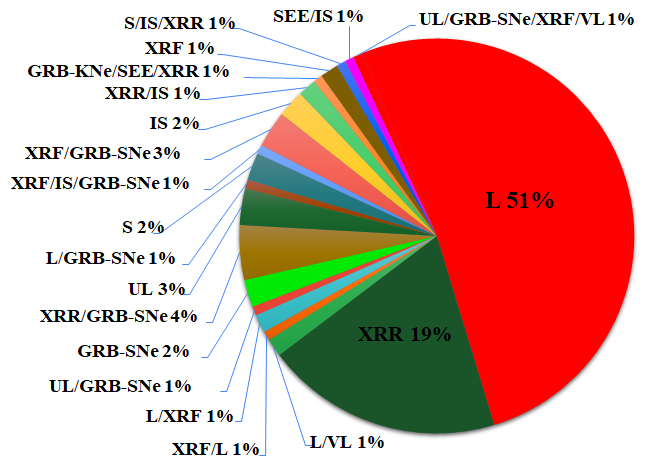} 
    \textbf{(b)}
    \centering
        \includegraphics[width=0.57\textwidth]{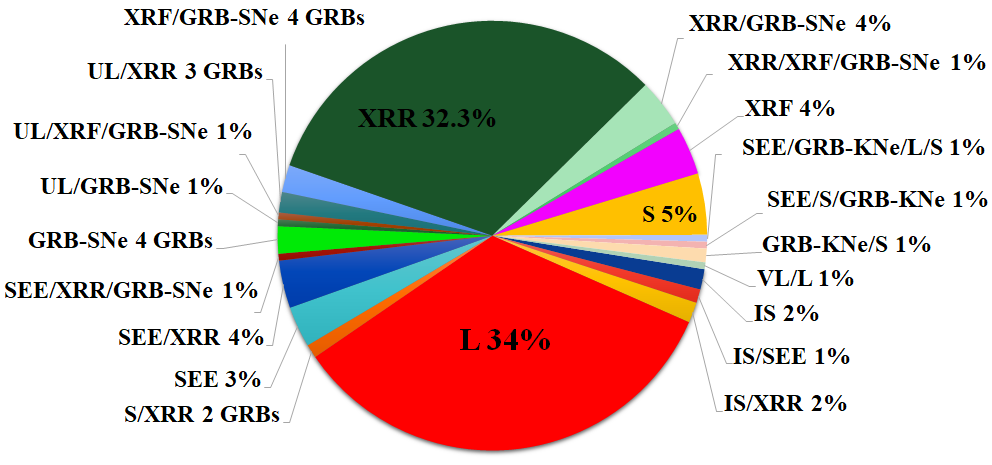}
\end{minipage}
\begin{minipage}\textwidth
    \textbf{(c)}
    \centering
        \includegraphics[
        width=0.43\textwidth,
        ]{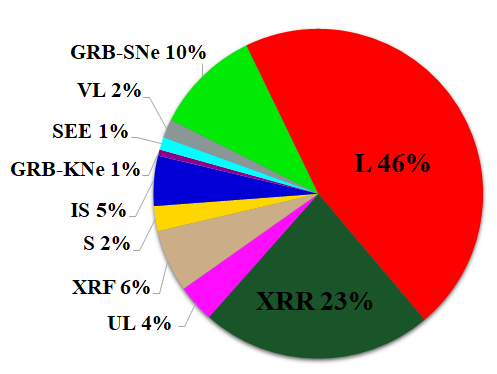} 
    \textbf{(d)}
    \centering
        \includegraphics[width=0.43\textwidth]{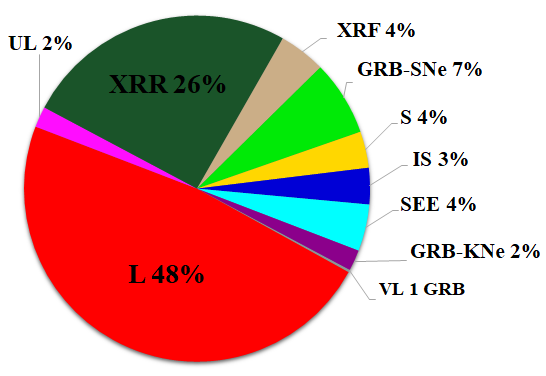}
\end{minipage} 
    \caption{\small
    a: Breakdown of all 134 optical GRBs into their different combinations of ten major classes (Table. \ref{classes}). b: Breakdown of all 203 X-ray GRBs into their different combinations of ten major classes (Table. \ref{classes}). c: Further breakdown of the optical GRBs into ten major classes used in our analysis: S (SGRBs), L (LGRBs), IS, SEE, GRB-SNe, UL, XRF, XRR, GRB-KNe, and VL. d: Further breakdown of the X-ray data into ten major classes used in our analysis: S (SGRBs), L (LGRBs), IS, SEE, GRB-SNe, UL, XRF, XRR, GRB-KNe, and VL.}
\label{fig:pie}
\end{figure*}

\subsection{X-ray data}
\label{xray data sample}

The X-ray sample consists of 222 GRBs with eleven properties, and it is taken from the X-ray \textit{Swift} data catalog \citep{2004ApJ...611.1005G, 2005SSRv..120..165B, Evans2007, Evans2009}, with the plateau properties obtained from Table 1 of \cite{Srinivasaragavan2020}. The dataset includes the following characteristics: 

\begin{itemize}
    \item \textit{$z$} - redshift of the GRB.
    \item \textit{$\log(F_{a,X})$} - base 10 logarithm of the flux at the end of the X-ray plateau phase.
    \item \textit{$\log(T_{a,X})$} - base 10 logarithm of the end time of the X-ray plateau.
    \item \textit{$\alpha_{X}$} - temporal decay power law index after the X-ray plateau.
    \item \textit{$\log(T_{90,X})$} - the base 10 logarithm of the duration in which 90$\%$ of the counts are detected from 5\% to 95\% of the total emission in the prompt phase.
    \item \textit{$\log(NH_{X})$} - base 10 logarithm of the observed Hydrogen column density.
    \item \textit{$\log(Fluence_{X})$} - logarithm in the base of 10 of the Fluence of the prompt X-ray emission.
    \item \textit{$\log(Peak_{X})$} - logarithm in the base of 10 of the peak flux of the prompt X-ray emission at 1 second.
    \item\textit{$PhotonIndex_{X}$} - Photon Index of the prompt X-ray emission, assuming a power-law.
    \item\textit{$\Gamma_{X}$} - spectral index parameter of the X-ray prompt emission .
    \item\textit{$\beta_{X}$} - spectral index parameter of the X-ray plateau emission.
\end{itemize}  

\begin{figure*}
    \includegraphics[width=\textwidth]{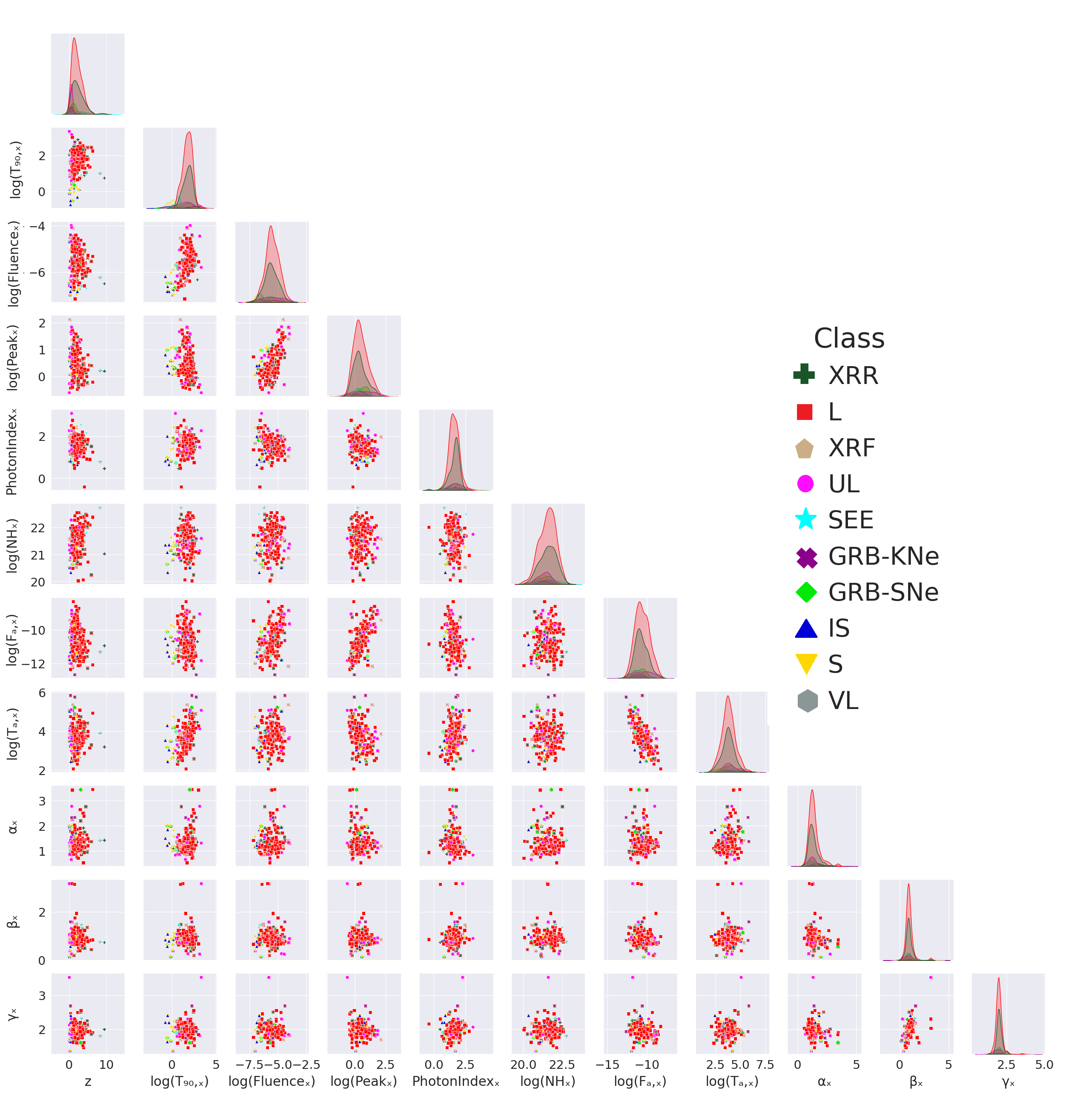}
    \caption{X-ray data sample distribution of 203 GRBs across all the parameters, with classes: XRR (dark green plus), L (red square), XRF (burlywood pentagon), UL (violet circle), SEE (cyan star), GRB-KNe (dark magenta cross), GRB-SNe (green diamond), IS (blue triangle-up), S (yellow triangle-down), and VL (gray hexagon).}
    \label{fig:xr_data}
\end{figure*}

Some of these parameters were converted to logarithmic forms to normalize them to a comparable order of magnitude. We dropped GRBs with $\log(NH_{X}$) $<$ 20 due to measurement errors and removed GRBs with NA values in any of their properties to avoid inconsistencies in our analysis. Thus, reducing the number of GRBs used in our analysis to 203. The dataset comprises ten distinct GRB classes, outlined in Table \ref{classes}. The diverse combinations of these classes within the dataset are visually depicted in Fig. \ref{fig:pie} b, revealing that many GRBs belong to multiple classes. To accommodate this, we have assigned each GRB to all the classes it belongs to.
This resulted in dividing the sample into ten major classes (Table \ref{classes}) for our analysis, as depicted in Fig. \ref{fig:pie} d. Table \ref{xr_table} displays a portion of this dataset, and a machine-readable table of the entire sample used here is also available online. Fig. \ref{fig:xr_data} showcases the relation between the eleven properties through a pair plot.

\begin{table}
\begin{center}
\begin{tabular}{cc}
\hline
Class & Abbreviation
\\ \hline
\\Long GRBs & L \\
\\Short GRBs & S\\
\\Ultra Long GRBs & UL\\
\\Very Long GRBs & VL\\
\\SGRB associated with a Kilonovae & GRB-KNe\\
\\Intrinsically Short GRBs & IS\\
\\Short GRBs with Extended Emission & SEE\\
\\X-ray flashes & XRF\\
\\X-ray rich & XRR\\
\\GRBs associated with Supernovae & GRB-SNe Ib/c (A, B, \\types (A, D, E, AB, DE) & C, D, E, AB, DE)
\end{tabular}
\caption{The table lists different classes of GRBs found in our data.}
\label{classes}
\end{center}
\end{table}

\begin{table*}
\begin{tabular}{ccccccccccccccc}
\hline
\begin{tabular}[c]{@{}c@{}}GRB name\end{tabular} & Class & $z$ & $\log(T_{90,X})$ &  $\log(Fl_{X})$ & $\log(Peak_{X})$ & $PI_{X}$ & $\log(NH_{X})$ & $\log(F_{a,X})$ & $\log(T_{a,X})$ & $\Gamma_{X}$ & $\alpha_{X}$ & $\beta_{X}$ \\ 
 &  &  & \scriptsize{(s)} & \scriptsize (erg cm$^{-2}$)   & \scriptsize(ph cm$^{-2}$ s$^{-1}$)  &  & \scriptsize(H atoms cm$^{-2}$) & \scriptsize(erg cm$^{-2}$ s$^{-1}$)    &\scriptsize(s) &  & &    \\
\hline
050318                                               & L    & 1.44     & 1.50     & -5.96        & 0.49         & 1.90        & 20.92   & -11.15  & 4.13    & 1.98  & 1.87  & 0.95 \\
050401                                               & L    & 2.90     & 1.52     & -5.08        & 1.02         & 1.40        & 22.12   & -9.65   & 3.12    & 1.78  & 0.98  & 0.99 \\
050505                                               & L    & 4.27     & 1.77     & -5.60        & 0.26         & 1.41        & 22.23   & -11.04  & 4.27    & 2.01  & 1.46  & 0.97 \\
050730                                               & L    & 3.96     & 2.19     & -5.62        & -0.25         & 1.53        & 21.73   & -10.09  & 4.05    & 1.61  & 2.38  & 0.56 \\
050802                                               & L    & 1.71     & 1.27     & -5.69        & 0.43         & 1.54        & 21.25   & -10.28  & 3.63    & 1.85  & 1.44  & 0.85 \\
050803                                               & L    & 3.50     & 1.94     & -5.66        & -0.01         & 1.38        & 21.34   & -10.86  & 4.21    & 2.11  & 1.52  & 1.21 \\
050826                                               & L    & 0.29     & 1.55     & -6.38        & -0.42         & 1.16        & 21.96   & -12.33  & 4.76    & 2.45  & 1.80  & 1.11 \\
050904                                               & L    & 6.9     & 2.24     & -5.31        & -0.20         & 1.25        & 22.39   & -12.06  & 4.93    & 1.90  & 3.44  & 0.84 \\
060115                                               & XRR  & 3.53     & 2.14     & -5.77        & -7.14         & 1.00        & 21.43   & -11.30  & 3.70    & 2.11  & 0.89  & 1.00 \\
060124                                               & XRR  & 3.53     & 2.29     & -6.33        & -0.05         & 1.84        & 21.91   & -10.68  & 4.47    & 2.00  & 1.36  & 0.99
\end{tabular}
\caption{\small 
This table represents a small portion of our X-ray data sample, showing the GRB name, class, and features used in the analysis. \textit{Fl$_{X}$} and \textit{PI$_{X}$} stand for \textit{Fluence$_{X}$} and \textit{PhotonIndex$_{X}$}, respectively.}
\label{xr_table}
\end{table*}

\subsection{Optical data}
\label{optical data sample}

The optical sample consists of 179 GRBs with ten properties. \cite{2022ApJS..261...25D} built and used an extended optical LC data sample by analyzing 500 GRB optical afterglows with recorded redshifts. These were obtained after a thorough search of various sources in the literature for GRB detections between 1997 May to 2021 May. These detections were made by different satellites, such as the Swift Ultraviolet/Optical Telescope (UVOT, \citep{2005SSRv..120...95R}), and ground-based telescopes/detectors, like $\gamma$-ray Burst Optical/Near-IR Detector (GROND, \citep{2008PASP..120..405G}), the MITSuME \citep{2005NCimC..28..755K}, the Subaru Telescope \citep{Subaru2004PASJ...56..381I},  Re-ionization and Transients InfraRed camera/telescope (RATIR, \citep{2012SPIE.8446E..10B, 10.1117/12.926927}), etc. Out of these, roughly 19$\%$ of GRBs from \cite{2022ApJS..261...25D} have incomplete data in their plateau properties, meaning we face the problem of missing values in several parameters. Therefore, we dropped GRBs with NA values in any of their properties. We also dropped those GRBs with $\log(NH_{X})$ $<$ 20 due to measurement errors. Thus, our final data set consists of 134 GRBs. The sample contains ten GRB classes, which are presented in Table \ref{classes}. The various classes within the sample and their different combinations are presented in Fig. \ref{fig:pie} a. In our sample, many GRBs belong to multiple classes, as illustrated in Fig. \ref{fig:pie} a. To account for this, we have assigned each GRB to all the classes it belongs to. Consequently, we have divided the sample into ten major classes (Table \ref{classes}) for our analysis, as shown in Fig. \ref{fig:pie} c. We use the following properties from the dataset:

\begin{itemize}
    \item \textit{$\log(F_{a,opt})$} - base 10 logarithm of flux at the end of the optical plateau.
    \item \textit{$\log(T_{a,opt})$} - base 10 logarithm of the end time of the optical plateau.
    \item $\alpha_{opt}$ - temporal decay power law index after the optical plateau.
    \item $\beta_{opt}$ - spectral index parameter of the optical plateau.
\end{itemize}

\begin{figure*}
    \includegraphics[width=\textwidth]{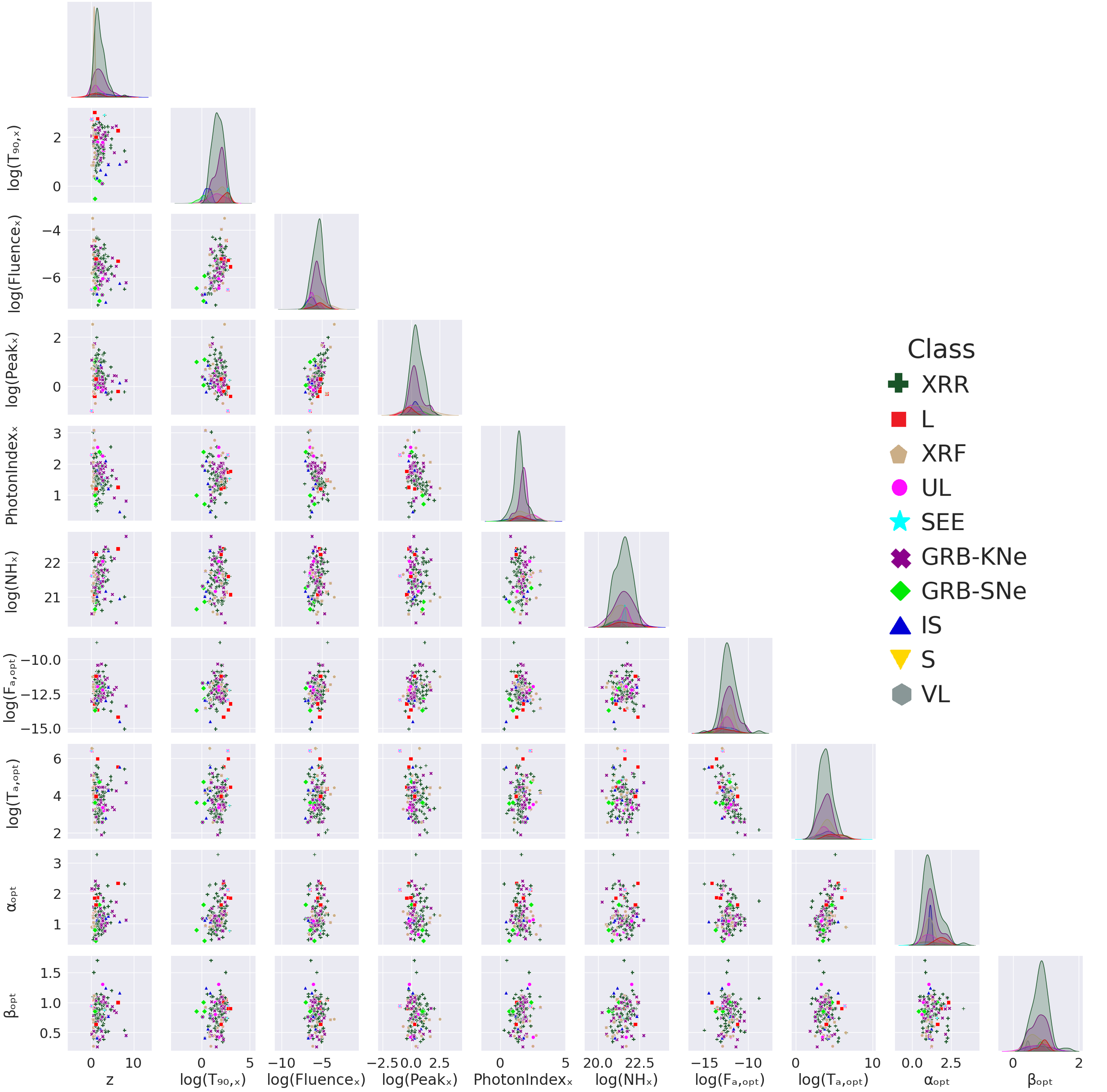}
    \caption{\small Optical data sample distribution of 134 GRBs across all the parameters, with classes: XRR (dark green plus), L (red square), XRF (burlywood pentagon), UL (violet circle), SEE (cyan star), GRB-KNe (dark magenta cross), GRB-SNe (green diamond), IS (blue triangle-up), S (yellow triangle-down), and VL (gray hexagon).}
    \label{fig:opt_data}
\end{figure*}

The additional six variables, as given in Table \ref{opt_table}, such as \textit{z}, \textit{PhotonIndex}$_{X}$, $\log(Fluence_X)$, $\log(NH_{X})$, $\log(Peak_{X})$ and $\log(T_{90,X}$), are obtained from the X-ray LC and combined here, along with the optical properties. A portion of our dataset is shown in Table \ref{opt_table}, and the complete sample used here is available as a machine-readable table. 
Fig. \ref{fig:opt_data} shows the pair plot of all the ten properties used in this analysis. 

For notation purposes, we denote LGRBs as L and SGRBs as S in the figures and tables. We also considered the rest frame or intrinsic $T_{90,X}$ ($T^*_{90,X}$) as an extra parameter for analyzing the optical and X-ray data sets. It is defined as follows:
\begin{equation}
    \mbox{($\log(T^{*}_{90,X})$)} = \log\biggl(\frac{(T_{90,X})}{1 + z}\biggl)
\end{equation}

\begin{table*}
\centering
\begin{tabular}{cccccccccccc}
\hline
GRB name & Class          & \multicolumn{1}{c}{$z$} & \multicolumn{1}{c}{$\log(F_{a, opt})$} & \multicolumn{1}{c}{$\log(T_{a, opt})$} & \multicolumn{1}{c}{$\alpha_{opt}$} & \multicolumn{1}{c}{$\beta_{opt}$} & \multicolumn{1}{c}{\textit{PI$_{X}$}} & \multicolumn{1}{c}{$\log(T_{90, X})$} & \multicolumn{1}{c}{$\log(NH_{X})$} & \multicolumn{1}{c}{$\log(Fl_{X})$} & \multicolumn{1}{c}{$\log(Peak_{X})$} \\
      &            &                 &   \scriptsize{(erg cm$^{-2}$ s$^{-1}$)}   & \scriptsize{(s)}  &  &    &      & \scriptsize{(s)} & \scriptsize{(H atoms cm$^{-2}$)} & \scriptsize{(erg cm$^{-2}$)} & \scriptsize{(ph cm$^{-2}$ s$^{-1}$)}   \\ \hline
100316B   & L              & 1.18                  & -11.75                    & 4.08                        & 1.96                      & 1.14                     & 2.23                            & 0.57                       & 21.41                     & -6.69                           & 0.11                        \\
100508A   & L              & 0.52                  & -12.87                    & 4.02                        & 0.87                      & 0.40                      & 1.23                            & 1.71                       & 20.84                     & -6.15                          & -0.39                        \\
100513A   & L              & 4.77                  & -13.09                    & 4.32                        & 1.38                      & 1.20                      & 1.62                            & 1.92                       & 22.05                     & -5.85                          & -0.22                       \\
100621A   & XRR              & 0.54                  & -11.87                    & 4.18                        & 1.55                      & 0.78                     & 1.90                            & 1.80                       & 22.28                      & -4.67                          & 1.10                         \\
100814A   & L              & 1.43                  & -12.91                    & 5.59                        & 2.25                      & 0.41                     & 1.47                            & 2.24                       & 20.89                      & -5.04                          & 0.39                         \\
100906A   & XRR              & 1.73                  & -10.41                    & 2.60                        & 1.09                      & 0.84                     & 1.78                            & 2.05                       & 21.96                      & -4.92                          & 1.00                      \\
110205A   & XRR              & 2.21                  & -13.04                    & 4.79                        & 1.95                      & 0.74                     & 1.80                            & 2.42                       & 21.37                      & -4.76                          & 0.55                      \\
110213A & L & 1.46 & -11.53 & 4.36 & 1.91 & 0.90 & 1.83 & 1.68 & 21.56 & -5.22 & 0.20 \\
110422A & L & 1.77 & -12.22 & 3.34 & 0.62 & 0.88 & 0.86 & 1.41 & 22.02 & -4.38 & 1.48 \\
110503A & L & 1.61 & -12.98 & 4.71 & 1.39 & 0.80 & 0.88 & 1.00 & 21.13 & -5.00 & 0.13
\end{tabular}

\caption{\small This table represents a small portion of the optical data sample, showing the GRB name, class, and other features used in the analysis. \textit{Fl$_{X}$} and \textit{PI$_{X}$} stand for \textit{Fluence$_{X}$} and \textit{PhotonIndex$_{X}$}, respectively.}
\label{opt_table}
\end{table*}

\section{Overview of the Methodologies}\label{methodologies}

{Below, we describe the clustering method, GMM, used in our analysis. GMM is implemented using the open-source statistical software \textsc{r} \citep{R}.}

\subsection{Gaussian Mixture Model}
\label{sec:GMM} 
Gaussian Mixture Model (GMM) \citep{bishop2007} is a probabilistic clustering technique represented by a combination of Gaussian functions based on the number of clusters \textit{K}, where three parameters define each Gaussian function:
\begin{itemize}
    \item Mean ($\mu$) defines the centroid
    of the Gaussian.
    \item Covariance matrix ($\Lambda$) defines the width of the Gaussian.
    \item Mixing probability ($\xi$) defines the weight of the Gaussian.
\end{itemize}

For a given data set (\textit{x}$_{1}$, \textit{x}$_{2}$, \textit{x}$_{3}$, . . . , \textit{x}$_{n}$) with \textit{n} observations, in a \textit{D}-dimensional space which can be represented by a matrix \textit{X} having \textit{n} $\times$ \textit{D} dimension, the algorithm tries to maximize the log-likelihood function (Eq. \ref{eq:GMM_1}) by obtaining the optimal values of the parameters $\mu$, $\Lambda$, and $\xi$.

\begin{equation}
    \label{eq:GMM_1}
    \ln p(X|\xi,\mu,\Lambda) = \sum_{i=1}^{n} \ln \Bigl\{\sum_{j=1}^{K}\xi_{j}N(x_{i}|\mu_{j}, \Lambda_{j})\Bigl\}.
\end{equation}

where $\ln p(X|\xi,\mu,\Lambda)$ represents the Gaussian probability density function,
$\xi$ lies between 0 and 1, along with $\sum_{j=1}^{K} \xi_{j}=1$, 
and $N(x_{i}|\mu_{j} \Lambda_{j})$ represents the Gaussian density of each component which is given by:

\begin{equation}
    \label{eq:GMM_3}
    N(x|\mu_{j}, \Lambda_{j}) = \frac{1}{(2\pi)^{D/2}(\lvert\Lambda_{j}\lvert)^{1/2}} exp\Bigl\{0.5(x - \mu_{j})^{T}\Lambda_{j}^{-1}(x - \mu_{j})\Bigl\}.
\end{equation}

For obtaining the optimal values of these parameters, GMM uses an iterative approach called the Expectation-Maximization (EM) algorithm as follows:
\begin{enumerate}
    \item Initialize the three parameters ($\mu$, $\Lambda$, $\xi$) accordingly and calculate the initial value of the log-likelihood using Eq. \ref{eq:GMM_1}.
    \item Expectation step (E-step): use the parameters' current values to compute each Gaussian component's responsibility ($\beta_{ij}$) using the following equation
    \begin{equation}
        \label{eq:EM_1}
        \beta_{ij} = \frac{\xi_{j}N(x_{i}|\mu_{j}, \Lambda_{j})}{\sum_{k=1}^{K}\xi_{k}N(x_{i}|\mu_{k}, \Lambda_{k})}.
    \end{equation}
    \item Maximization step (M-step): evaluate new values of the parameters ($\mu$, $\Lambda$, $\xi$) using the current responsibilities obtained from the E-step.
    \item Re-calculate the log-likelihood (Eq. \ref{eq:GMM_1}) using the updated values of the parameters ($\mu$, $\Lambda$, $\xi$) obtained in the M-step. E-step and M-step are repeated until convergence is reached either for the log-likelihood or the parameters.
\end{enumerate}

To implement GMM, we used the \textsc{mclust} package in \textsc{r}, which natively calculates Bayesian Information Criteria (BIC) explained below.

Since GMM is a probabilistic approach, it can accurately assign a data point that falls precisely between two cluster centers to the appropriate cluster. Statistically, GMM provides better interpretability as it gives the optimal number of clusters based on BIC. It also allows the user to restrict the number of clusters produced to achieve a more distinct distribution or to let the algorithm automatically decide the optimal number of clusters based on BIC. BIC is a statistical criterion \citep{10.1214/aos/1176344136} which decides the optimal parameters of a model to be fitted with a set of observations. BIC is calculated as BIC = -2ln(L) + \textit{q}ln(m), where \textit{q} is the number of free parameters in the model, \textit{m} is the number of observations in the data set, and \textit{L} is the likelihood function for the model and the data. However, in the \textsc{mclust} package, BIC is calculated as the negative of what is defined above. Hence, a higher value of BIC indicates a better fit \citep{10.1093/comjnl/41.8.578, doi:10.1198/016214502760047131}.

\subsection{ Multiple Imputation by Chained Equations (MICE)}
\label{MICE}

MICE is an iterative method for imputing missing values in multivariate data, utilizing information from the other variables in the dataset \citep{schafer2002missing}. These values are assumed to be missing at random \citep{rubin1976inference}. MICE fills in the absent entries using the complete variables in the data set iteratively. Generally, the imputation is performed ten times \citep{van2011mice}. However, we impute the missing variables 20 times for each MICE iteration to reduce the randomness of the imputation. Among the several MICE tools, we have used "midastouch," a predictive mean matching method \citep{little2019statistical}. It initializes missing features to their mean and estimates them by training a model on the rest of the complete data.

\section{Results}
\label{results}

\subsection{Optical GRB Data}
\label{opt_res}

We feed the GMM algorithm different combinations of optical parameters to check their effects on clustering. We have tried all possible combinations among these parameters, starting with all ten parameters, and iteratively reduced them to three parameters covering different combinations, {resulting in a total of 968 combinations}. However, only 914 combinations gave us at least two clusters. We present and discuss a subset of these results generated by the various parameter combinations focusing on plateau properties, which are listed below. In this study, we focus on the plateau parameters because the plateau emission is most likely driven by a magnetar scenario. Thus, we expect similar properties {to characterize} the GRBs in this region of LC. These combinations showcase some interesting microtrends, which we have highlighted in Sec \ref{opt_mcirotrends}.

\begin{enumerate}
    \item All ten parameters with $\log(T^{*}_{90,X})$.
    \item We here start to reduce the parameter space to six variables focusing on the variables of the plateau emission and changing each by each one of the afterglow or of the prompt emission. Thus, we use $\log(F_{a,opt})$, $\log(T_{a,opt})$, $\alpha_{opt}$, $\beta_{opt}$, and we have added for the prompt parameters $\log(T_{90,X})$,  and $\log(NH_{X})$.
    \item To ii), we just changed the variable $\log(T_{90,X})$ with $\log(Fluence_{X})$.
    \item To iii), we just changed the variable $\beta_{opt}$ with $\log(Peak_{X})$.
    \item Now we reduce to five combinations with $\log(T_{90,X})$, $\log(F_{a,opt})$, $\log(T_{a,opt})$, $\alpha_{opt}$, and $\beta_{opt}$.
    \item To v), we have changed the variable $\log(T_{90,X})$ with $\log(NH_{X})$.
    \item To vi), we change parameter $\log(T_{a,opt})$ with $\log(Fluence_{X})$.
    \item To vi), we change the parameter $\log(F_{a,opt})$ with $PhotonIndex_{X}$.
    \item We reduce to a combination of four: $\log(Peak_{X})$, $\log(NH_{X})$, $\log(T_{a,opt})$, $\alpha_{opt}$.
    \item We here reduce to a parameter of three plateau variables $\log(F_{a,opt})$, $\log(T_{a,opt})$, $\alpha_{opt}$.
    \item We reduce to a parameter of three plateau variables $\log(F_{a,opt})$, $\alpha_{opt}$, $\beta_{opt}$.
    \item We here reduce to a parameter of three plateau variables $\log(T_{a,opt})$, $\alpha_{opt}$, $\beta_{opt}$.
\end{enumerate}

\begin{figure*}

    \begin{minipage}{\textwidth}
        \textbf{(a)}
        \centering
        \includegraphics[width=0.3\textwidth]{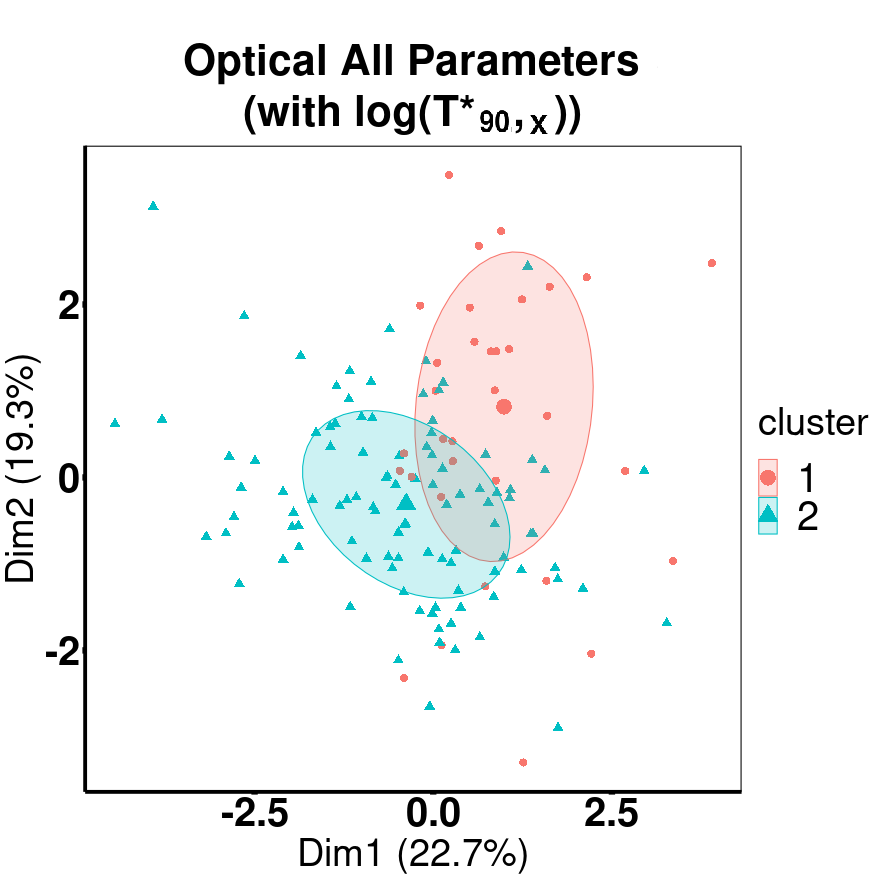}
        \includegraphics[width=0.46\textwidth]{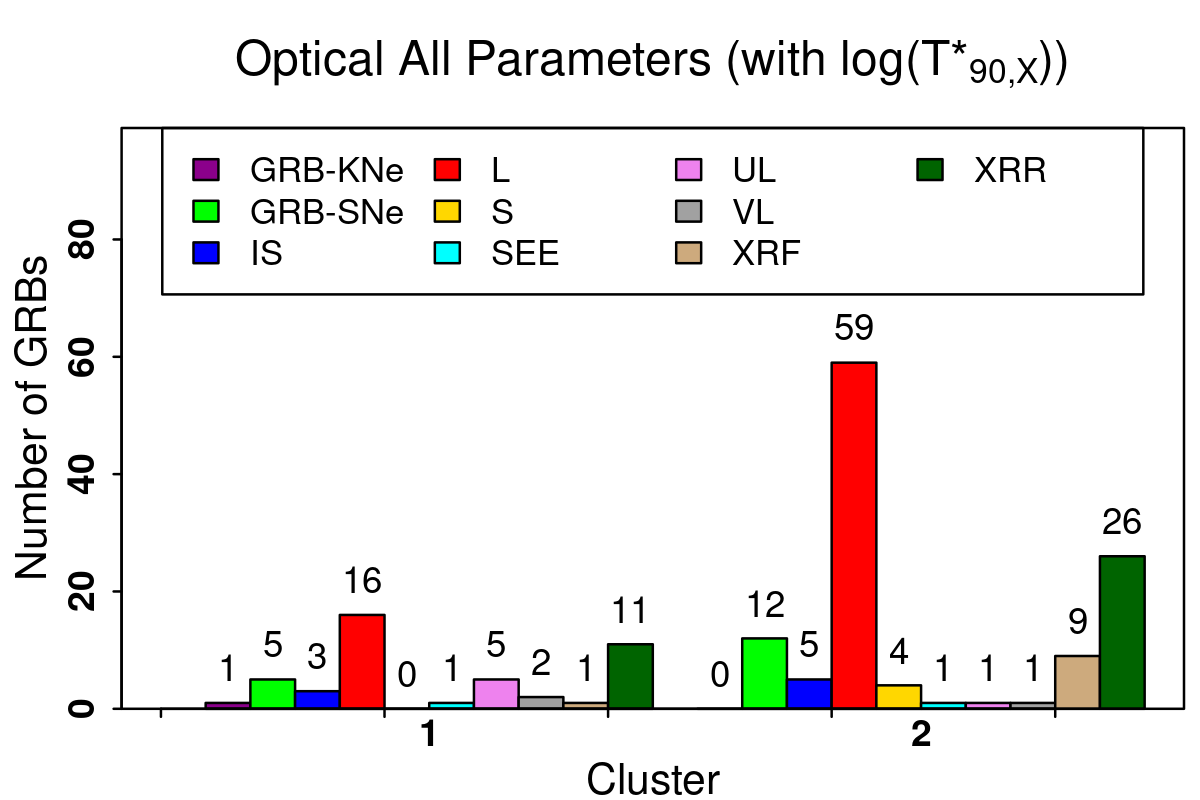}
    \end{minipage}

    \begin{minipage}{\textwidth}
    \textbf{(b)}
        \centering
        \includegraphics[width=0.3\textwidth]{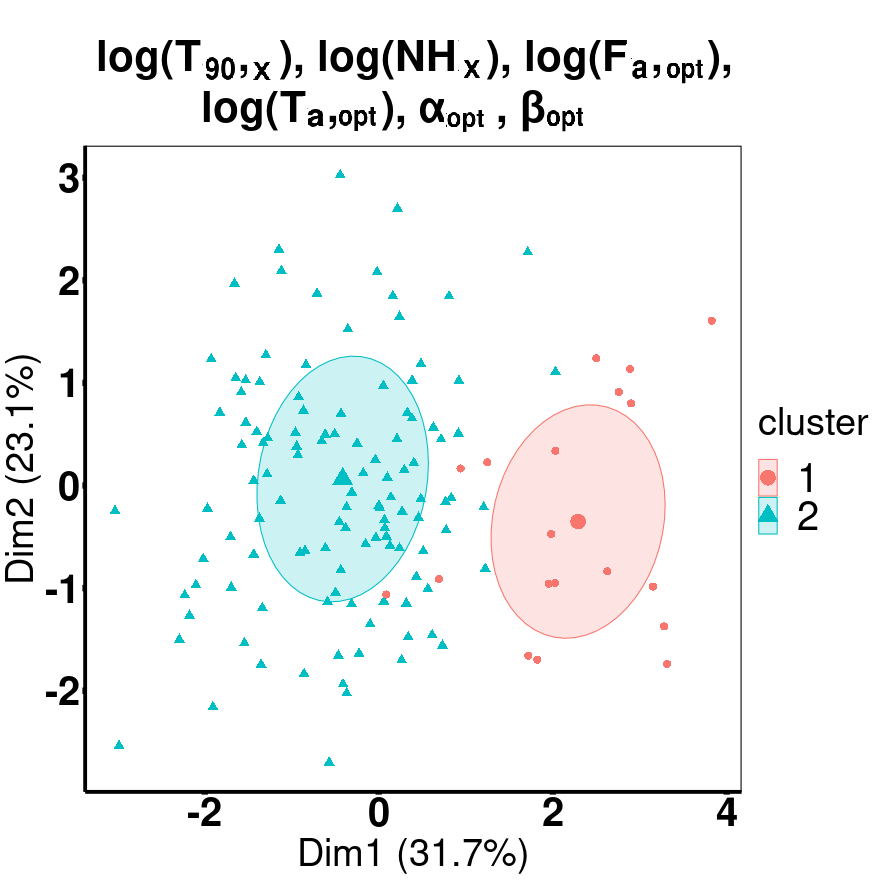} 
        \includegraphics[width=0.46\textwidth]{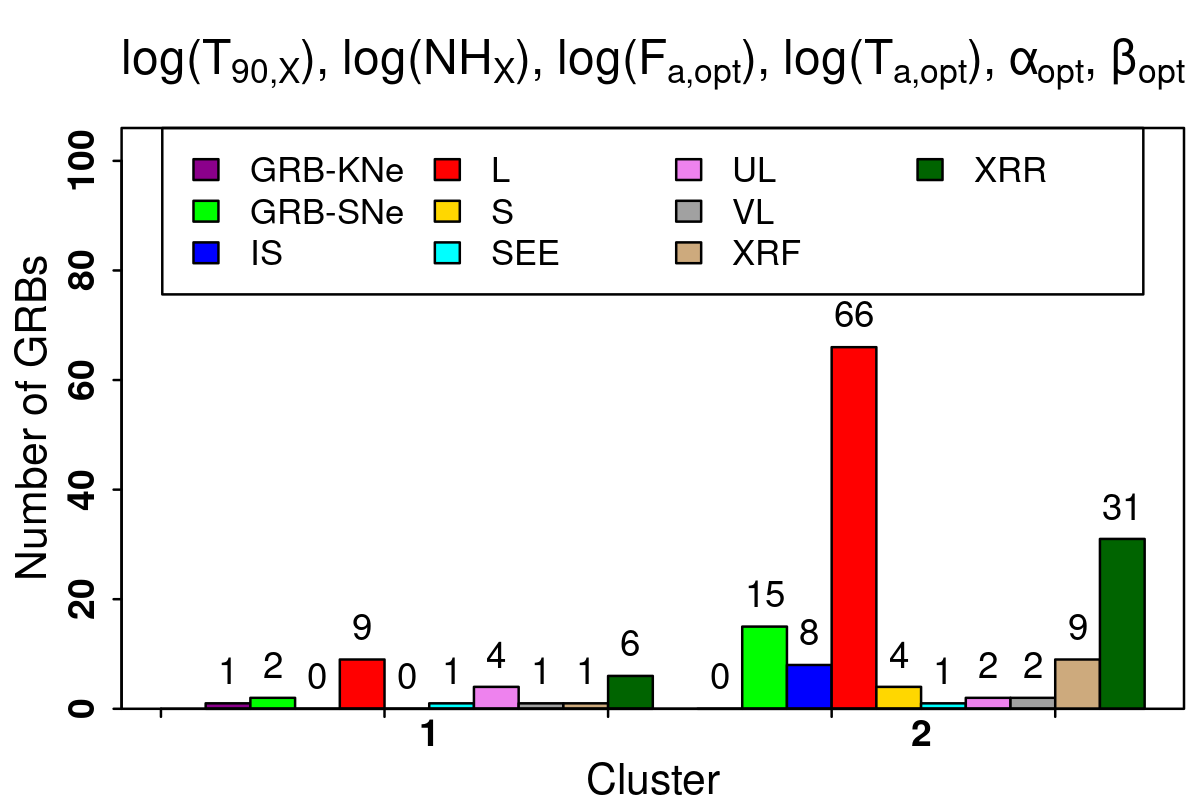}
    \end{minipage}
    
    \begin{minipage}{\textwidth}
    \textbf{(c)}
        \centering       
        \includegraphics[width=0.3\textwidth]{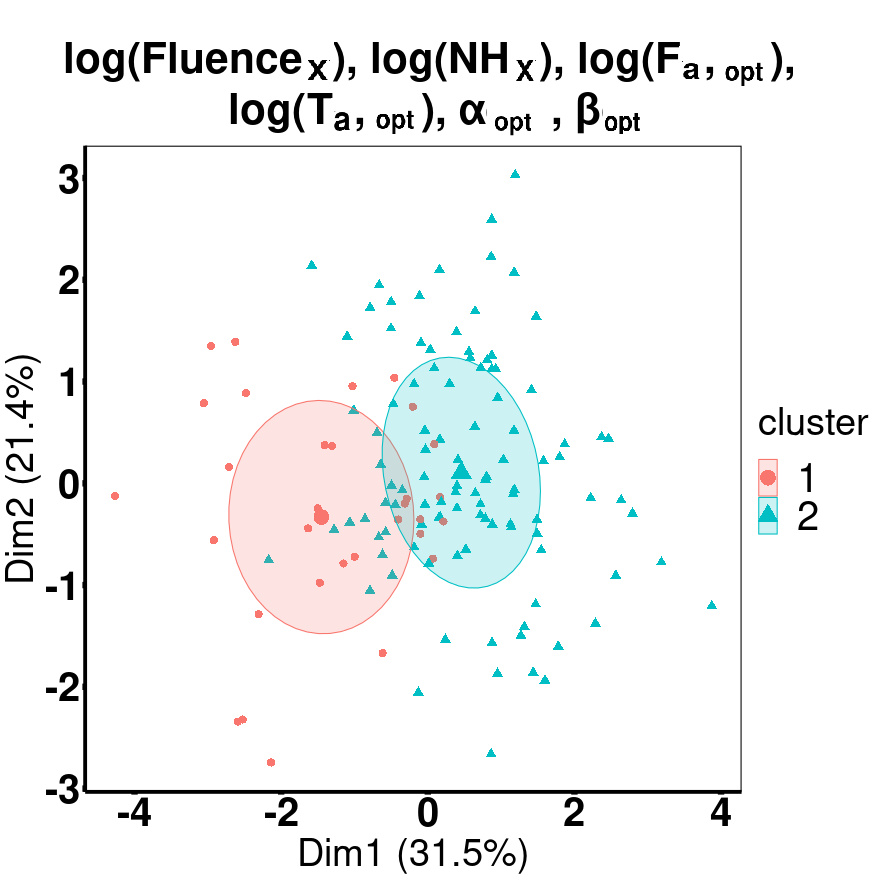}
        \includegraphics[width=0.46\textwidth]{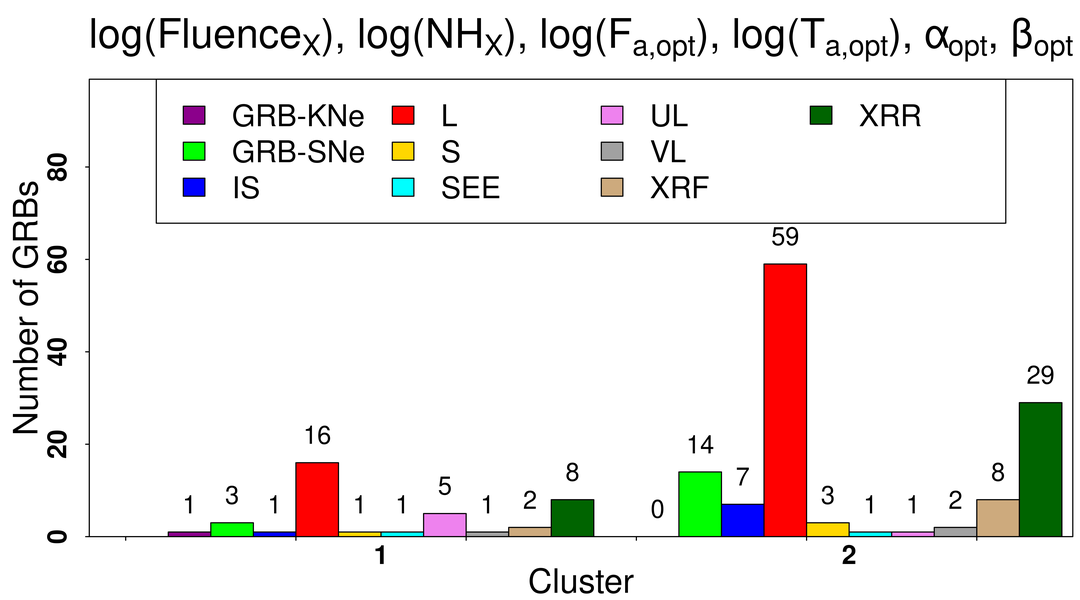}
    \end{minipage}
    
    \begin{minipage}{\textwidth}
    \textbf{(d)}
        \centering
        \includegraphics[width=0.3\textwidth]{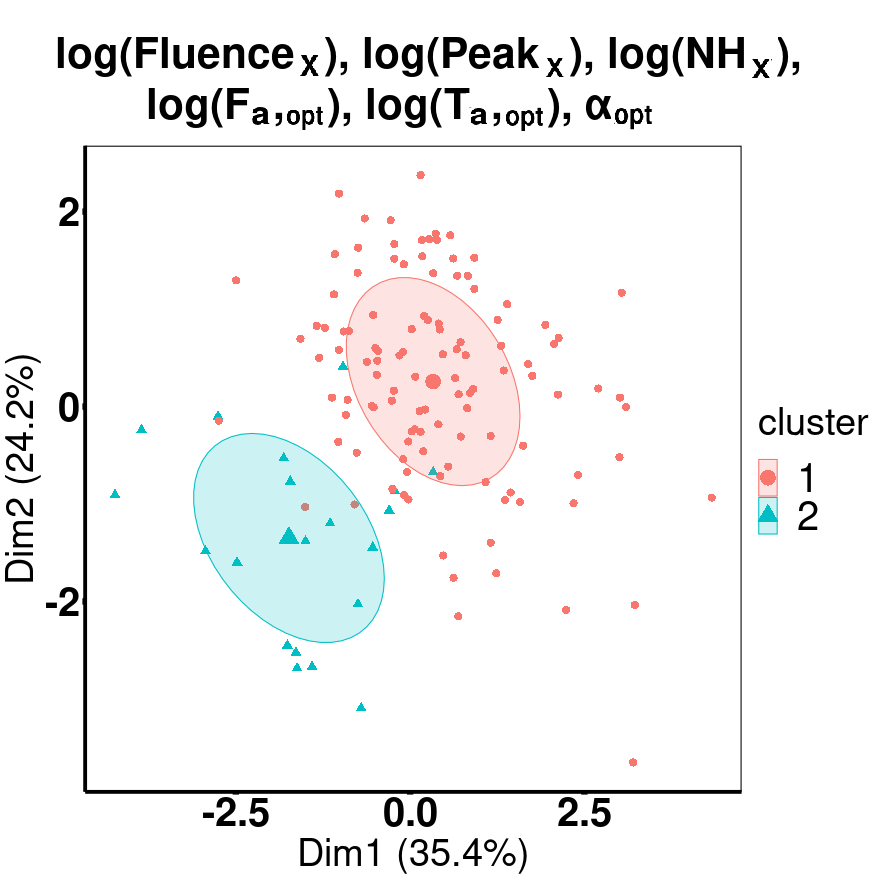}
        \includegraphics[width=0.46\textwidth]{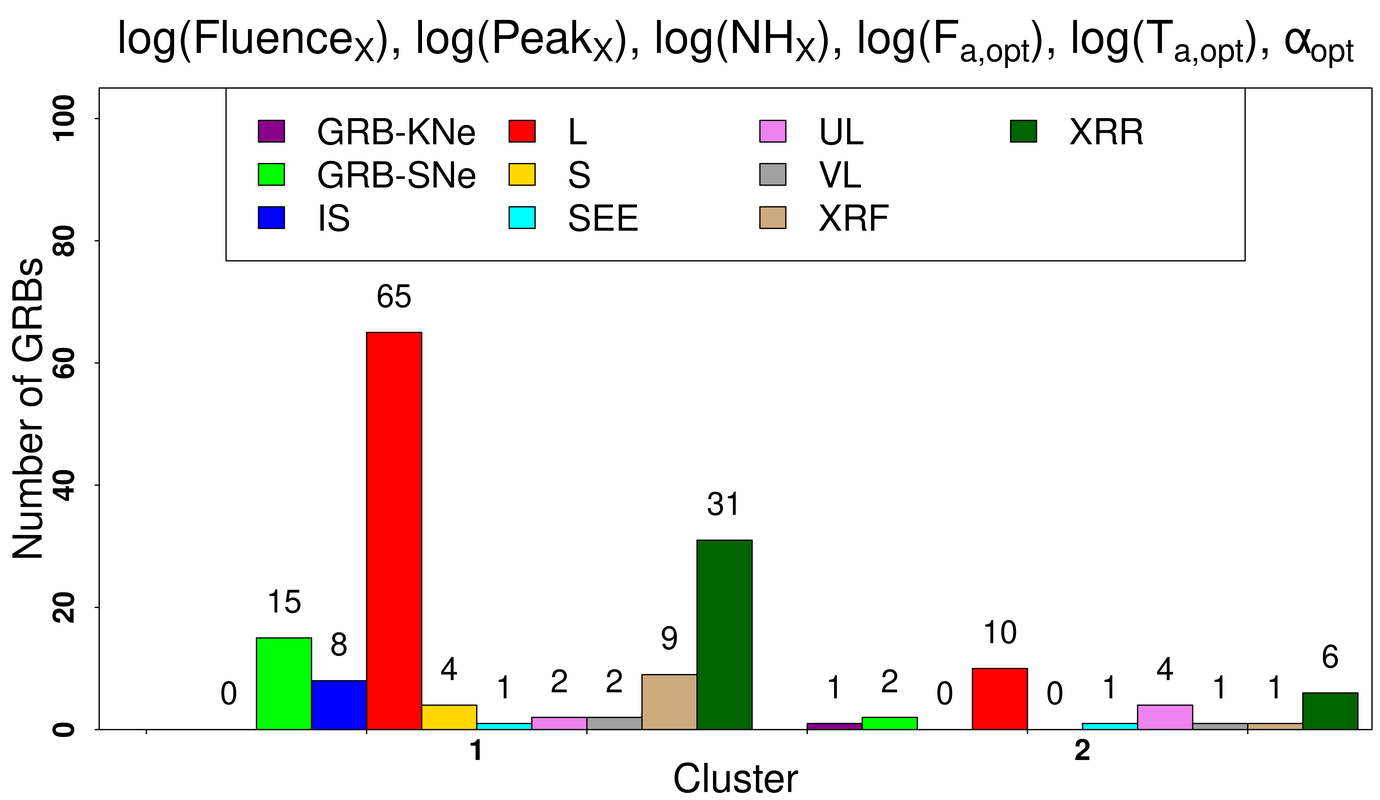}
    \end{minipage}
    \caption{\small 
    This figure shows the first four sets of clustering results obtained for the optical sample using GMM (Sec. \ref{optical data sample}). The left column shows the scatter plot of the clusters obtained for the first four parameter space mentioned in Sec. \ref{opt_res}. The right column shows the distribution of the GRBs within each cluster based on its ten major classes (Fig. \ref{fig:pie} a). `Dim1' and `Dim2' are the first two principal components used to describe the multi-parameter clustering graphically. First row: clustering of GRBs using all ten parameters (with $\log(T^{*}_{90,X})$). Second row: clustering of GRBs using a combination of $\log(T_{90,X})$, $\log(NH_{X})$, $\log(F_{a,opt})$, $\log(T_{a,opt})$, $\alpha_{opt}$, and $\beta_{opt}$. Third row: clustering of GRBs using a combination of $\log(Fluence_{X})$, $\log(NH_{X})$, $\log(F_{a,opt})$, $\log(T_{a,opt})$, $\alpha_{opt}$, and $\beta_{opt}$. Fourth row: clustering of GRBs using a combination of $\log(Fluence_{X})$, $\log(Peak_{X})$, $\log(NH_{X})$, $\log(F_{a,opt})$, $\log(T_{a,opt})$, and $\alpha_{opt}$.}
    \label{fig:opt_all}
\end{figure*}

\begin{figure*}
    \begin{minipage}{\textwidth}
        \textbf{(a)}
        \centering
        \includegraphics[width=0.3\textwidth]{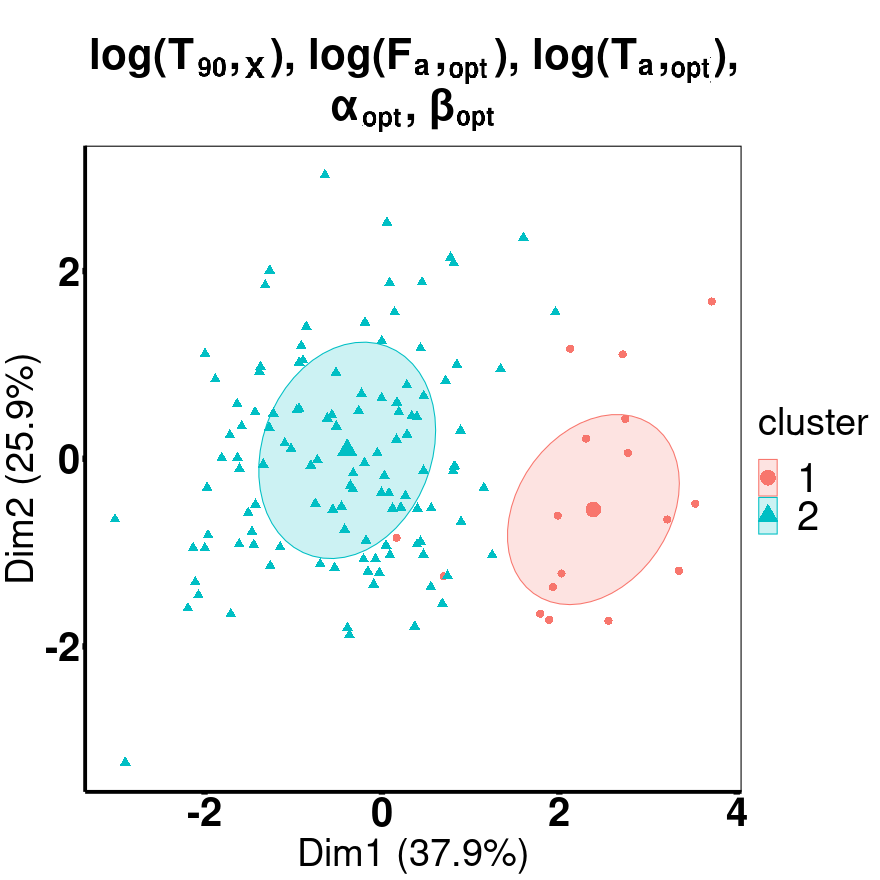}
        \includegraphics[width=0.53\textwidth]{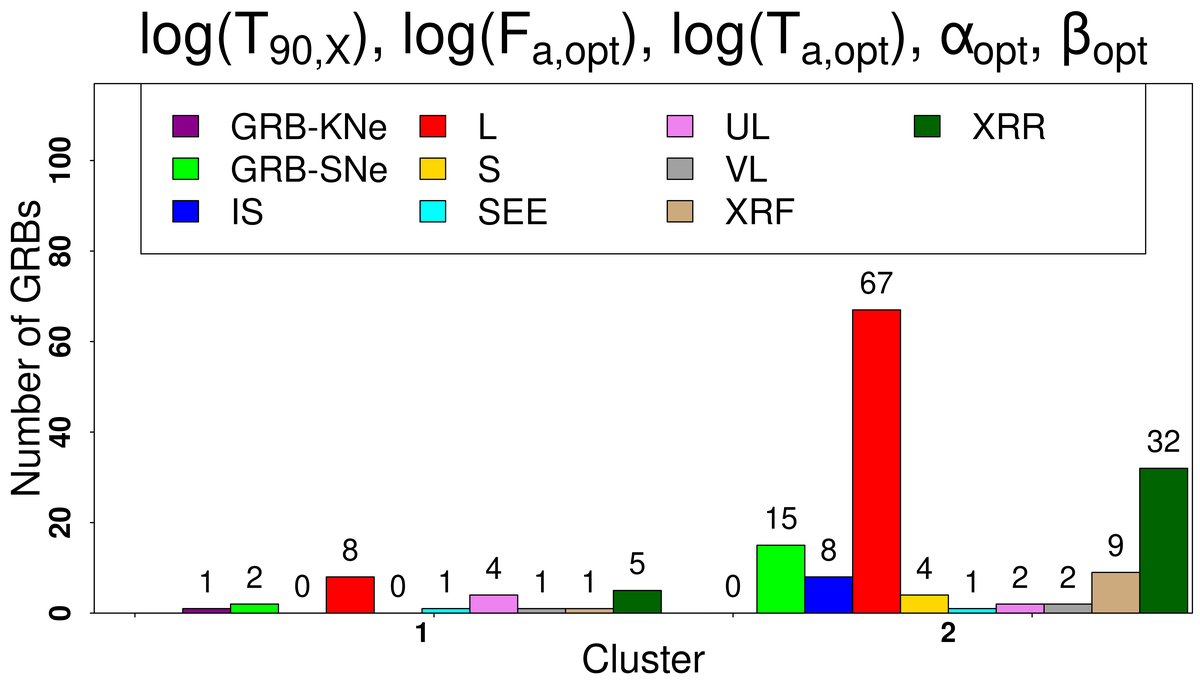}  
        \end{minipage}
    \begin{minipage}{\textwidth}
        \textbf{(b)}
        \centering
        \includegraphics[width=0.3\textwidth]{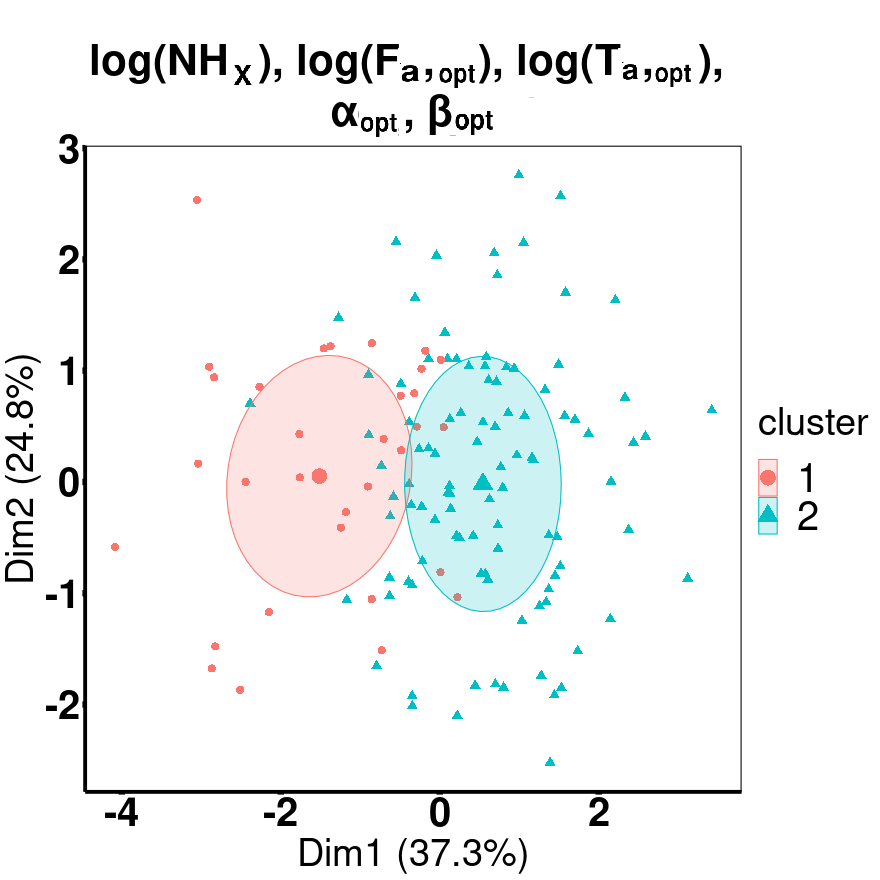}
        \includegraphics[width=0.53\textwidth]{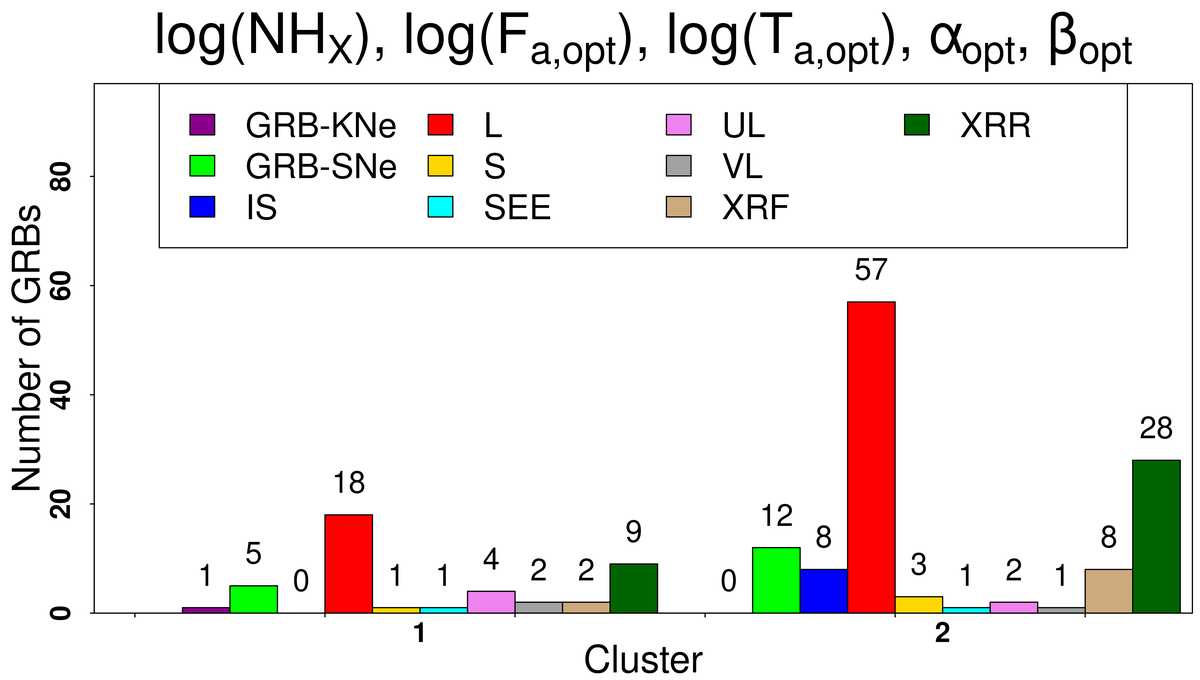}  
    \end{minipage}
    \begin{minipage}{\textwidth}
        \textbf{(c)}
        \centering
        \includegraphics[width=0.3\textwidth]{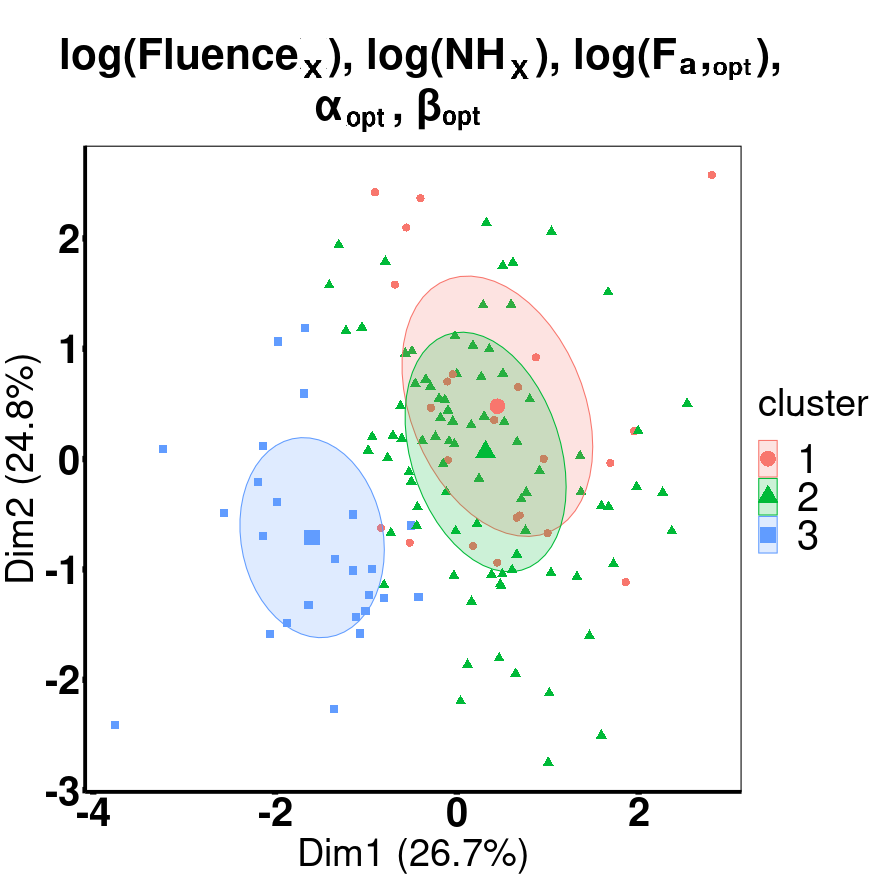}
        \includegraphics[width=0.53\textwidth]{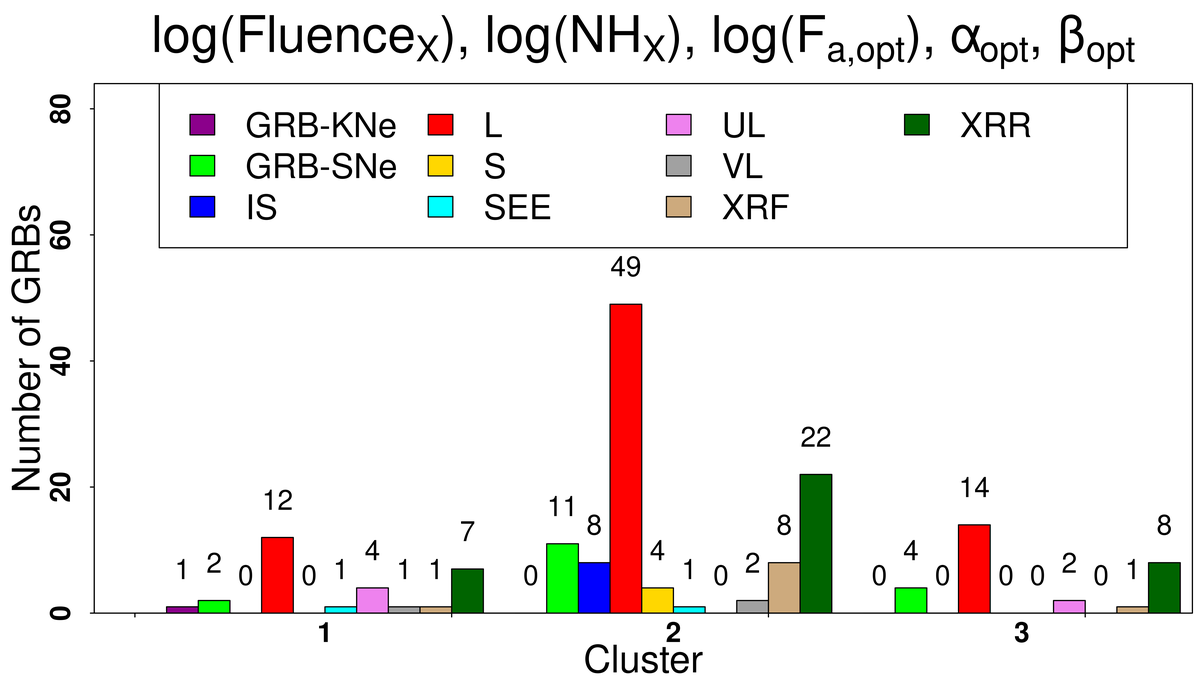}  
    \end{minipage}
    \begin{minipage}{\textwidth}
        \textbf{(d)}
        \centering
        \includegraphics[width=0.3\textwidth]{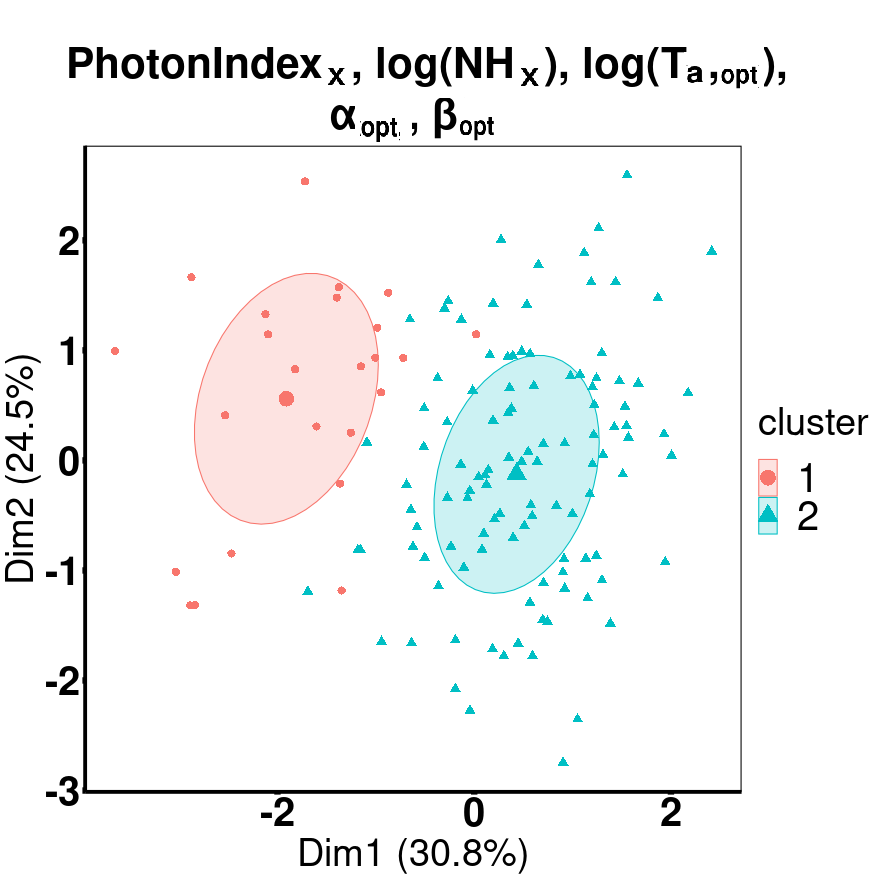}
        \includegraphics[width=0.53\textwidth]{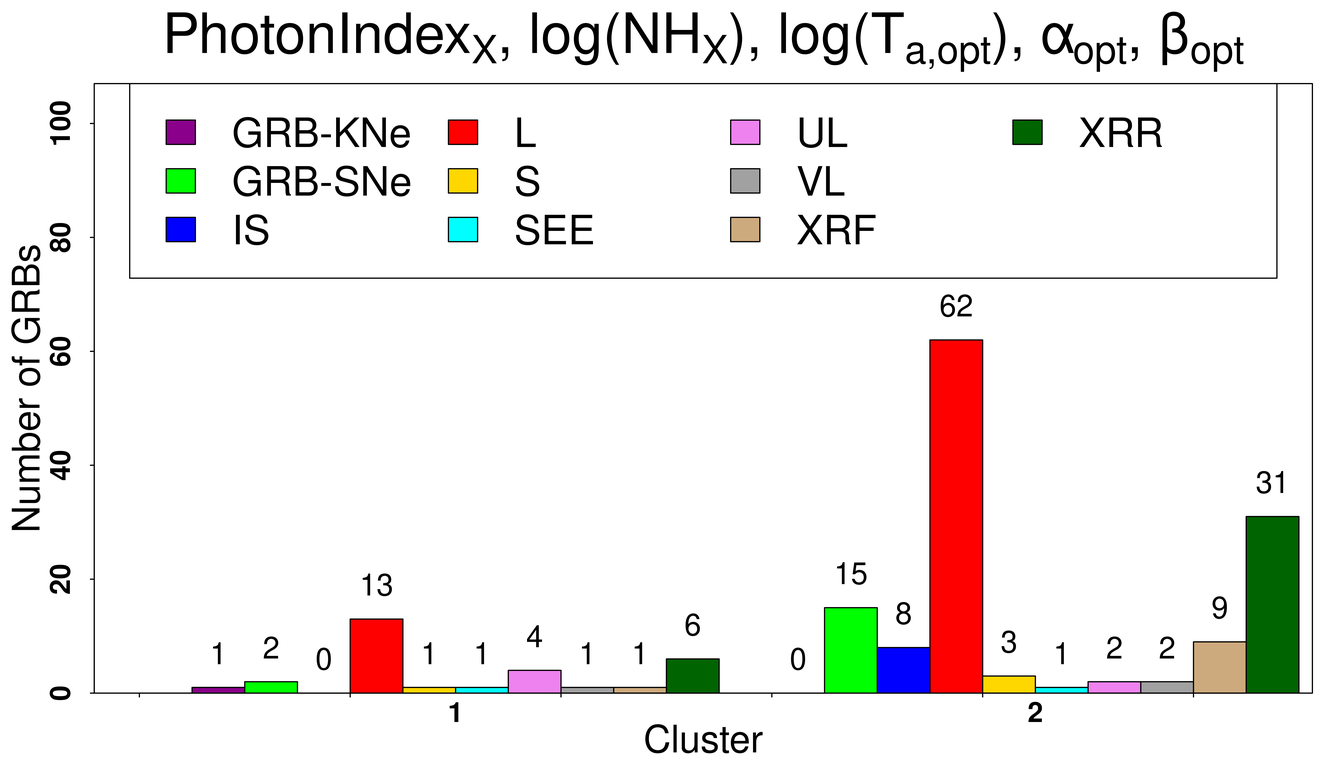}  
    \end{minipage}
    \caption{\small   
    In this set of figures, we reduce the parameters space to five variables in the optical sample as stated in (v), (vi), (vii), and (viii) of Sec. \ref{opt_res}.
    First row: clustering of GRBs using a combination of $\log(T_{90,X})$, $\log(F_{a,opt})$, $\log(T_{a,opt})$, $\alpha_{opt}$, and $\beta_{opt}$.
    Second row: clustering of GRBs using a combination of $\log(NH_{X})$, $\log(F_{a,opt})$, $\log(T_{a,opt})$, $\alpha_{opt}$, and $\beta_{opt}$.
    Third row: clustering of GRBs using a combination of $\log(Fluence_{X})$, $\log(NH_{X})$, $\log(F_{a,opt})$, $\alpha_{opt}$, and $\beta_{opt}$.
    Fourth row: clustering of GRBs using a combination of $PhotonIndex_{X}$, $\log(NH_{X})$, $\log(T_{a,opt})$, $\alpha_{opt}$, and $\beta_{opt}$.
    }
    \label{fig:opt_gls}
\end{figure*}

\begin{figure*}
    \begin{minipage}{\textwidth}
        \textbf{(a)}
        \centering
        \includegraphics[width=0.3\textwidth]{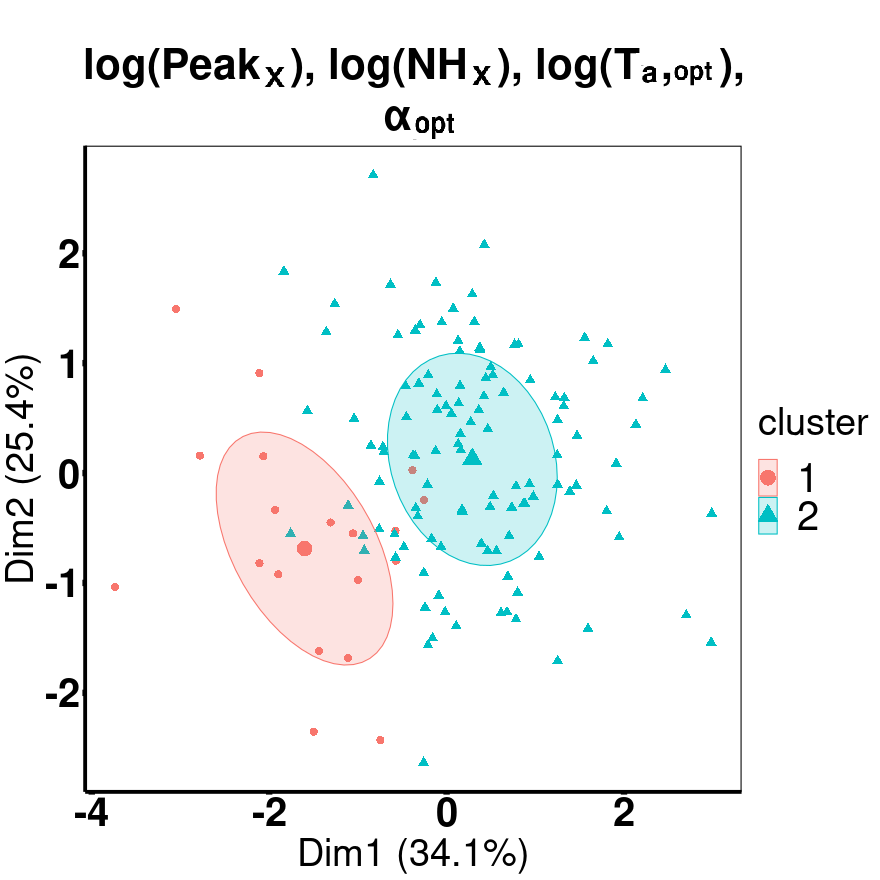}
        \includegraphics[width=0.53\textwidth]{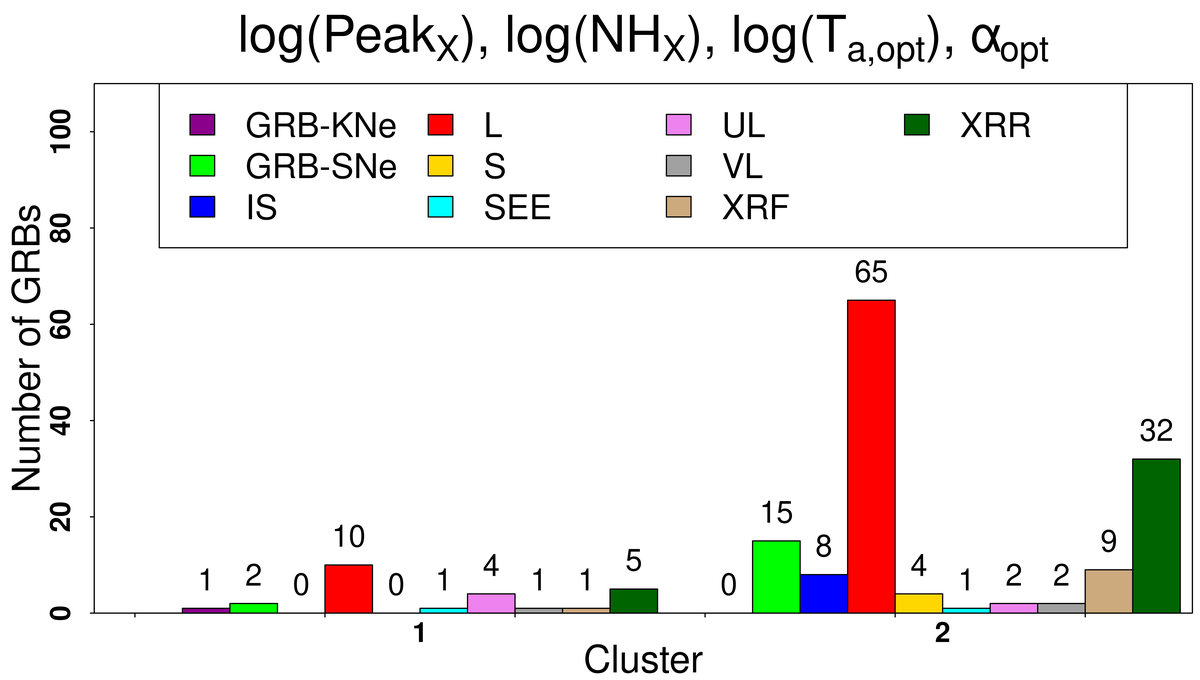}  
    \end{minipage}
    \begin{minipage}{\textwidth}
        \textbf{(b)}
       \centering
        \includegraphics[width=0.3\textwidth]{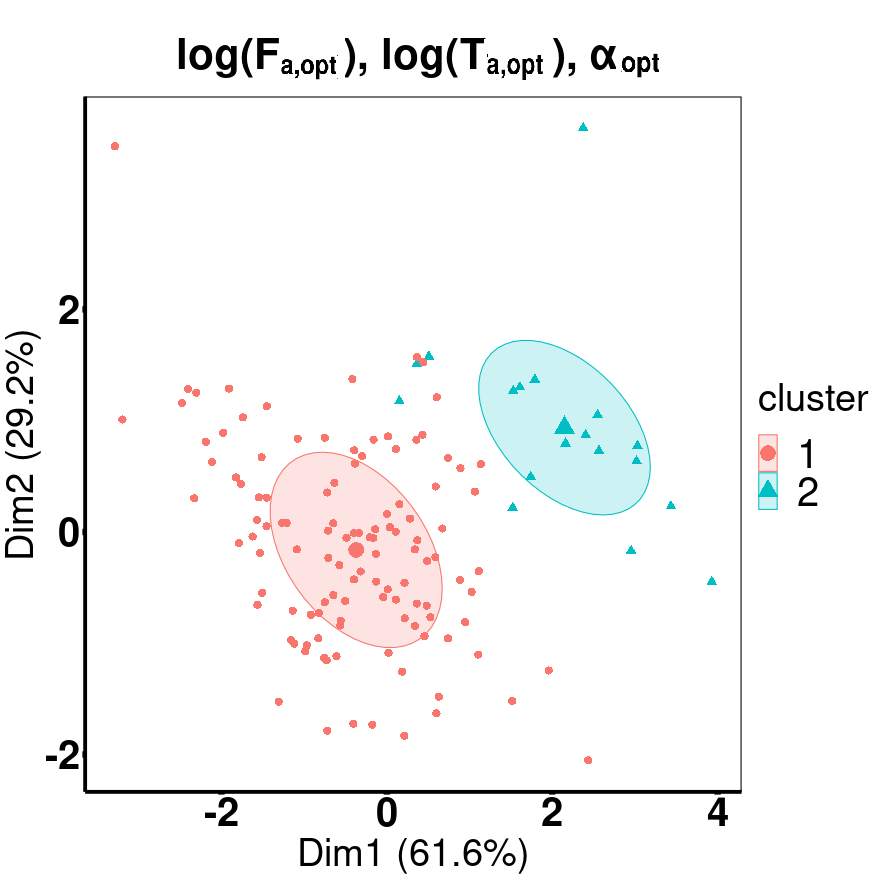}
        \includegraphics[width=0.53\textwidth]{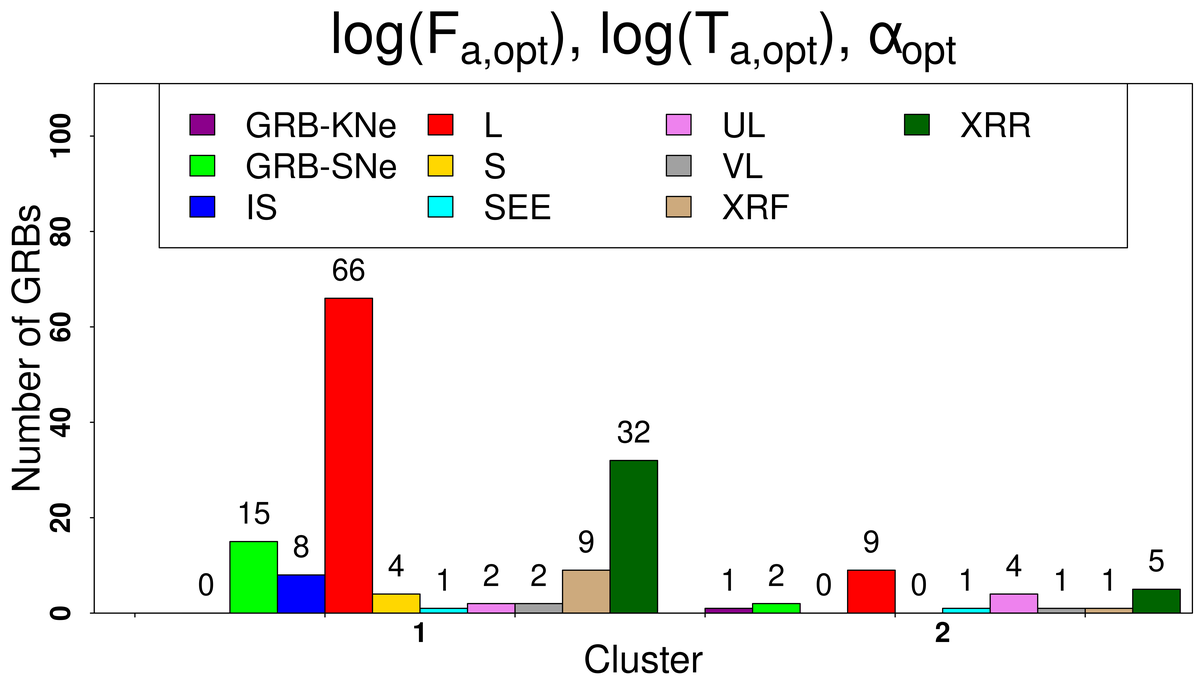}  
        \end{minipage}
    \begin{minipage}{\textwidth}
        \textbf{(c)}
       \centering
        \includegraphics[width=0.3\textwidth]{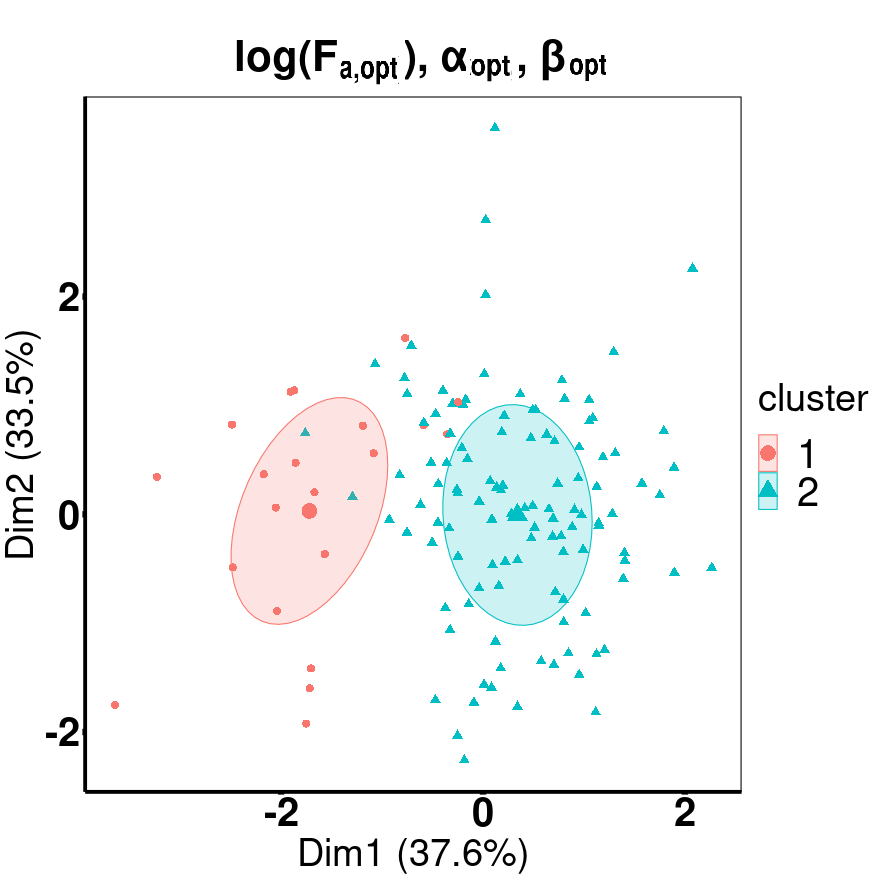}
        \includegraphics[width=0.53\textwidth]{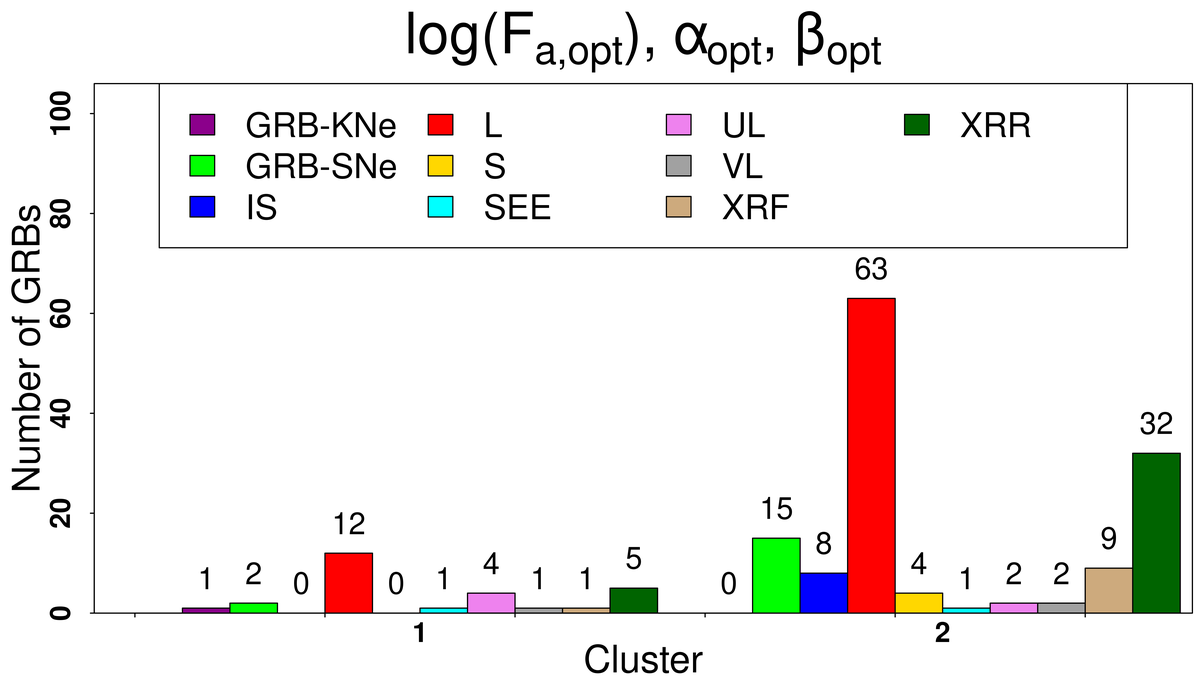}  
        \end{minipage}
    \begin{minipage}{\textwidth}
        \textbf{(d)}
       \centering
        \includegraphics[width=0.3\textwidth]{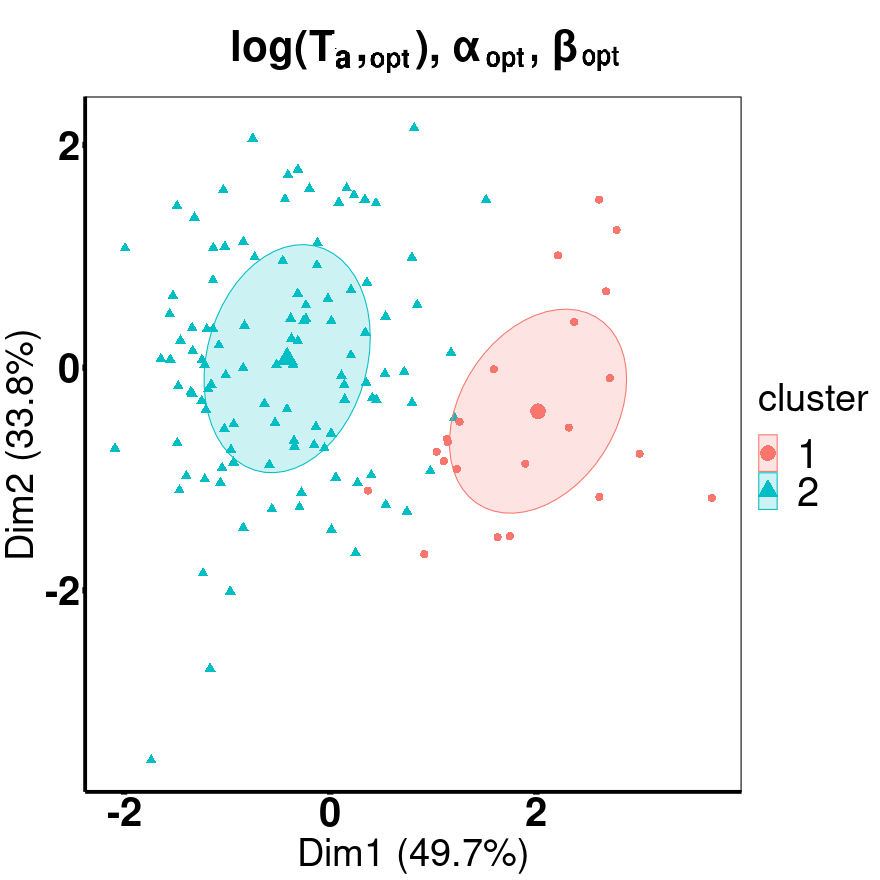}
        \includegraphics[width=0.53\textwidth]{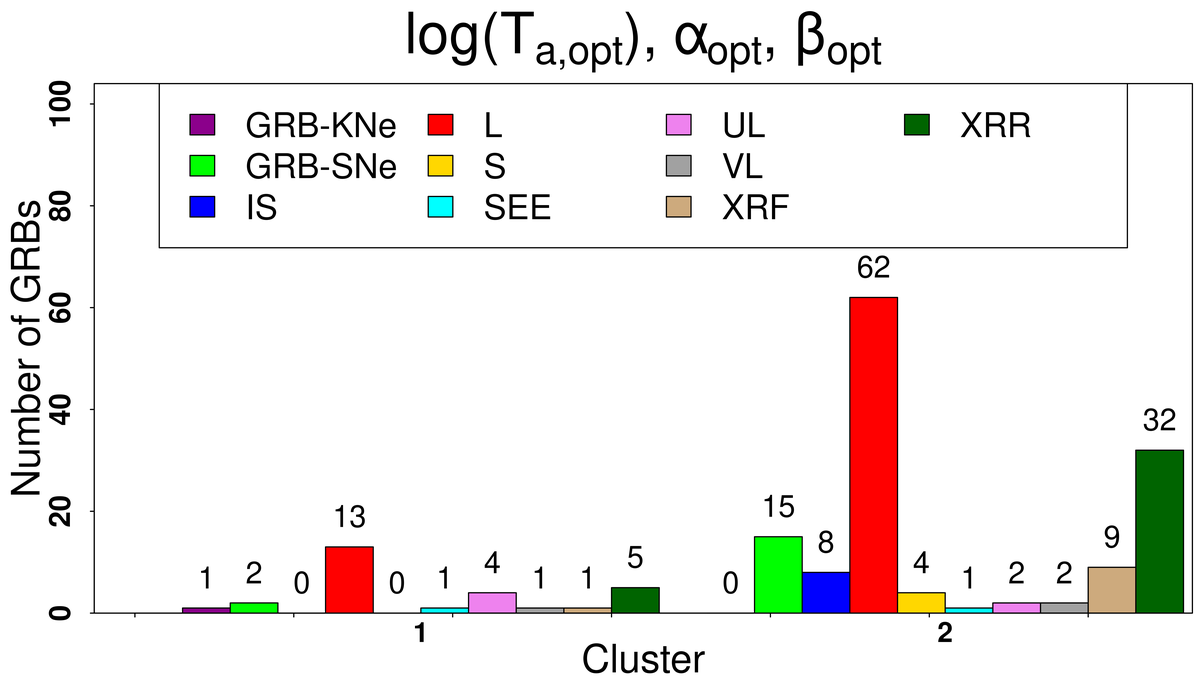}  
    \end{minipage}
    \caption{\small 
    This figure shows the third set of clustering results obtained for the optical sample using GMM for the last four combinations stated in Sec. \ref{opt_res}.
    First row: clustering of GRBs using a combination of $\log(Peak_{X})$, $\log(NH_{X})$, $\log(T_{a,opt})$, and $\alpha_{opt}$.
    Second row: clustering of GRBs using only three plateau parameters - $\log(F_{a,opt})$, $\log(T_{a,opt})$, and $\alpha_{opt}$.
    Third row: clustering of GRBs using only three plateau parameters - $\log(F_{a,opt})$, $\alpha_{opt}$, and $\beta_{opt}$.
    Fourth row: clustering of GRBs using only three plateau parameters - $\log(T_{a,opt})$, $\alpha_{opt}$, and $\beta_{opt}$.
    }
    \label{fig:opt_gls_9}
\end{figure*}

We have started the analysis by clustering GRBs from our dataset having ten major classes: S (SGRBs), L (LGRBs), IS, SEE, GRB-SNe, UL, XRF, XRR, GRB-KNe, and VL (Fig. \ref{fig:pie} c). The left panels in Fig. \ref{fig:opt_all}, Fig. \ref{fig:opt_gls}, and Fig. \ref{fig:opt_gls_9} show the cluster plot for the parameter combinations stated above. In these scatter plots, the \textit{x} and \textit{y} axes are the first two principal components derived from Principal Component Analysis (PCA) for the respective parameter combination. The optimal number of clusters obtained for each case is based on the maximum BIC value (see Table \ref{BIC Table}). For the majority of the cases, the optimal number of clusters is two, regardless of the number of classes we have in our dataset. To interpret these cluster plots and to check which GRB class is grouped in which cluster, we plotted the distribution corresponding to each case. These distribution plots are shown in the right panels of Fig. \ref{fig:opt_all}, \ref{fig:opt_gls}, and \ref{fig:opt_gls_9}, respectively. These plots show a skewed and non-uniform distribution of different classes of GRBs among the clusters. However, we spotted some micro trends in these results, which we explain further in Sec \ref{opt_mcirotrends}.

\begin{table}
\begin{center}
\begin{tabular}{|l|l|l|l|}
\hline
\multicolumn{1}{|l|}{Parameter Combination} &
\multicolumn{2}{|l|}{Optimal number of clusters} & \\ \cline{2-3}
\multicolumn{1}{|l|}{} & Optical & X-ray \\
\hline
(i) All parameters (with $\log(T^{*}_{90})$) & 2 & 3 \\
(ii) $\log(T_{90})$, $\log(NH)$, & 2 & 3 \\
$\log(F_{a})$, $\log(T_{a})$, $\alpha$, and $\beta$ & & \\
(iii) $\log(Fluence)$, $\log(NH)$, & 2 & 4 \\
$\log(F_{a})$, $\log(T_{a})$, $\alpha$, and $\beta$ &  &  \\
(iv) $\log(Fluence)$, $\log(Peak)$ & 2 & 4 \\
$\log(NH)$, $\log(F_{a})$, $\log(T_{a})$, and $\alpha$ & & \\
(v) $\log(T_{90})$, $\log(F_{a})$, & 2 & 7 \\
$\log(T_{a})$, $\alpha$, and $\beta$ & & \\
(vi) $\log(NH)$, $\log(F_{a})$, & 2 & 5 \\
$\log(T_{a})$, $\alpha$, and $\beta$ & & \\
(vii) $\log(Fluence)$, $\log(NH)$, & 3 & 4 \\
$\log(F_{a})$, $\alpha$, and $\beta$ & & \\
(viii) $PhotonIndex$, $\log(NH)$, & 2 & 5 \\
$\log(T_{a})$, $\alpha$, and $\beta$ & & \\
(ix) $\log(Peak)$, $\log(NH)$, & 2 & 4 \\
$\log(T_{a})$, and $\alpha$ & & \\
(x) $\log(F_{a})$, $\log(T_{a})$, and $\alpha$ & 2 & 3 \\
(xi) $\log(F_{a})$, $\alpha$, and $\beta$ & 2 & 4 \\
(xii) $\log(T_{a})$, $\alpha$, and $\beta$ & 2 & 6 \\
\hline
\end{tabular}
\caption{The table presents the optimal number of clusters utilized by GMM based on the maximum BIC values for the parameter combinations employed in this work for both optical and X-ray data. Since the same parameter space was utilized for both the optical and X-ray data, the parameter space presented in the table is generic and does not include any specific subscripts.}
\label{BIC Table}
\end{center}
\end{table}

\subsubsection{Microtrends in optical}\label{opt_mcirotrends}

Even though clustering does not provide a clear distinction in the overall distribution, it reveals repetitive grouping of specific GRB classes, which we have highlighted below. 

\begin{enumerate}
    \item {For instance, the majority of XRR, XRF, and GRB-SNe can be found within the cluster with the largest occurrence of the LGRBs, as evident in the right panels of Fig. \ref{fig:opt_all} a, b, c, d, Fig. \ref{fig:opt_gls} a, b, c, d, and Fig. \ref{fig:opt_gls_9} a, b, c, d.}
    \item {It is worth noting that in most cases, the majority of ULs are found in a different cluster that has a lower number of LGRBs instead of being grouped into the cluster that has the highest number of LGRBs (right panels of Fig. \ref{fig:opt_all} a, b, c, d, Fig. \ref{fig:opt_gls} a, b, c, d, and Fig. \ref{fig:opt_gls_9} a, b, c, d).}
    \item {Counter-intuitively to what we anticipated, we can see that the majority of the four SGRBs almost always are grouped within the cluster that contains most of the LGRBs (right panels in Fig. \ref{fig:opt_all} a, b, c, d, Fig. \ref{fig:opt_gls} a, b, c, d, and Fig. \ref{fig:opt_gls_9} a, b, c, d).}
    \item {The majority of IS samples also follow the same trend as stated above, i.e., they also go along the same cluster with the highest LGRBs (right panels in Fig. \ref{fig:opt_all} a, b, c, d, Fig. \ref{fig:opt_gls} a, b, c, d, and Fig. \ref{fig:opt_gls_9} a, b, c, d).}
    \item For the VL GRBs, in 10 cases (Fig. \ref{fig:opt_all} b, c, d, Fig. \ref{fig:opt_gls} a, c, d and Fig. \ref{fig:opt_gls_9} a, b, c, d) majority belong to the cluster with the largest number of LGRBs. However, in 2 cases (Fig. \ref{fig:opt_all} a and Fig. \ref{fig:opt_gls} b) majority of VLs belong to the cluster with fewer LGRBs.
\end{enumerate}
These microtrend findings, as stated in (i), (ii), (iii), (iv), and (v), point toward some interesting conclusions, which we have highlighted in Sec.~\ref{summary and conclusion}.

\subsection{X-ray GRB Data}
\label{xray_res}

We employed the same methodology as in Sec \ref{opt_res} to analyze the X-ray data. Similar to the optical data analysis, we started with all eleven parameters and iteratively reduced them to three parameters covering different combinations (1981 in total). All 1981 combinations yielded at least two clusters.

For the X-ray sample as well, the reason for primarily focusing on plateau parameters is to investigate the characterization of GRBs based on the magnetar properties. We present the same parameter combinations as in the optical case listed below.

\begin{enumerate}
    \item All eleven parameters with $\log(T^{*}_{90,X})$.
    \item We here start to reduce the parameter space to six variables, focusing on the variables of the plateau emission and changing each by each one of the afterglow or of the prompt emission. Thus, we use $\log(F_{a,X})$, $\log(T_{a,X})$, $\alpha_{X}$, $\beta_{X}$, and we have added the prompt parameters $\log(T_{90,X})$,  and $\log(NH_{X})$.
    \item To ii), we just changed the variable $\log(T_{90,X})$ with $\log(Fluence_{X})$.
    \item To iii), we just changed the variable $\beta_{X}$ with $\log(Peak_{X})$.
    \item Now we reduce to five combinations with $\log(T_{90,X})$, $\log(F_{a,X})$, $\log(T_{a,X})$, $\alpha_{X}$, and $\beta_{X}$.
    \item To v), we have changed the variable $\log(T_{90,X})$ with $\log(NH_{X})$.
    \item To vi), we change parameter $\log(T_{a,X})$ with $\log(Fluence_{X})$.
    \item To vi), we change the parameter $\log(F_{a,X})$ with $PhotonIndex_{X}$.
    \item We reduce to a combination of four: $\log(Peak_{X})$, $\log(NH_{X})$, $\log(T_{a,X})$, $\alpha_{X}$.
    \item We here reduce to a parameter of three plateau variables $\log(F_{a,X})$, $\log(T_{a,X})$, $\alpha_{X}$.
    \item We reduce to a parameter of three plateau variables $\log(F_{a,X})$, $\alpha_{X}$, $\beta_{X}$.
    \item We here reduce to a parameter of three plateau variables $\log(T_{a,X})$, $\alpha_{X}$, $\beta_{X}$.
\end{enumerate}

We started the analysis by clustering GRBs from our dataset, having ten major classes: S (SGRBs), L (LGRBs), IS, SEE, GRB-SNe, UL, XRF, XRR, GRB-KNe, and VL (Fig. \ref{fig:pie} d). The left panel in Fig. \ref{fig:xr_all}, \ref{fig:xr_gls}, and \ref{fig:xr_gls_balanced} shows the cluster plot for the above parameter combinations. For the same parameter combinations mentioned above, we only obtain two (in one case three, see Table \ref{BIC Table}) clusters based on the maximum BIC value with the optical data (Sec. \ref{opt_res}). While with the X-ray data, the optimal number of clusters varies from three to seven based on the maximum BIC value (see Table \ref{BIC Table}). For example, the left panel in Fig. \ref{fig:xr_all} d shows four clusters. Whereas for the same parameters as in Fig. \ref{fig:xr_all} d, in the optical data (left panel of Fig. \ref{fig:opt_all} d), the number of clusters is two. To interpret these cluster plots and to check which GRB classes are grouped into which clusters, we plotted the distributions corresponding to each case, as shown in the right panels of Fig. \ref{fig:xr_all}, \ref{fig:xr_gls}, and \ref{fig:xr_gls_balanced}, respectively. We again spotted some micro trends in these results, almost similar to the optical microtrends (Sec. \ref{opt_mcirotrends}), which highlight Sec \ref{xray_microtrends}.

\subsubsection{Microtrends in X-rays} \label{xray_microtrends}

For the X-ray dataset, we also pinpoint the same major repetitive grouping of specific GRB classes, as found in the optical data (Sec. \ref{opt_mcirotrends}), {outlined} below.

\begin{figure*}
    \begin{minipage}{\textwidth}
        \textbf{(a)}
        \centering
        \includegraphics[width=0.3\textwidth]{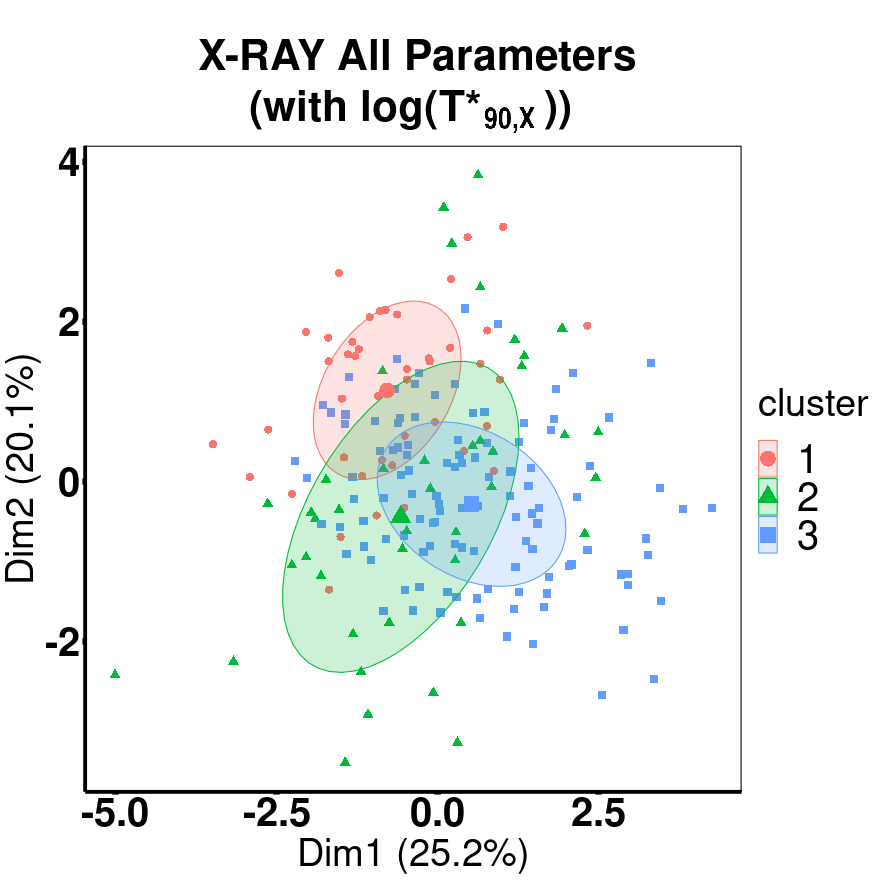}
        \includegraphics[width=0.46\textwidth]{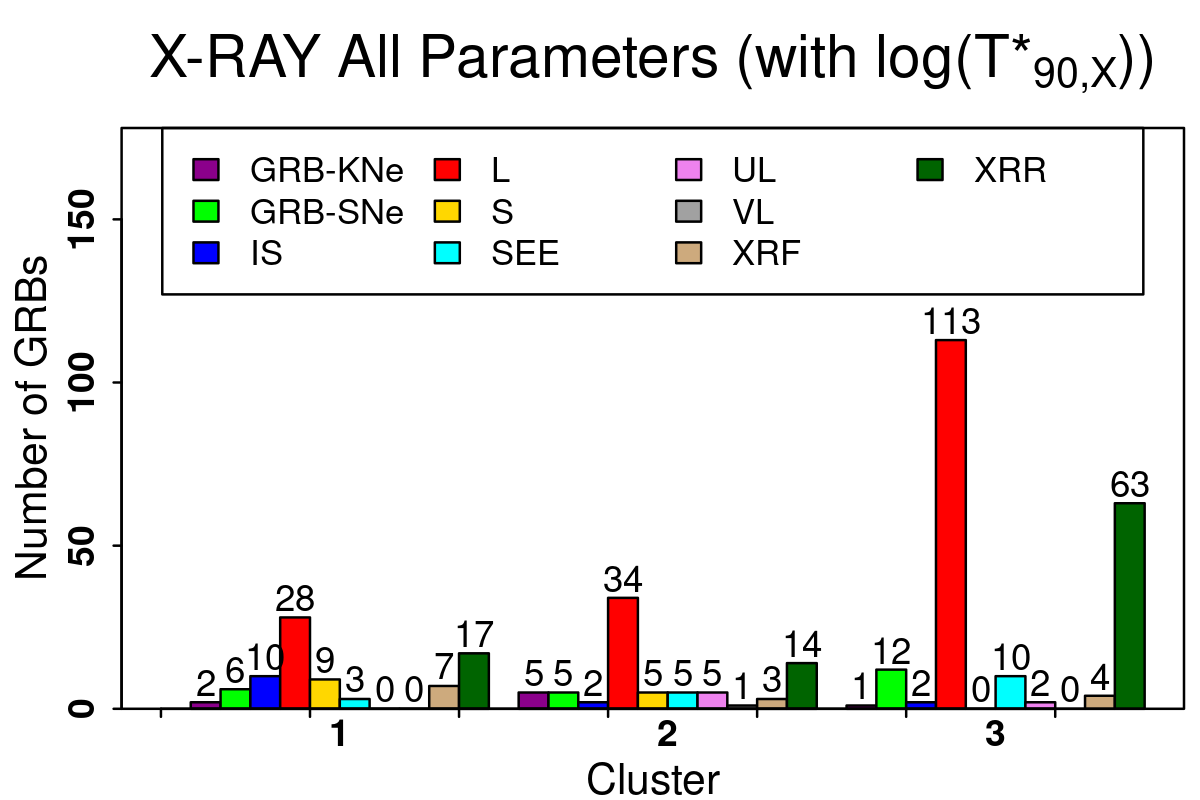}
    \end{minipage}

    \begin{minipage}{\textwidth}
    \textbf{(b)}
        \centering
        \includegraphics[width=0.3\textwidth]{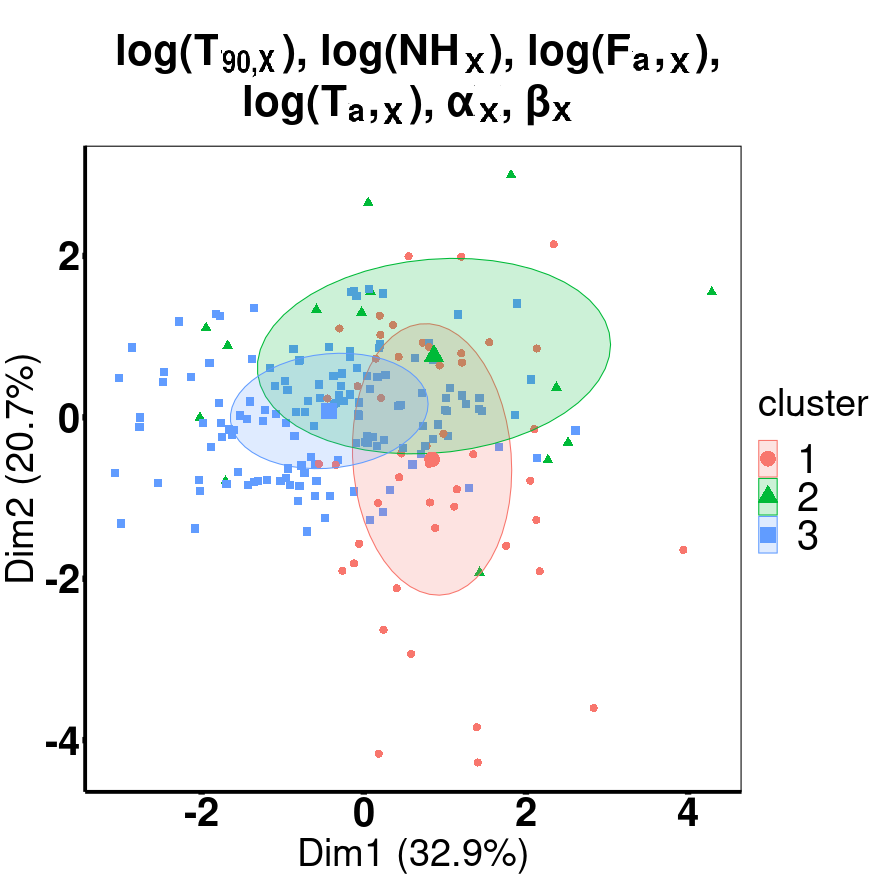} 
        \includegraphics[width=0.46\textwidth]{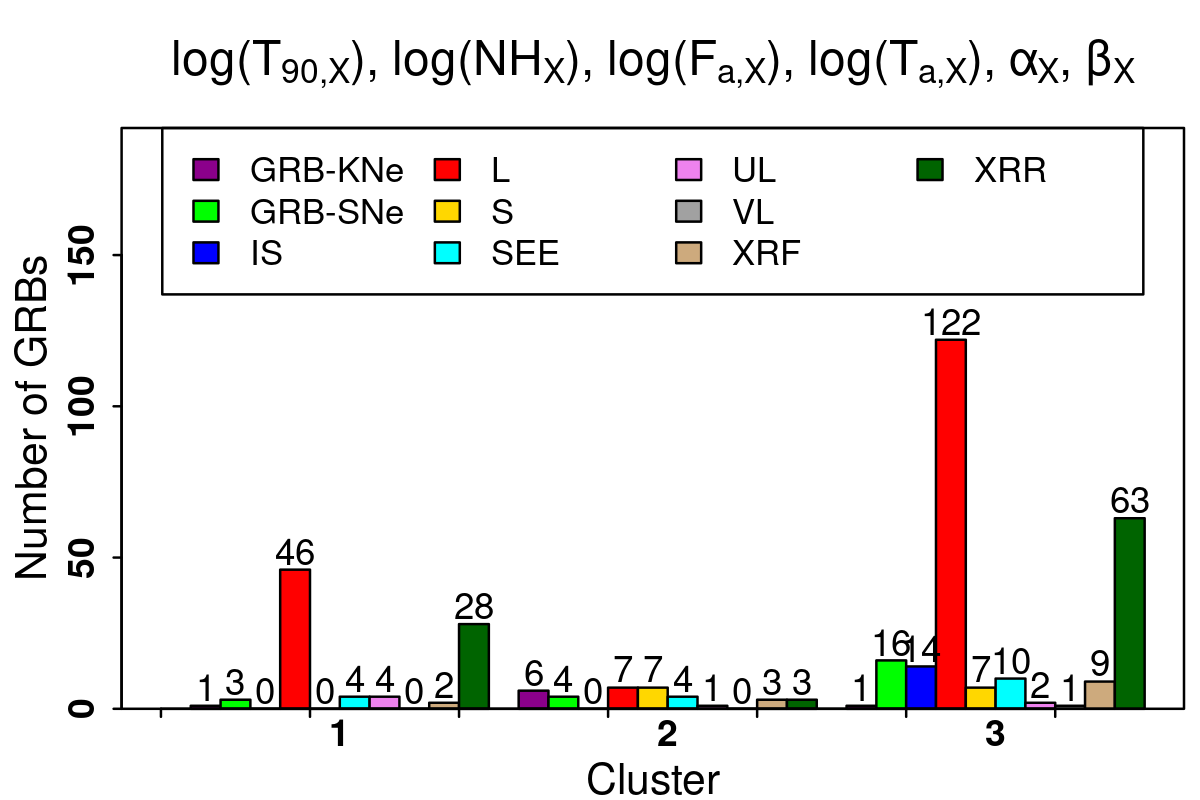}
    \end{minipage}
    
    \begin{minipage}{\textwidth}
    \textbf{(c)}
        \centering       
        \includegraphics[width=0.3\textwidth]{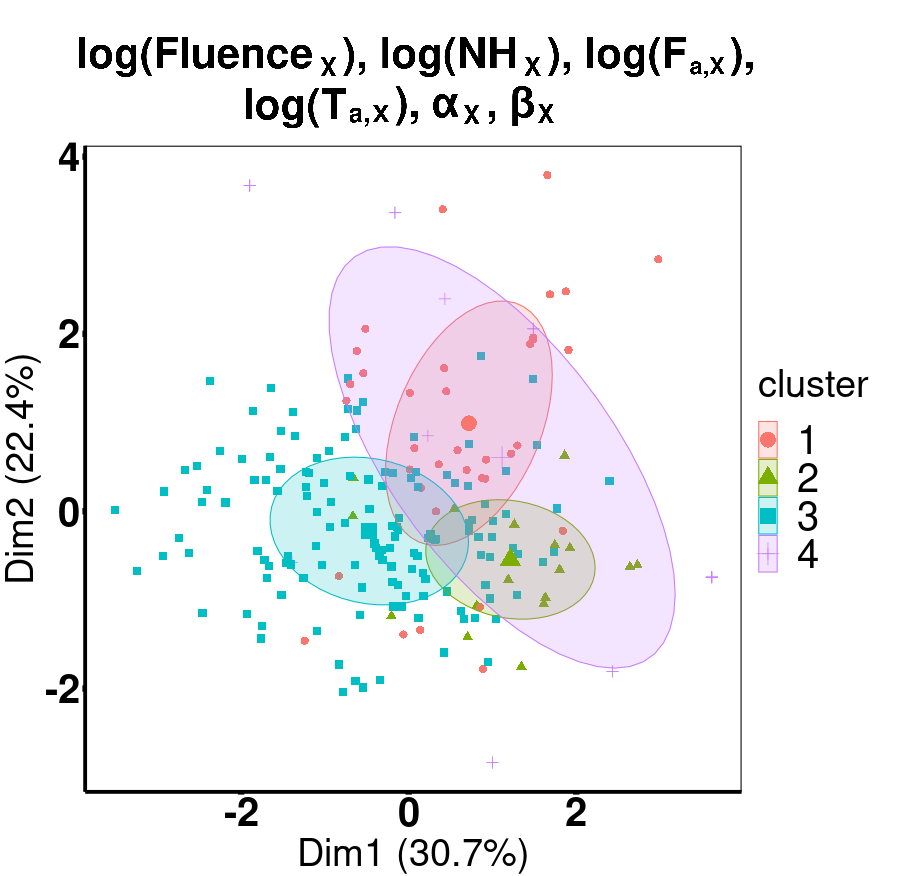}
        \includegraphics[width=0.46\textwidth]{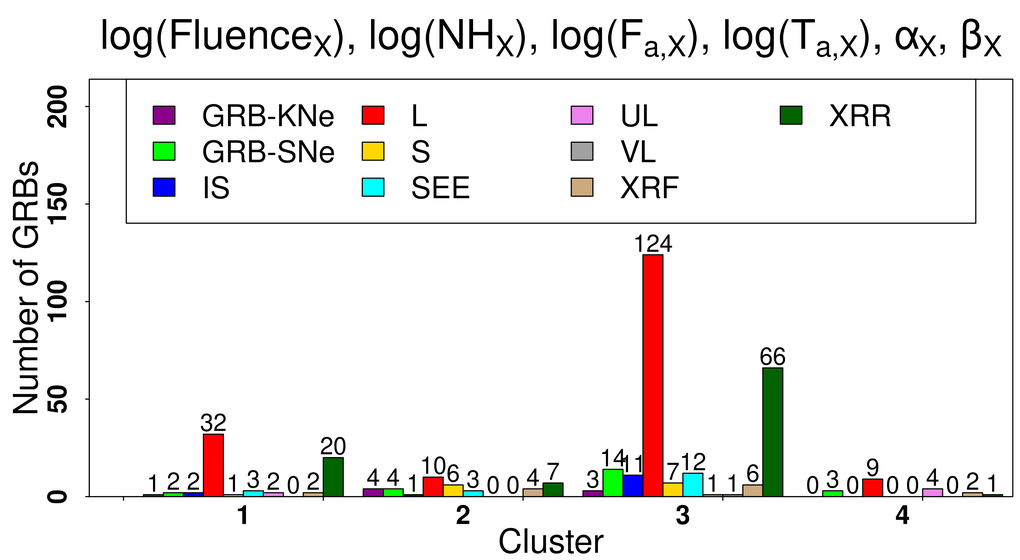}
    \end{minipage}
    
    \begin{minipage}{\textwidth}
    \textbf{(d)}
        \centering
        \includegraphics[width=0.3\textwidth]{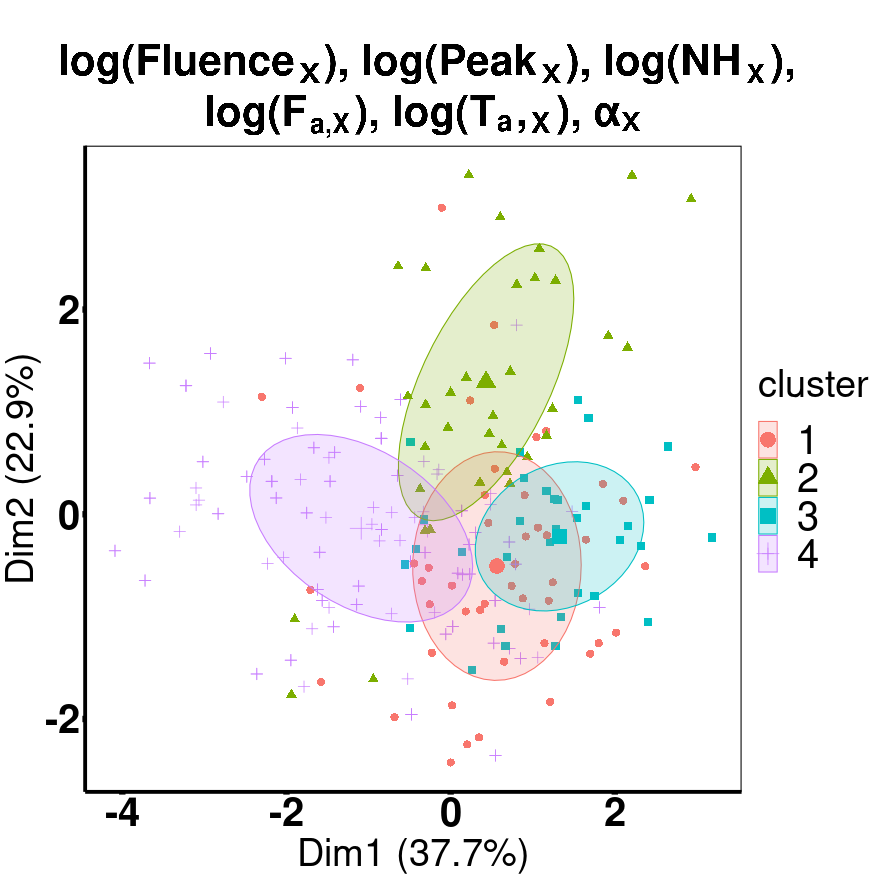}
        \includegraphics[width=0.46\textwidth]{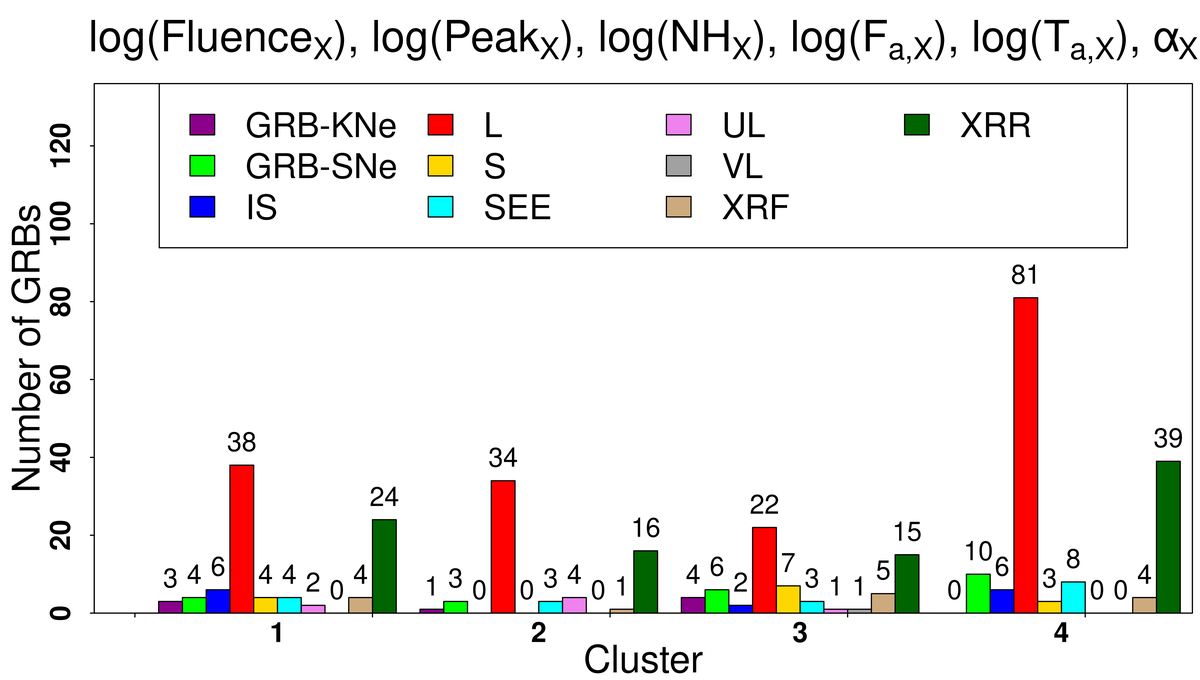}
    \end{minipage}
    \caption{\small 
    This figure shows the first set of four clustering results obtained for the X-ray sample using GMM. The left column shows the scatter plot of the clusters obtained for the first four parameter space mentioned in Sec. \ref{xray_res}. The right column shows the distribution of the GRBs within each cluster based on its eight major classes (Fig. \ref{fig:pie} b). `Dim1' and `Dim2' are the first two principal components used to describe the multi-parameter clustering graphically. First row: clustering of GRBs using all eleven parameters (with $\log(T^{*}_{90,X})$). Second row: clustering of GRBs using a combination of $\log(T_{90,X})$, $\log(NH_{X})$, $\log(F_{a,X})$, $\log(T_{a,X})$, $\alpha_{X}$, and $\beta_{X}$. Third row: clustering of GRBs using a combination of $\log(Fluence_{X})$, $\log(NH_{X})$, $\log(F_{a,X})$, $\log(T_{a,X})$, $\alpha_{X}$, and $\beta_{X}$. Fourth row: clustering of GRBs using a combination of $\log(Fluence_{X})$, $\log(Peak_{X})$, $\log(NH_{X})$, $\log(F_{a,X})$, $\log(T_{a,X})$, and $\alpha_{X}$.}
    \label{fig:xr_all}
\end{figure*}

\begin{figure*}
    \begin{minipage}{\textwidth}
        \textbf{(a)}
        \centering
        \includegraphics[width=0.3\textwidth]{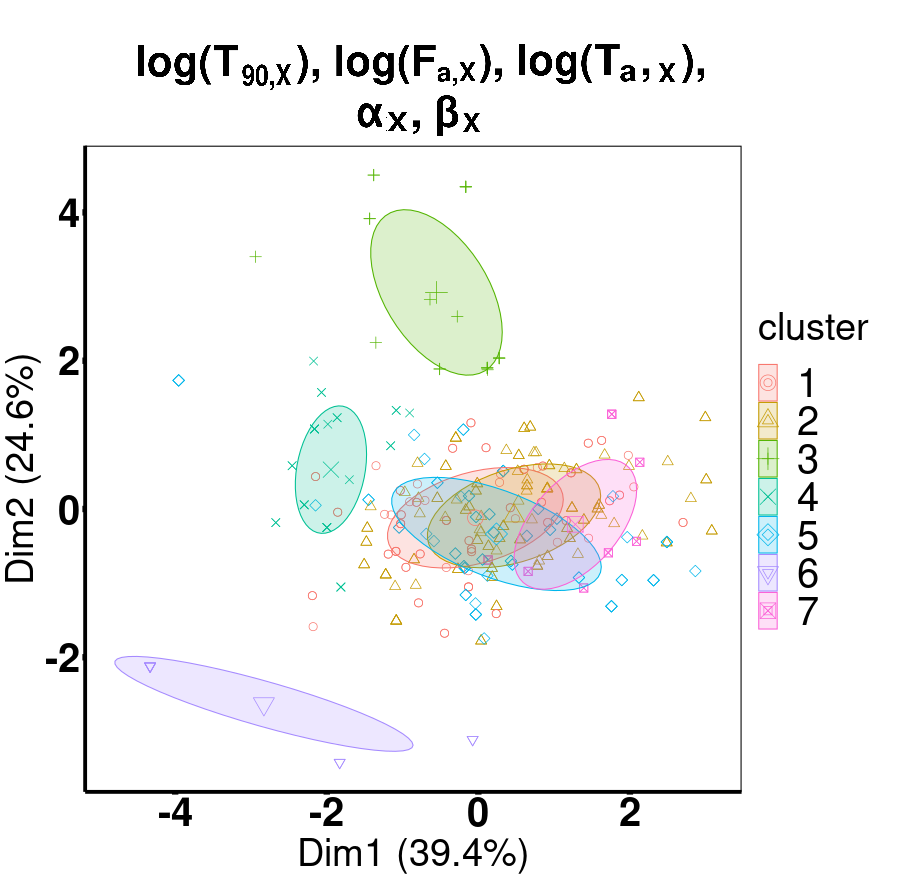}
        \includegraphics[width=0.50\textwidth]{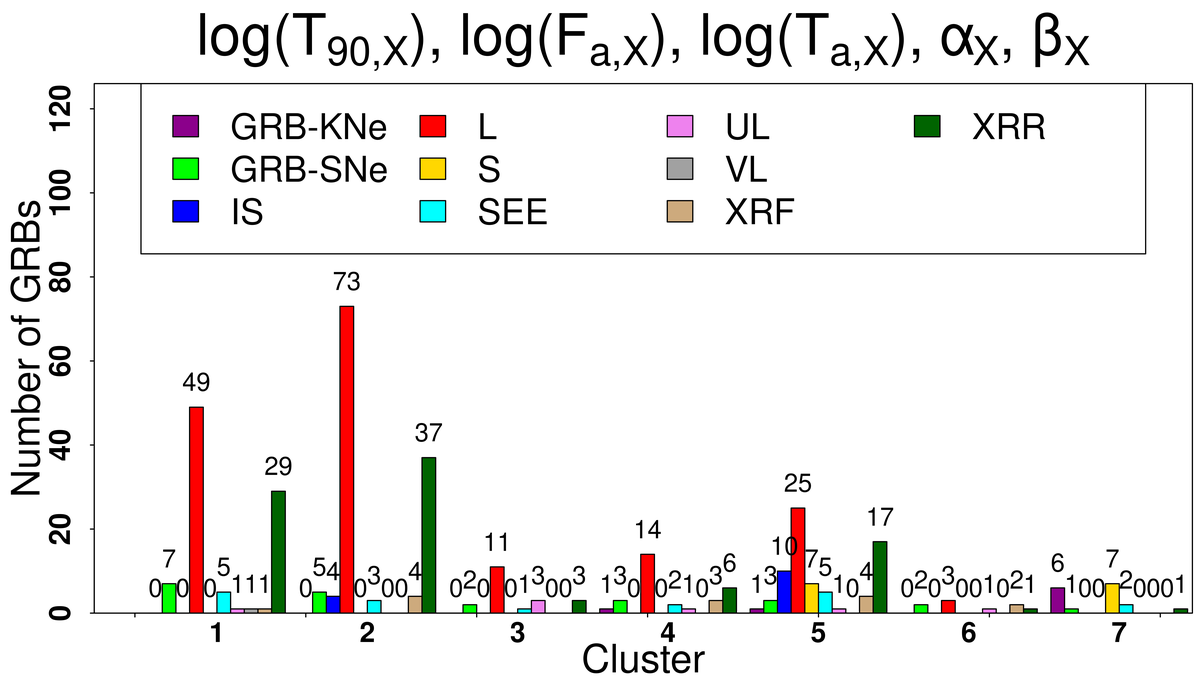}
    \end{minipage}

    \begin{minipage}{\textwidth}
    \textbf{(b)}
        \centering
        \includegraphics[width=0.3\textwidth]{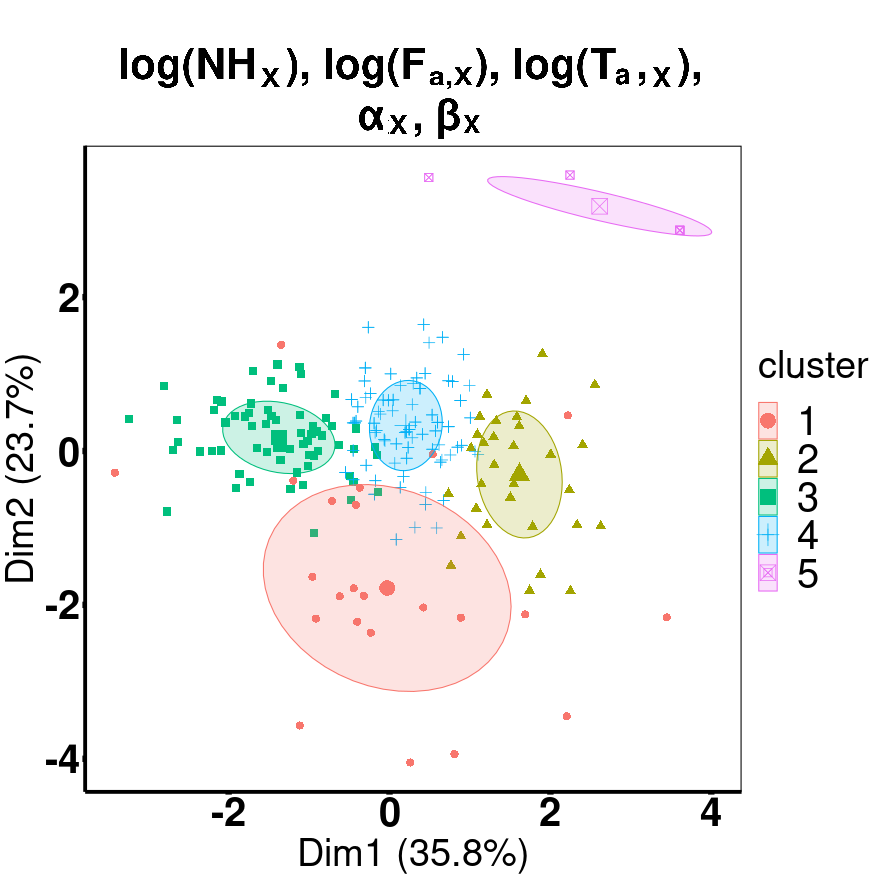} 
        \includegraphics[width=0.50\textwidth]{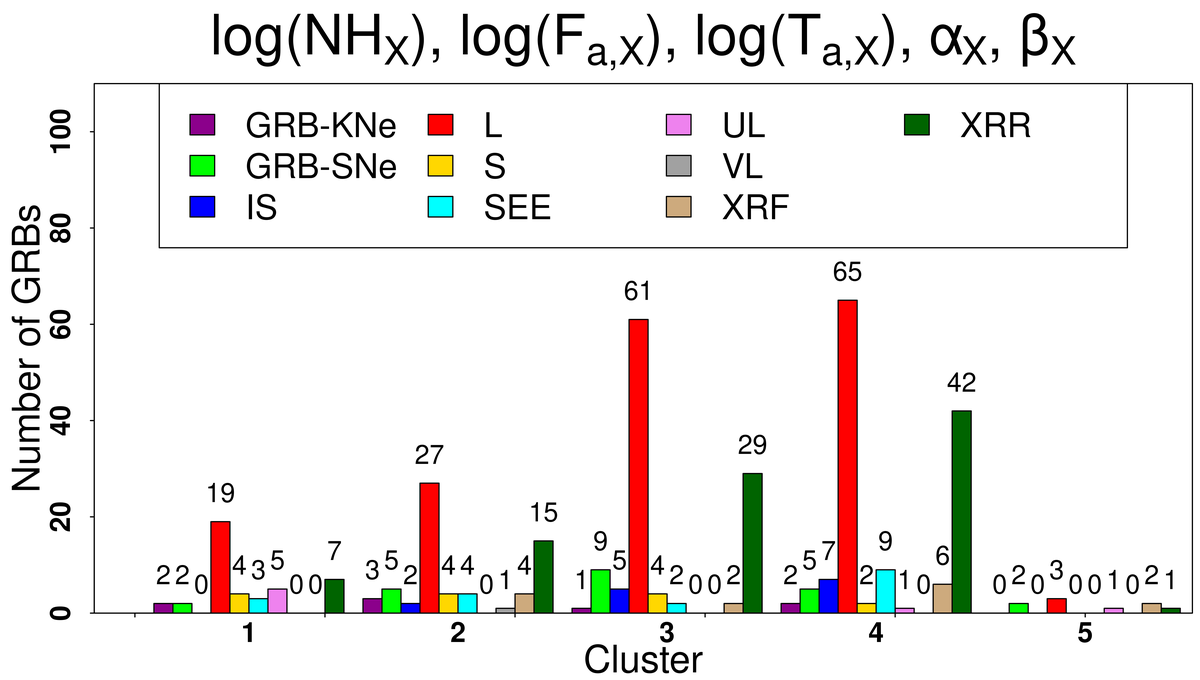}
    \end{minipage}
    
    \begin{minipage}{\textwidth}
    \textbf{(c)}
        \centering       
        \includegraphics[width=0.3\textwidth]{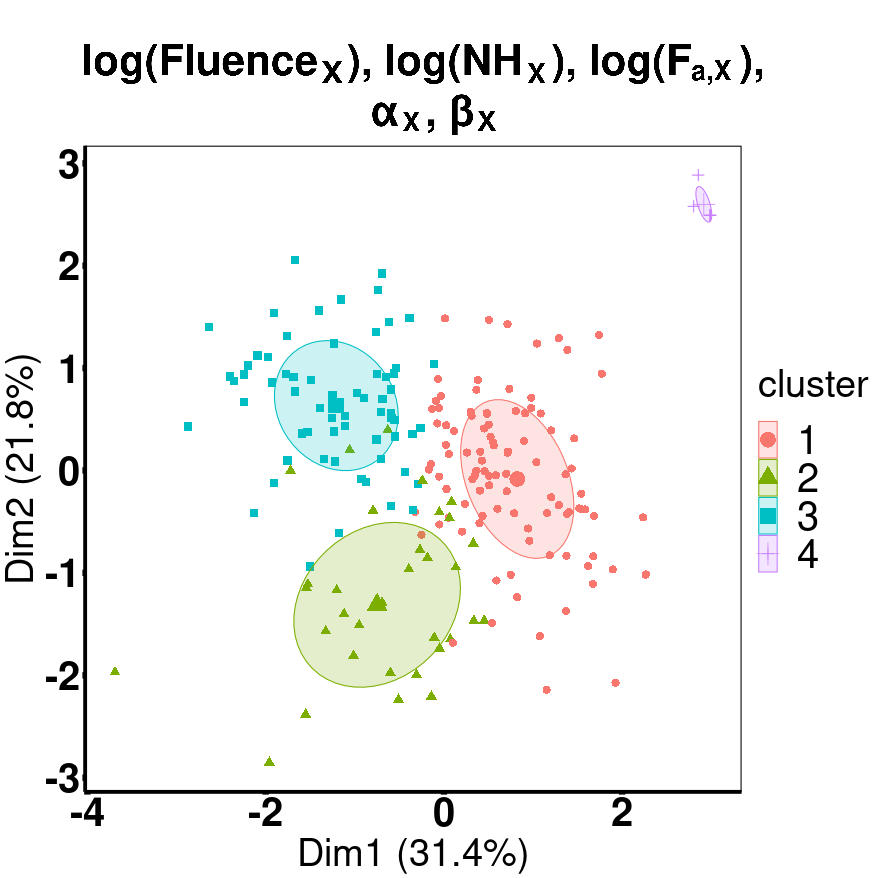}
        \includegraphics[width=0.50\textwidth]{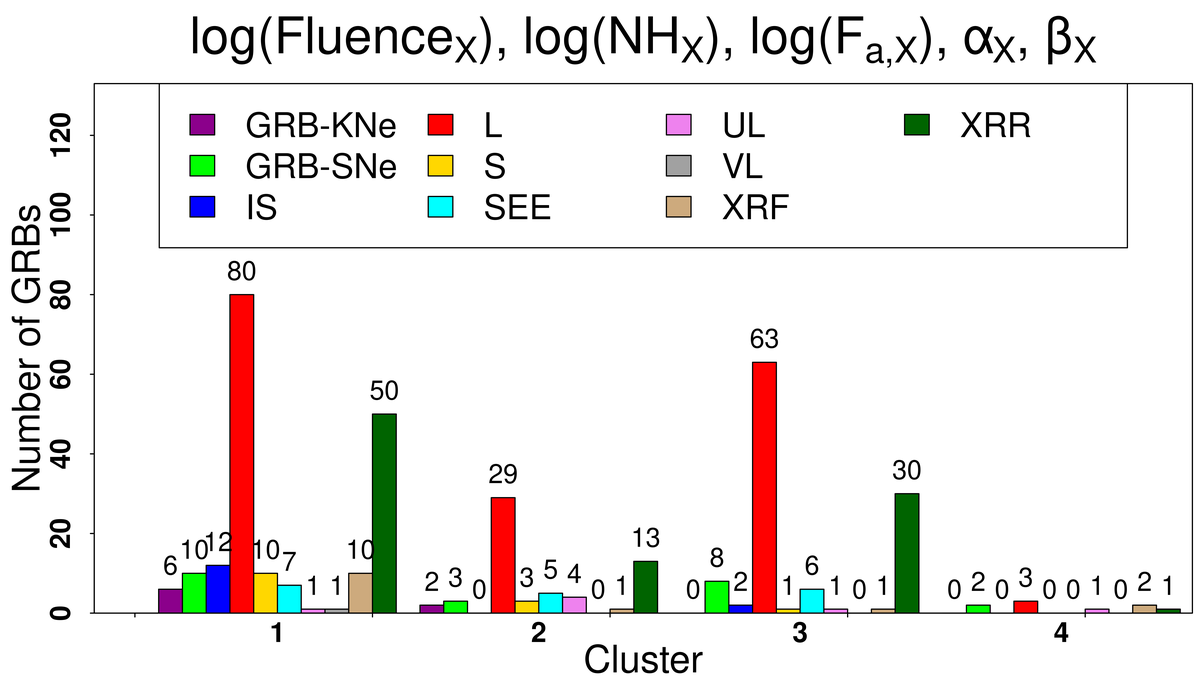}
    \end{minipage}
    
    \begin{minipage}{\textwidth}
    \textbf{(d)}
        \centering
        \includegraphics[width=0.3\textwidth]{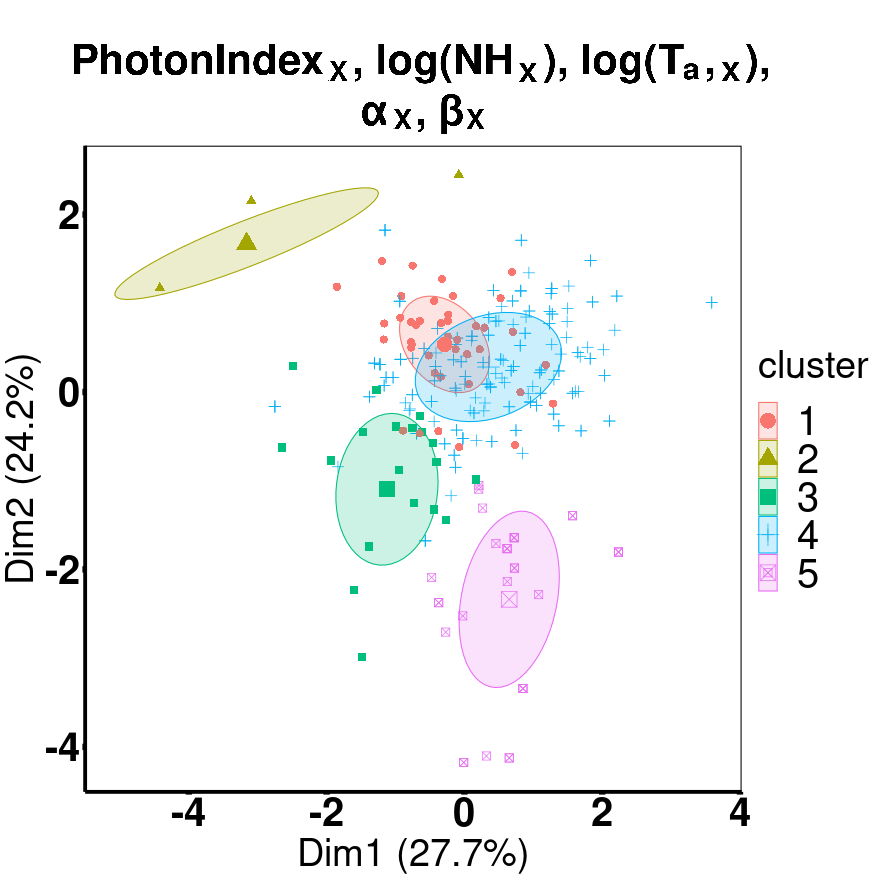}
        \includegraphics[width=0.50\textwidth]{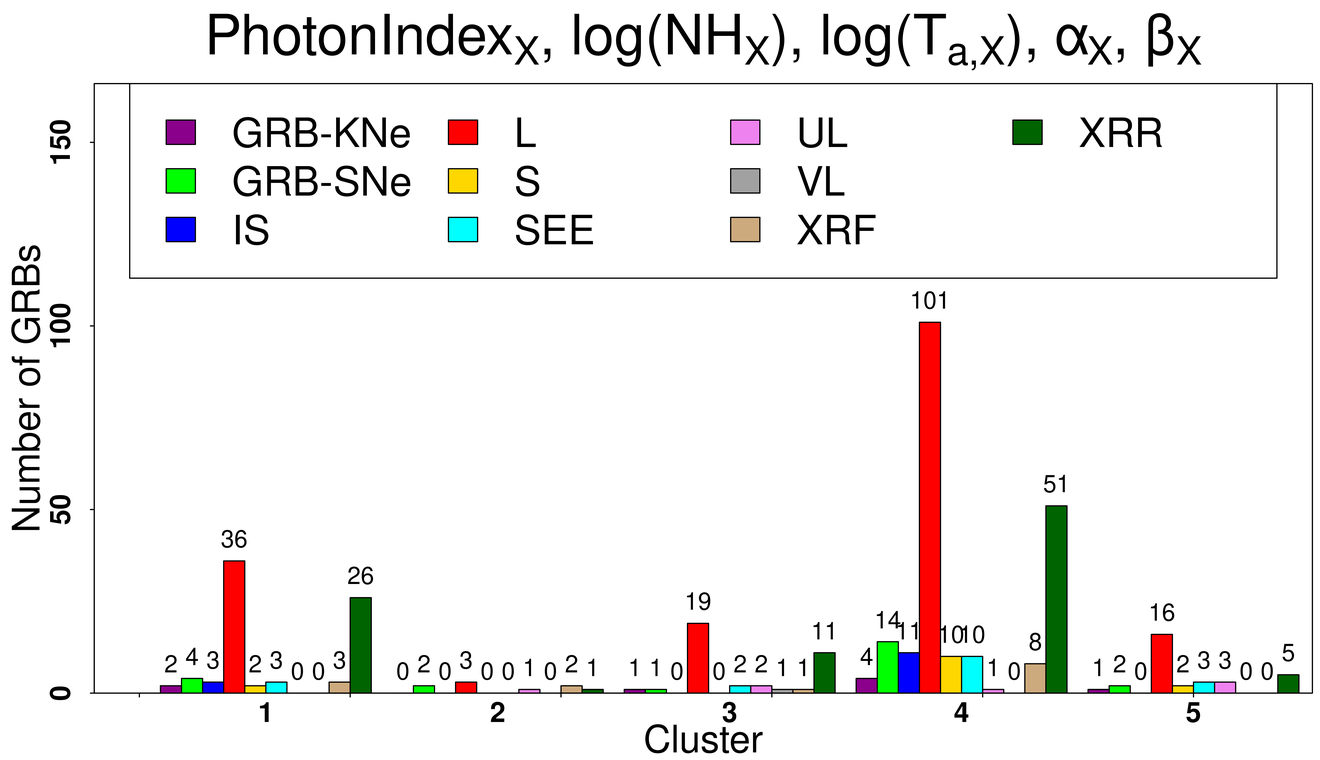}
    \end{minipage}
    \caption{\small
    This figure shows the second set of four clustering results obtained for the X-ray sample using GMM for the parameter space mentioned in (v), (vi), (vii), and (viii) of Sec. \ref{xray_res}. First row: clustering of GRBs using a combination of $\log(T_{90,X})$, $\log(F_{a,X})$, $\log(T_{a,X})$, $\alpha_{X}$, and $\beta_{X}$. Second row: clustering of GRBs using a combination of $\log(NH_{X})$, $\log(F_{a,X})$, $\log(T_{a,X})$, $\alpha_{X}$, and $\beta_{X}$. Third row: clustering of GRBs using a combination of $\log(Fluence_{X})$, $\log(NH_{X})$, $\log(F_{a,X})$, $\alpha_{X}$, and $\beta_{X}$. Fourth row: clustering of GRBs using a combination of $PhotonIndex_{X}$, $\log(NH_{X})$, $\log(T_{a,X})$, $\alpha_{X}$, and $\beta_{X}$.}
    \label{fig:xr_gls}
\end{figure*}

\begin{figure*}
    \begin{minipage}{\textwidth}
        \textbf{(a)}
        \centering
        \includegraphics[width=0.3\textwidth]{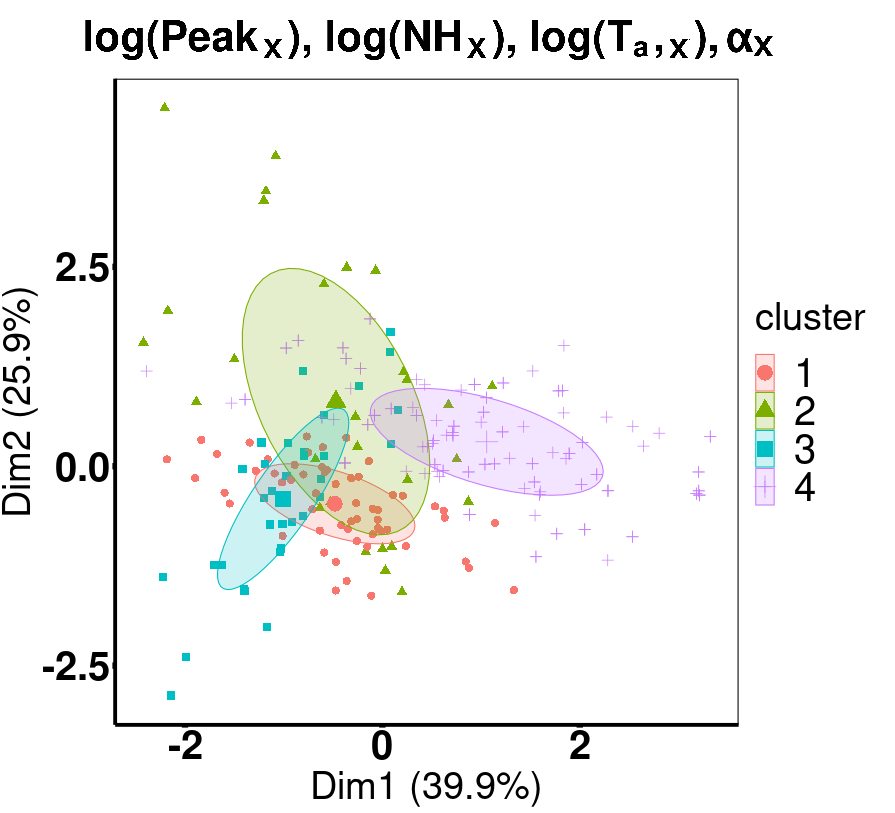}
        \includegraphics[width=0.52\textwidth]{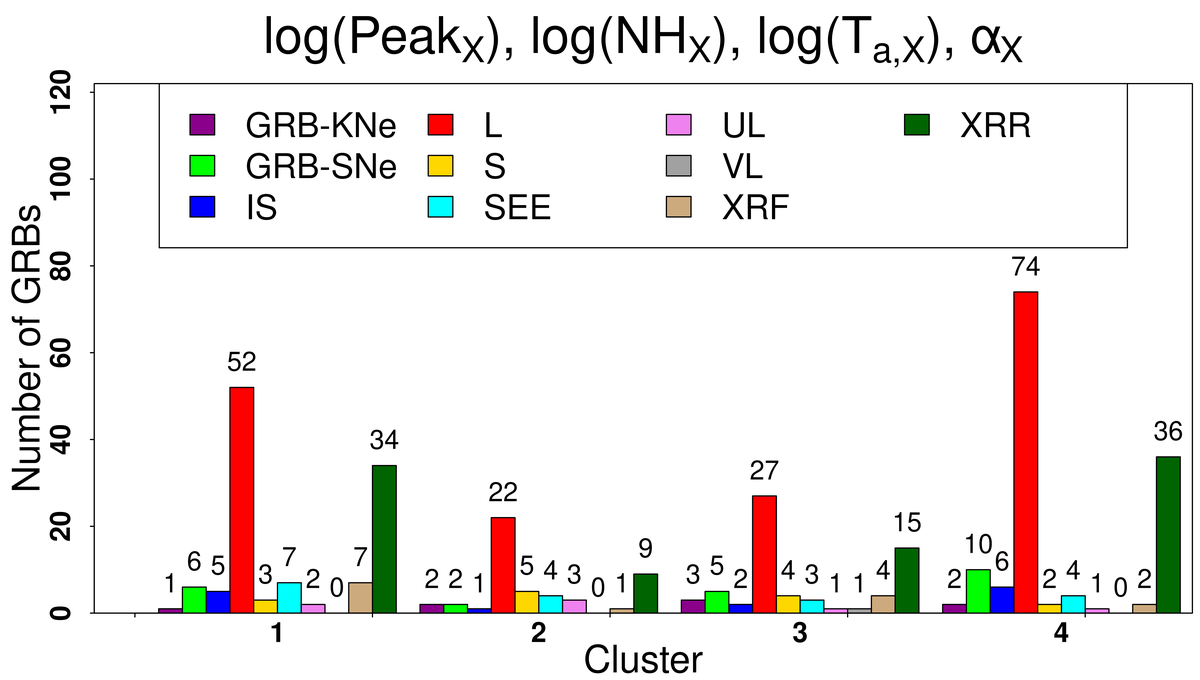}  
    \end{minipage}
    \begin{minipage}{\textwidth}
        \textbf{(b)}
       \centering
        \includegraphics[width=0.3\textwidth]{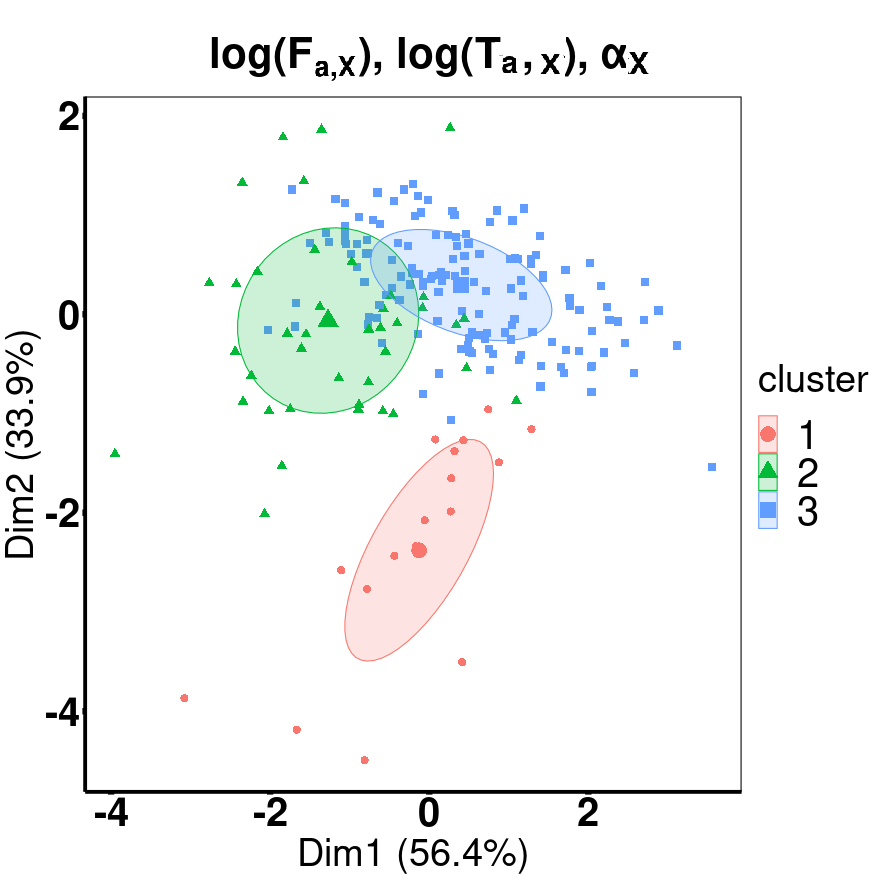}
        \includegraphics[width=0.52\textwidth]{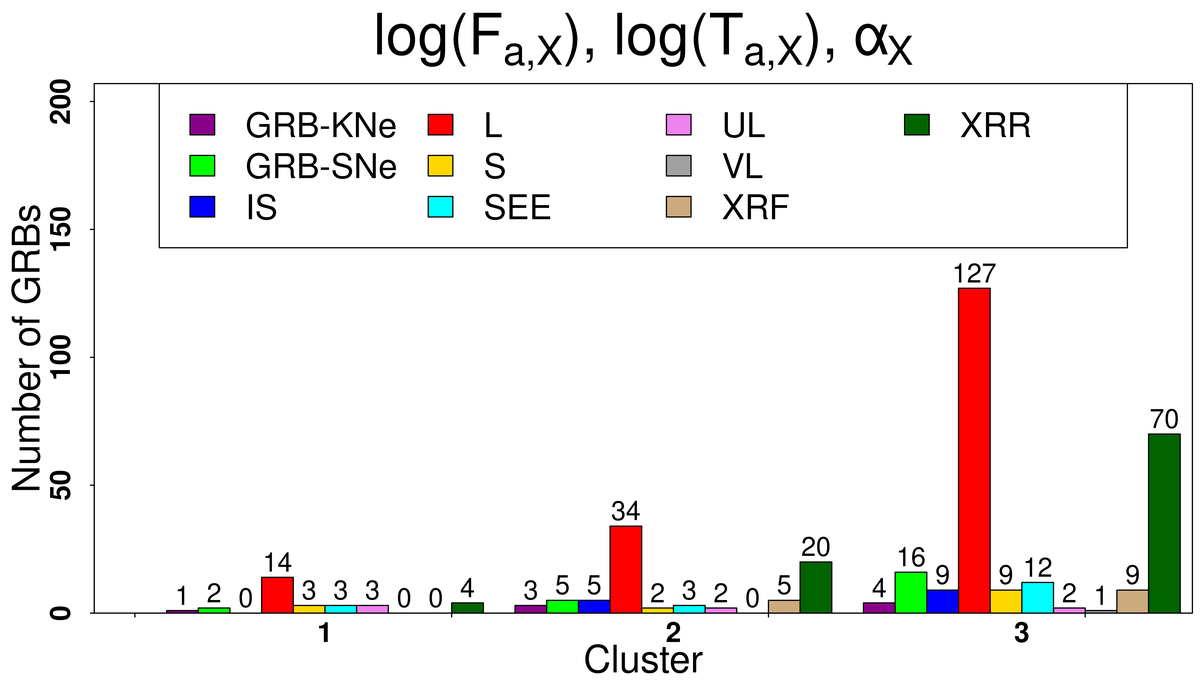}  
        \end{minipage}
    \begin{minipage}{\textwidth}
        \textbf{(c)}
       \centering
        \includegraphics[width=0.3\textwidth]{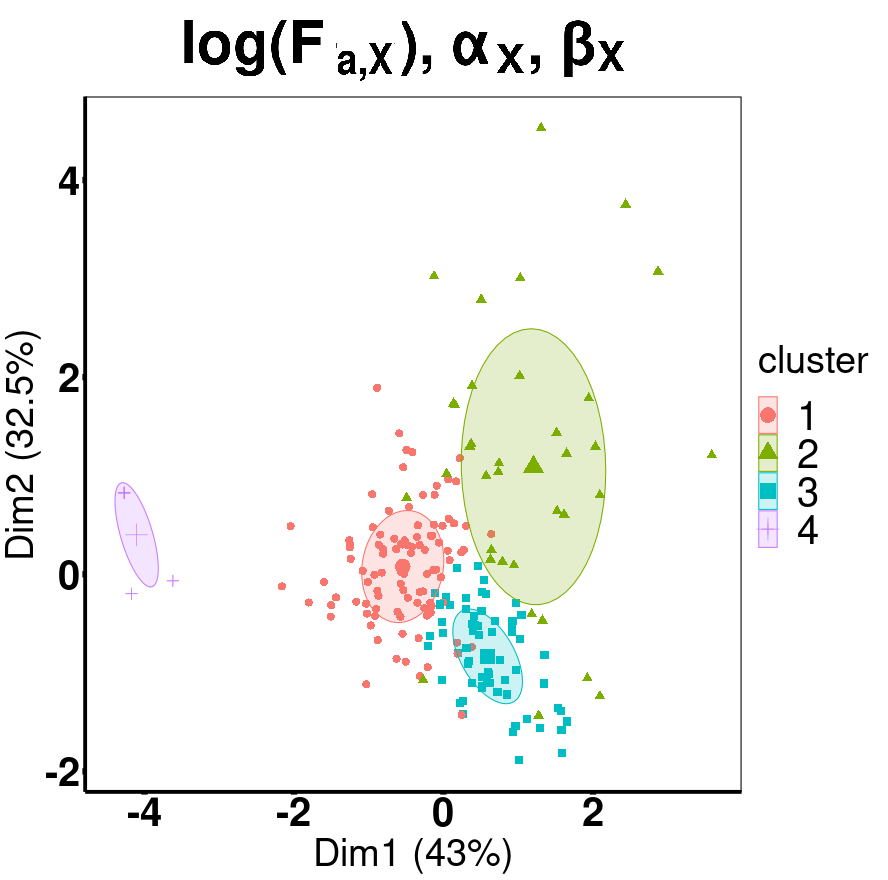}
        \includegraphics[width=0.52\textwidth]{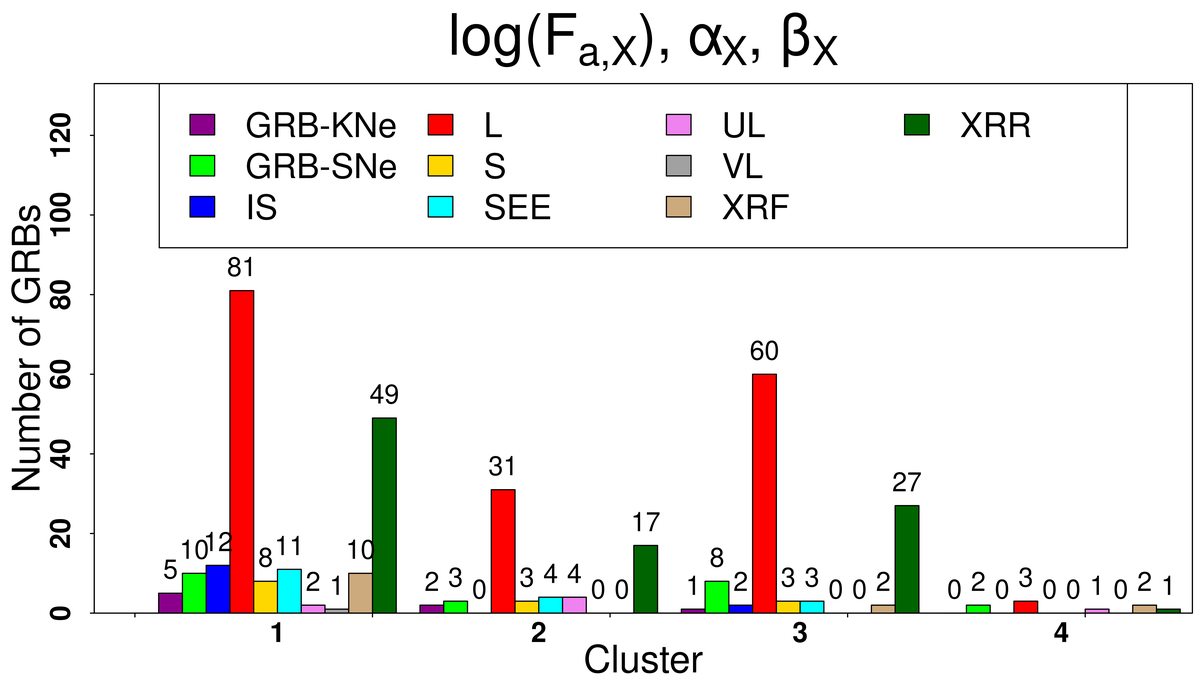}  
        \end{minipage}
    \begin{minipage}{\textwidth}
        \textbf{(d)}
       \centering
        \includegraphics[width=0.3\textwidth]{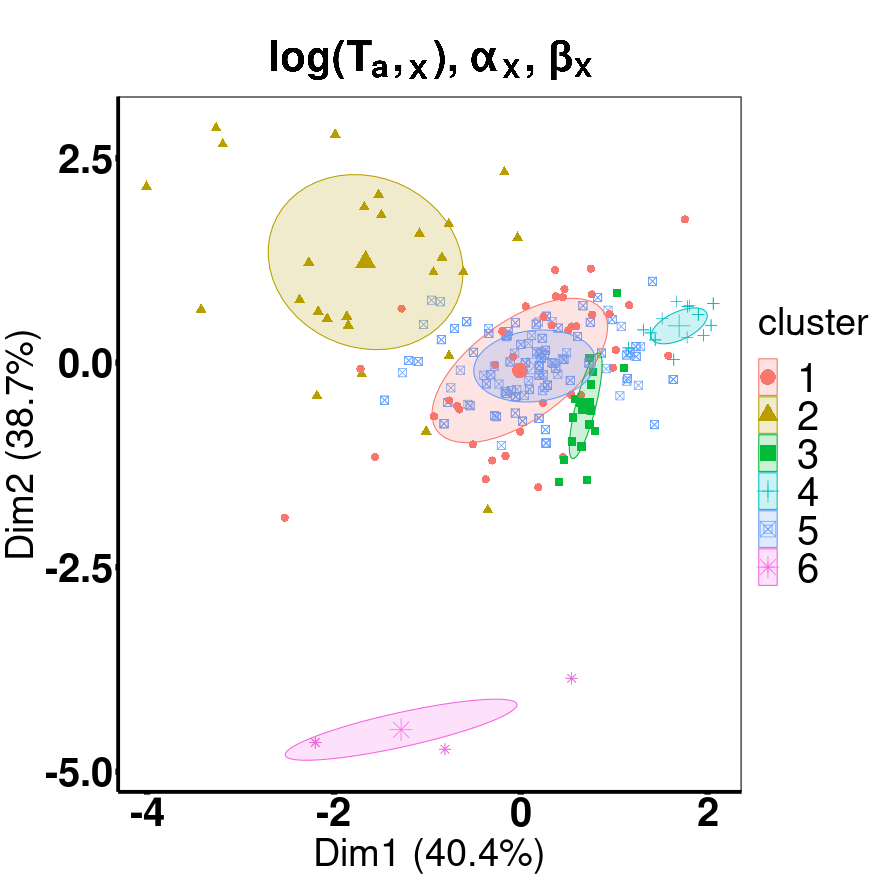}
        \includegraphics[width=0.52\textwidth]{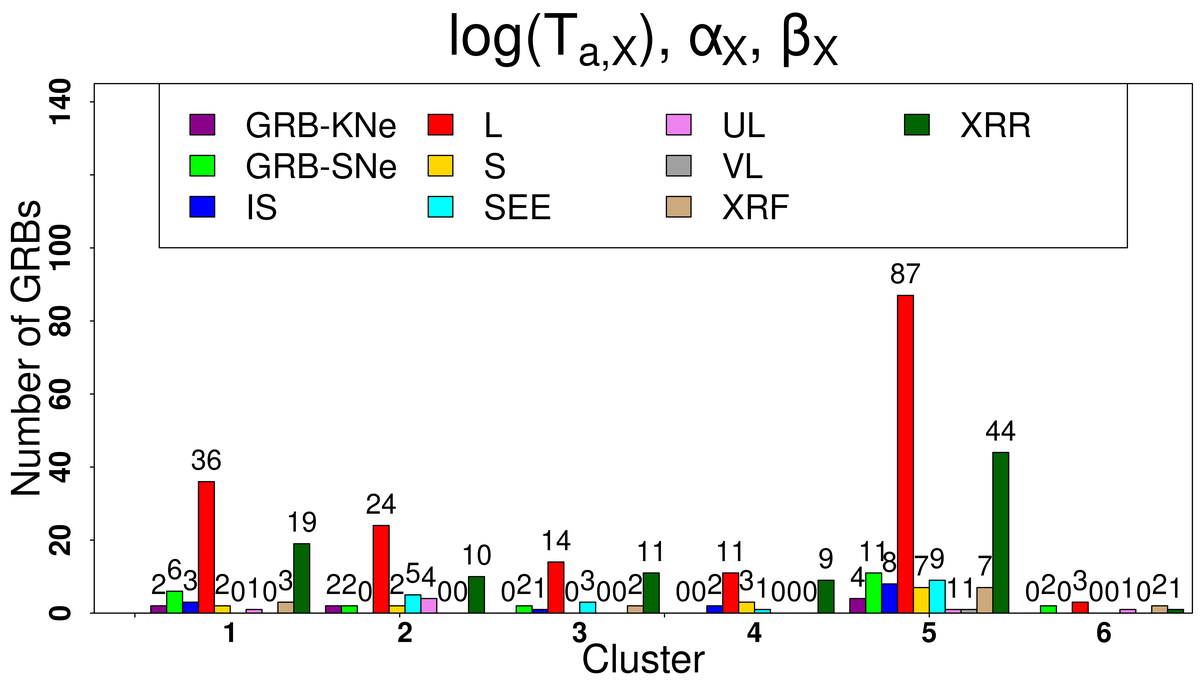}  
    \end{minipage}
    \caption{\small  
    This figure shows the third set of four clustering results obtained for the X-ray sample using GMM for the last four combinations stated in Sec. \ref{xray_res}. First row: clustering of GRBs using a combination of $\log(Peak_{X})$, $\log(NH_{X})$, $\log(T_{a,X})$, and $\alpha_{X}$. Second row: clustering of GRBs using only plateau parameters - $\log(F_{a,X})$, $\log(T_{a,X})$, $\alpha_{X}$. Third row: clustering of GRBs using only plateau parameters - $\log(F_{a,X})$, $\alpha_{X}$, $\beta_{X}$. Fourth row: clustering of GRBs using only plateau parameters - $\log(T_{a,X})$, $\alpha_{X}$, $\beta_{X}$.
    } 
    \label{fig:xr_gls_balanced}
\end{figure*}

\begin{enumerate}
    \item The cluster with the highest occurrence of LGRBs exhibits a majority of XRR, as depicted in the right panels of Fig. \ref{fig:xr_all} a, b, c, d, Fig. \ref{fig:xr_gls} a, b, c, d, and Fig. \ref{fig:xr_gls_balanced} a, b, c, d. The same trend can be seen in XRF as well (right panels of Fig. \ref{fig:xr_all} b, c, Fig. \ref{fig:xr_gls} b, c, d, and Fig. \ref{fig:xr_gls_balanced} b, c, d). This trend does not appear in Fig. \ref{fig:xr_all} a, d, Fig. \ref{fig:xr_gls} a, and Fig. \ref{fig:xr_gls_balanced} a. GRB-SNe also follows the same trend as XRR and XRF (right panels of Fig. \ref{fig:xr_all} a, b, c, d, Fig. \ref{fig:xr_gls} c, d, and Fig. \ref{fig:xr_gls_balanced} a, b, c, d). This GRB-SNe trend does not match in Fig. \ref{fig:xr_gls} a, b.
    \item {Interestingly, it is noteworthy that most ULs are not clustered together, with the highest number of LGRBs in the majority of cases. Instead, they tend to be predominantly found within a distinct cluster that does not exhibit the highest occurrence of LGRBs (right panels of Fig. \ref{fig:xr_all} a, b, c, d, Fig. \ref{fig:xr_gls} a, b, c, d, and Fig. \ref{fig:xr_gls_balanced} a, b, c, d).}
    \item {In agreement with our expectation SGRBs are found in a cluster that presents the lowest number of LGRBs (right panels of \ref{fig:xr_all} a, d, Fig. \ref{fig:xr_gls} a, and Fig. \ref{fig:xr_gls_balanced} a). In Fig. \ref{fig:xr_all} b and Fig. \ref{fig:xr_gls} b SGRBs are equally spread between the clusters. Contrary to our expectation, in Fig. \ref{fig:xr_all} c, Fig. \ref{fig:xr_gls} c, d, and \ref{fig:xr_gls_balanced} b, c, d, we notice that most SGRBs are found in a cluster with the highest LGRBs.}
    \item {In addition, most IS samples are also primarily found in the cluster with the highest concentration of LGRBs (see the right plots of Fig. \ref{fig:xr_all} b, c, Fig. \ref{fig:xr_gls} b, c, d, and Fig. \ref{fig:xr_gls_balanced} a, b, c, d). In Fig. \ref{fig:xr_all} d, it is equally spread in the highest and second-highest cluster. However, in agreement with our expectation, in Fig. \ref{fig:xr_all} a, and Fig. \ref{fig:xr_gls} a, most IS are primarily found in the cluster that does not have the highest number of LGRBs.} 
    \item {It is interesting to note that most SEE samples are always grouped within the cluster containing most of the LGRBs, which is counter-intuitive in accordance with our expectations because, in the literature, they are claimed to be similar to the SGRBs \citep{barkov2011model}. This trend can be clearly noticed in the right panels of Fig. \ref{fig:xr_all} a, b, c, d, Fig. \ref{fig:xr_gls} b, c, d, and Fig. \ref{fig:xr_gls_balanced} b, c, d. Agreeing with our intuition in Fig. \ref{fig:xr_gls} a and Fig. \ref{fig:xr_gls_balanced} a, we do not see SEE clustered with the highest LGRBs.}
\end{enumerate}

The microtrend findings (i), (ii), (iii), (iv), and (v) presented above lead to intriguing conclusions that have been emphasized in Sec.~\ref{summary and conclusion}.

\subsection{The microtrends using MICE}\label{mice_results}

To further validate our findings and explore potential microtrends, we expanded our analysis by utilizing the MICE imputation technique (Sec. \ref{MICE}) to increase the sample size. We present a subset of the same parameter combinations presented previously for the optical and X-ray samples.

\begin{figure*}
   \centering
        \includegraphics[width=0.3\textwidth]{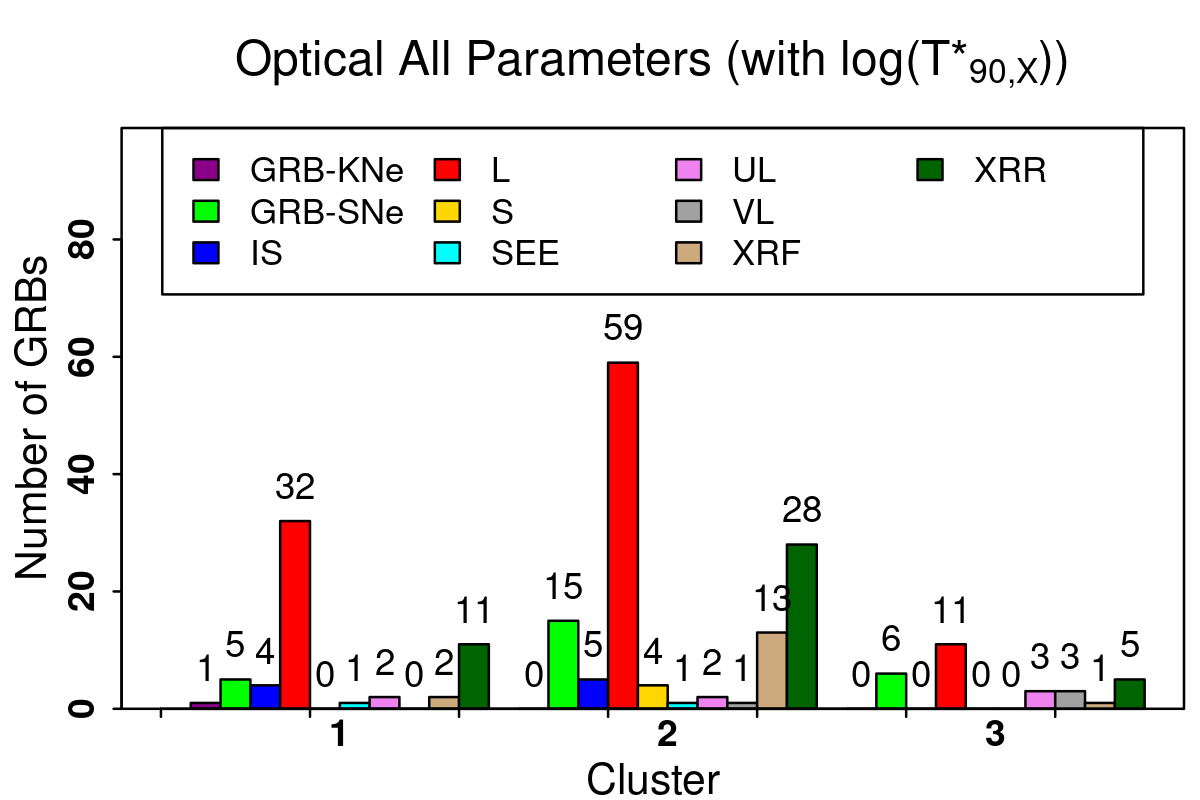}
        \includegraphics[width=0.34\textwidth]{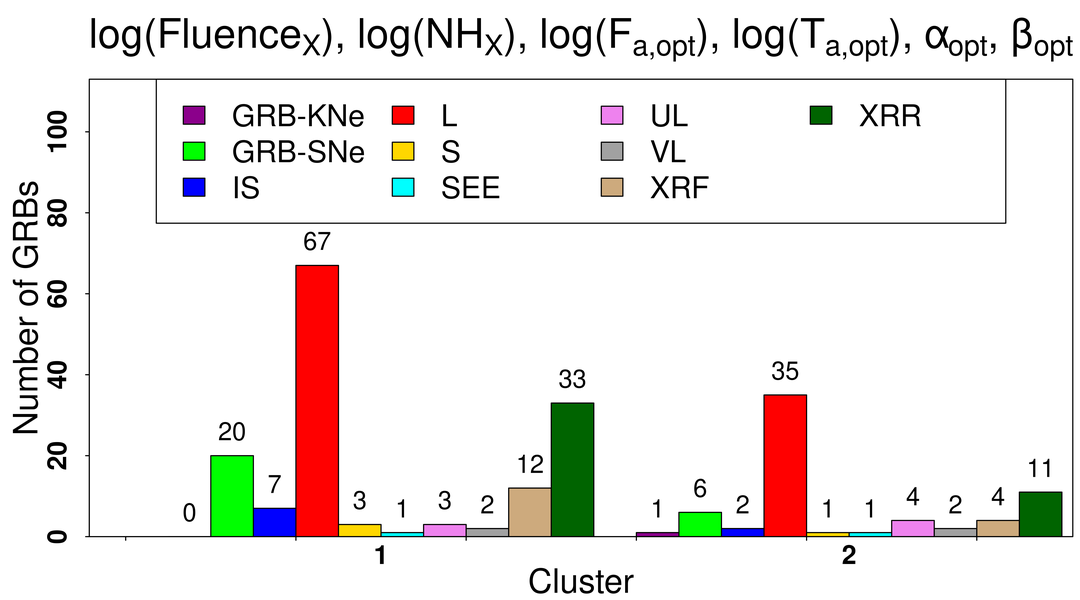} 
        \includegraphics[width=0.34\textwidth]{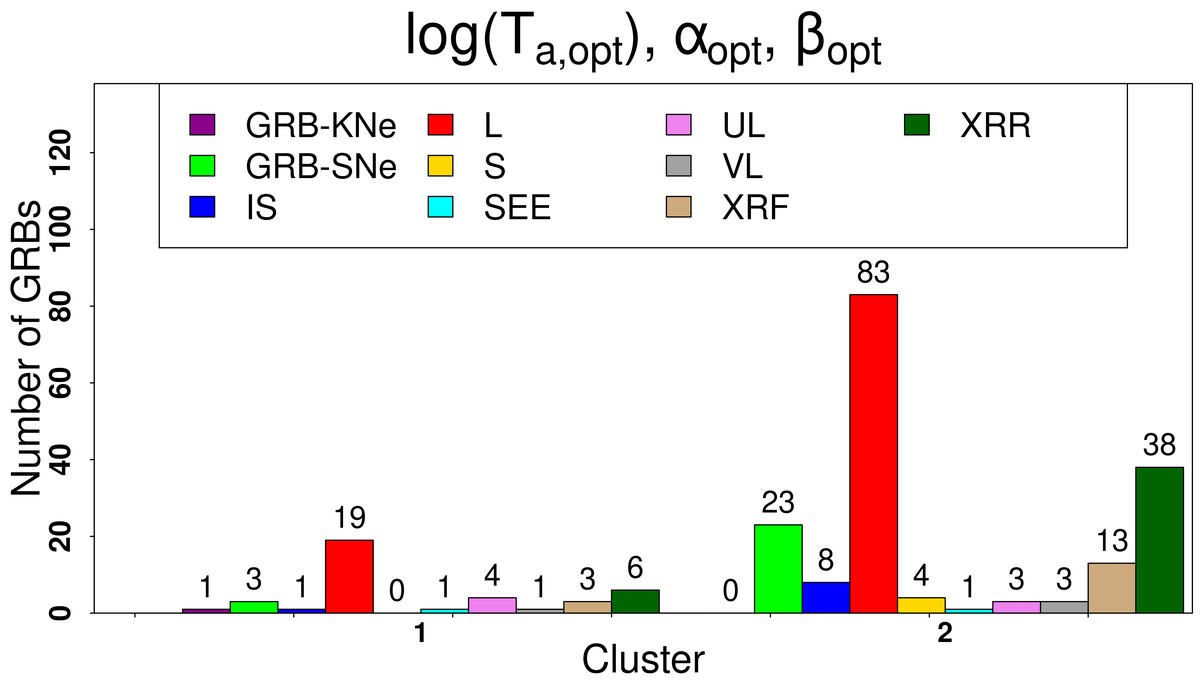}  
   \centering
        \includegraphics[width=0.3\textwidth]{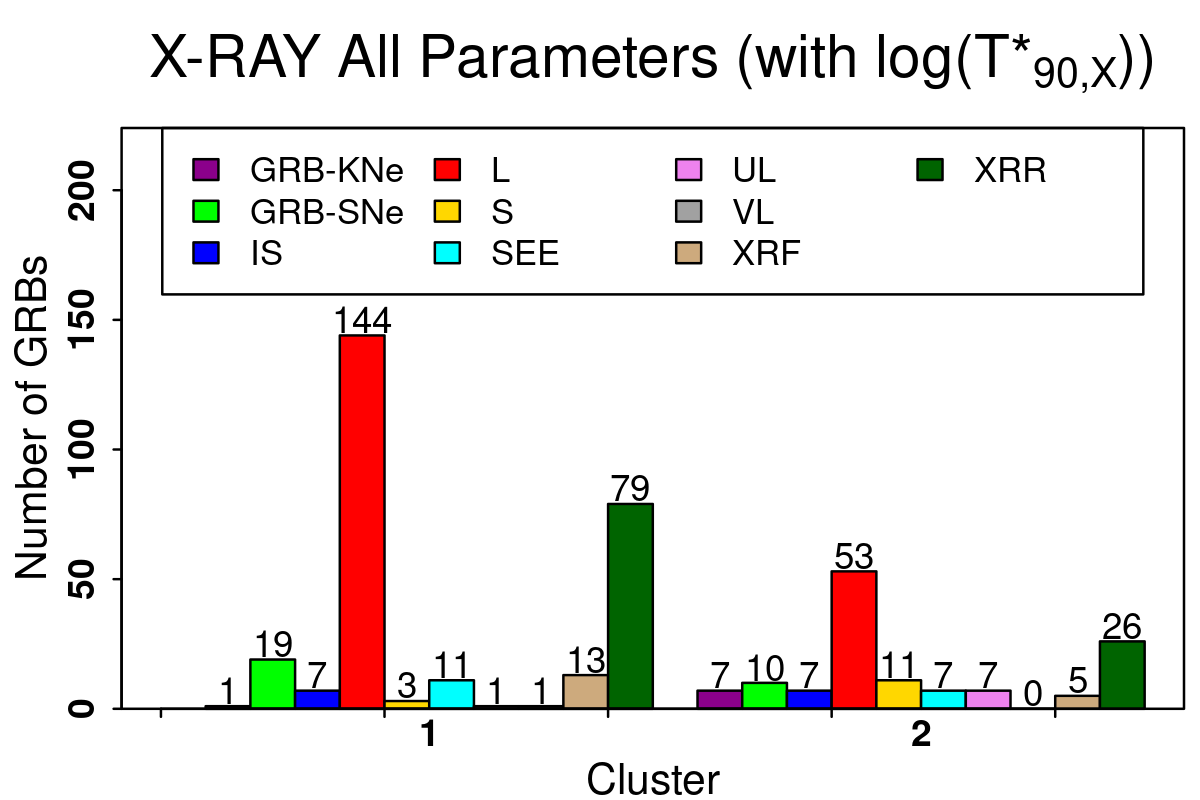}
        \includegraphics[width=0.34\textwidth]{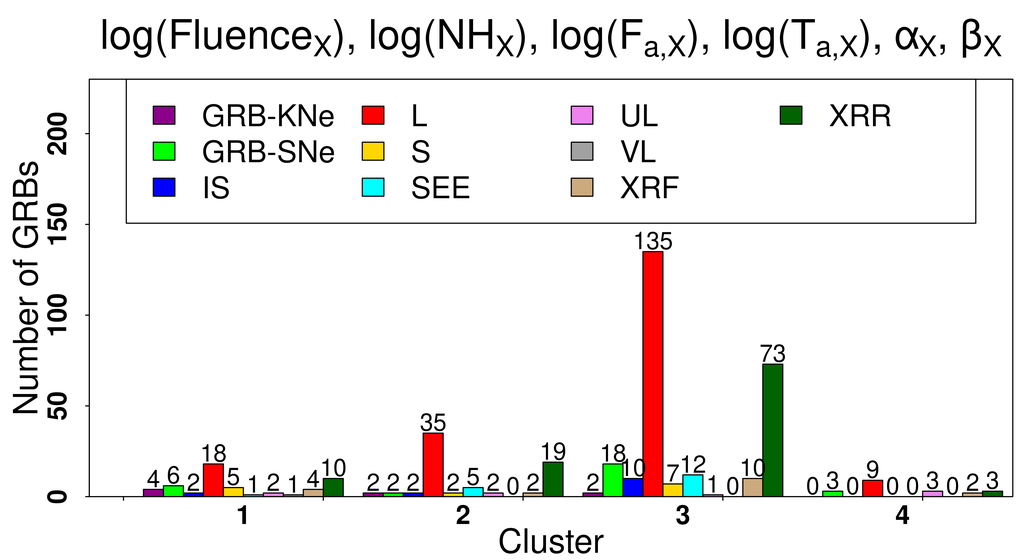} 
        \includegraphics[width=0.34\textwidth]{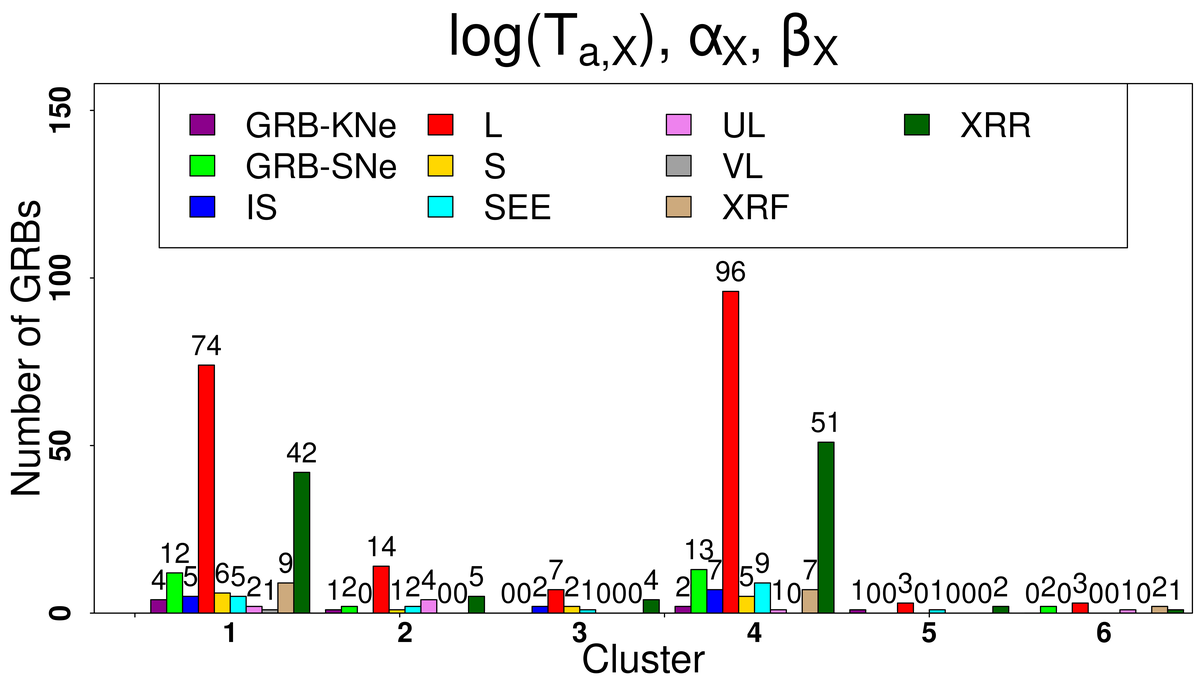} 
    \caption{\small  
    This figure shows the histogram distribution of GRB classes with the MICE-imputed data. The top three panels show the MICE-imputed optical data results. The bottom three panels show the MICE-imputed X-ray data results.} 
    \label{fig:mice_histograms}
\end{figure*}

\begin{enumerate}

    \item {For our optical sample, we imputed 45 GRBs (24.14\% of the full data set) that have four missing properties:  $\log(NH_{X})$, $\log(Peak_{X})$, $\log(Fluence_{X})$, and the $PhotonIndex_{X}$. The results for the clustering are depicted in the upper three panels of Fig. \ref{fig:mice_histograms}. We observe almost the same microtrends depicted in Sec. \ref{opt_mcirotrends} even after increasing our sample size, namely, most of the XRR, XRF, and GRB-SNe belongs to the cluster with the majority of LGRBs. However, contrary to our expectations, most IS and four SGRBs are clustered with the majority of LGRBs. On the other hand, most of the ULs are found in a different cluster from the one with the highest LGRBs. They are found in the cluster with the lowest LGRBs.}
    \item We repeat the same imputation for the X-ray data, where we imputed 19 GRBs (8.56\% of the full data set) which had $\log(Peak_{X})$ and $\log(NH_X)$ missing (as detailed in Sec. \ref{xray data sample}). The results for this data set are presented in the lower three panels of Fig. \ref{fig:mice_histograms}. We observe almost the same microtrends (Sec. \ref{xray_microtrends}) here. Namely, most XRR and GRB-SNe are found in the cluster with the majority of LGRBs. In addition, the left and middle figures in the bottom panel show that most of the XRFs are also found in the cluster which has the highest number of LGRBs. In contrast, SEE samples are also grouped within the cluster with most LGRBs. SGRBs cluster with the lowest occurrence of LGRBs in the left figure of the bottom panel. IS cluster with the highest LGRBs, except for the left figure in the bottom panel, where it is equally distributed among the two clusters. Interestingly, most ULs are concentrated in a cluster distinct from the cluster with the highest number of LGRBs. In the left figure of the bottom panel, ULs are found in the cluster with the lowest LGRBs.
\end{enumerate}

\section{Discussion}
\label{discussion}

Here we highlight some interesting points that can be inferred from our results (Sec. \ref{results}).

\begin{enumerate}
   \item GMM's application on our optical and X-ray datasets results in two or more than two clusters (Table \ref{BIC Table}). 
   However, in both the data samples, we do not obtain a scenario where we have the same number of clusters as the morphological classes presented in our data (Fig. \ref{fig:pie} c, d).
   \item The X-ray cluster plots (Sec. \ref{opt_res}) show a higher number of clusters than the optical results (Sec. \ref{xray_res}) for the same parameter space presented here. 
   It is worth noting that a higher number of clusters does not necessarily imply a distinct distribution of the various GRB classes.
   \item Point (i) stated in Sec. \ref{opt_mcirotrends} and Sec. \ref{xray_microtrends} suggests a potential similarity between the progenitors of XRR, XRF, and GRB-SNe with those of LGRBs.
   \item Point (ii) stated in Sec. \ref{opt_mcirotrends} and Sec. \ref{xray_microtrends} may indicate that the progenitors of ULs differ from those of LGRBs.
   \item The counter-intuitive results stated in points (iii) and (iv) of Sec. \ref{opt_mcirotrends} could be attributed to the influence of a small sample size, or different techniques may be adopted. Similar results are also mentioned in points (iii) and (iv) of Sec. \ref{xray_microtrends}.
   \item Our study uses the parameters from the plateau phase derived by fitting the LCs of the GRBs. However, testing the LCs in greater detail to understand the origin of the plateau phase is beyond the scope of this paper. Based on our microtrend results, XRRs and XRFs are always clustered according to the highest number of LGRBs. This clustering tendency suggests a potential inclination towards off-axis emission scenarios during the plateau phase for both XRR and XRF GRBs \citep{2009A&A...499..439G, 2019ApJ...871..123F}. Since ULs are always found in a different cluster than the highest LGRBs, thereby indicating a possible preference for sub-relativistic materials (such as Cocoons, etc., \citep{Piro2014, 2019ApJ...871..123F}) as a plausible scenario during the plateau phase for ULs.
   \item Distinguishing the central engine mechanism and pointing toward which scenario, the fallback accretion of matter onto a black hole or the spin-down luminosity from a newborn magnetar, for GRBs that do not satisfy the closure relations, is challenging since the data sample would be even more reduced than the current one.
   \item According to our analysis and micro trend results, the high-energy gamma-ray plateau properties sample can also be used for this study. The problem here is that we have only 4 GRBs with gamma-ray plateau properties \citep{2020ApJ...905..112F, 2021ApJS..255...13D} from 2008 until May 2016 taken from the second Fermi-LAT GRB Catalog (2FLGC, \citep{2019ApJ...878...52A}). We need a significantly large data sample to study and explore if the high energy emission has the same origin as the optical and X-ray plateau properties. In this analysis, we have not added this sample because our main goal is to keep the sample as much as possible homogeneous.   
\end{enumerate}

\section{Summary and Conclusion}
\label{summary and conclusion}

It has been established in the literature that there are many different types of GRBs and that they differ in their physical properties (Sec. \ref{sec:intro}). In this analysis, we applied GMM to optical and X-ray GRB samples, exploring multiple combinations of the plateau and prompt parameters in order to analyze which GRB properties can give a clear and distinct clustering. Our investigation here {aims to find} statistical methods that can better classify {GRBs, thus leading} to a deeper understanding of their physical properties utilizing the plateau properties for the first time. Ideally, a clustering algorithm would generate `N' clusters, where `N' would reliably represent the number of physical classes in the data set. However, performing unsupervised clustering of our optical and X-ray data with GMM, we obtained no definitive clustering results where the GRBs were separated into distinct clusters based on their classes. Probing further, we observe certain microtrends, which we have highlighted in Sec. \ref{opt_mcirotrends} and Sec. \ref{xray_microtrends}. However, with these microtrends, we cannot pinpoint which GRB property plays an important role and which does not in driving these microtrends and can be investigated further. The conclusions that can be drawn from our results are presented below:

\begin{itemize}
    \item From our analysis, considering both the cluster and distribution plots, we do not see a clear clustering of GRBs in either the optical or X-ray GRB samples (Sec. \ref{opt_res} and Sec. \ref{xray_res}). However, we found some GRBs belonging to the same classes appear in different clusters (Sec. \ref{opt_mcirotrends} and Sec. \ref{xray_microtrends}). This indicates that despite the advantages of GMM, it could not adequately infer information from our datasets to obtain optimal results.
    \item We found that XRR, XRF, and GRB-SNe are clustered with LGRBs in both optical and X-ray results (Sec. \ref{opt_mcirotrends} (i), and Sec. \ref{xray_microtrends} (i), respectively). This indicates that XRR, XRF, and GRB-SNe may originate from the same progenitor as of LGRBs or the same progenitors as of LGRBs but with different environments. This indeed tells us that they are sub-classes of LGRBs, as already stated in the literature \citep{soderberg2004redshift, Woosley2006ARA&A, 2007PASJ...59..695A, chincarini2010MNRAS.406.2113C, 2014A&A...567A..29M, 2016ApJS..224...20Y, bi2018statistical, 2019ApJ...884...59L}.
    \item We also observe ULs clustered separately from LGRBs in most optical and X-ray results (Sec. \ref{opt_mcirotrends} (ii) and Sec. \ref{xray_microtrends} (ii), respectively). This highlights that ULs might be a separate subclass of GRBs with a different progenitor that is highlighted using plateau properties. \citet{Gendre_2013}, \citet{Levan2014}, \citet{Piro2014}, \citet{Boër_2015}, \citet{2018ApJ...859...48P}, and \citet{aloy2021MNRAS.500.4365A} also claim that ULs might have a different progenitor than conventional LGRBs. They are thought to originate from the low metallicity blue supergiant or the collapse of stars with much larger radii than those attributed to GRB progenitors. This leads to the point that plateau properties can distinguish between LGRBs and ULs.
    \item Applying GMM to the MICE imputed optical and X-ray samples, we see similar microtrends described in Sec. \ref{mice_results}. This gives credence to the fact that these microtrends potentially reflect underlying physics that can be explored using plateau properties. With this, we can infer which plateau properties are essential in conjunction with the prompt properties in classifying GRBs.
    \item Previous literature \citep{tsutsui2014,zhang2016,2017MNRAS.469.3374C} have already showcased results using GMM. However, the novelty of our work is that we also cross-verified and validated the results obtained from GMM with the observationally assigned GRB classes, which has hardly been explored in previous literature. Thus, these results can serve as a benchmark for further studies in this field and contribute to the improvement of clustering techniques.
    \item We replicated the results of \cite{tsutsui2014} and found that GMM can cluster small datasets with a significant number of each class distinctly. However, a dataset whose composition is highly skewed (Fig. \ref{fig:pie}), as in our case, leads to outliers and overlapping clusters, resulting in uncertainty over the classes of the clustered GRBs (Sec. \ref{results}). Thus, to gain a clear understanding of GRB clustering based on plateau properties, a larger dataset with a more balanced distribution of samples in each GRB category is necessary. We expect more extensive and accurate samples of GRBs from future satellite and ground-based surveys. The new data will be available with additional ML analysis to infer the redshift information and to determine the plateau emission properties more precisely by implementing precise LC reconstruction methods \citep{DainottiLCsubmitted}.
    \item On the other hand, using a different set of ML techniques with additional parameter combinations might also reveal hidden patterns in GRB classes that GMM is currently unable to discover. 
\end{itemize}

\section*{Acknowledgements}

This research was supported by the Visibility and Mobility module of the Jagiellonian University (Grant number: WSPR.WSDNSP.2.5.2022.5) and the NAWA STER Mobility Grant (Number: PPI/STE/2020/1/00029/U/00001). A.N. and M.G.D. are grateful to the Exploratory Research Fund for the financial support for the visit of A.N. to the Division of Science at the National Astronomical Observatory of Japan (NAOJ). A.N. is also extremely grateful to NAOJ for providing all the essential amenities during the visit to the NAOJ. S.B. is grateful to the NAOJ Associate Researcher Fellowship for supporting this research. This work was also supported by the Polish National Science Centre grant UMO-2018/30/M/ST9/00757 and by the Polish Ministry of Science and Higher Education grant DIR/WK/2018/12.

\section*{Data Availability}

Data in this paper have been downloaded from the Swift BAT+XRT repository. This work made use of data provided by the UK Swift Science Data Centre located at the University of Leicester. The data are taken from the following two articles: \cite{Srinivasaragavan2020} and \cite{2022ApJS..261...25D}


\bibliographystyle{mnras}
\bibliography{ref} 

\bsp	
\label{lastpage}
\end{document}